\documentclass[sigconf]{acmart}
\AtBeginDocument{%
  }

\setcopyright{cc}
\copyrightyear{2026}
\acmYear{2026}
\acmDOI{XXXXXXX.XXXXXXX}
\acmConference[arXiv '26]{arXiv '26}{September,
  2026}{}
\acmBooktitle{arXiv '26}

\usepackage{caption}
\usepackage{subcaption}

\usepackage{hyperref}
\usepackage[capitalise,noabbrev]{cleveref}

\begin{document}

\title{MultiGait: A Multi-Sensor Multi-Perspective Multi-Session Biometric Inference Benchmark and its Dataset}

\author{Julian Todt}
\affiliation{%
 \department{KASTEL Security Research Labs}
 \institution{Karlsruhe Institute of Technology}
 \streetaddress{Am Fasanengarten 5}
 \city{}
 \state{}
 \country{}
 \postcode{76131}
}
\email{julian.todt@kit.edu}

\author{Felix Morsbach}
\affiliation{%
 \department{KASTEL Security Research Labs}
 \institution{Karlsruhe Institute of Technology}
 \streetaddress{Am Fasanengarten 5}
 \city{}
 \state{}
 \country{}
 \postcode{}
}
\email{felix.morsbach@kit.edu}

\author{Philip Dissert}
\affiliation{%
 \department{KASTEL Security Research Labs}
 \institution{Karlsruhe Institute of Technology}
 \streetaddress{Am Fasanengarten 5}
 \city{}
 \state{}
 \country{}
 \postcode{}
}
\email{philip.dissert@student.kit.edu}

\author{Thorsten Strufe}
\affiliation{%
 \department{KASTEL Security Research Labs}
 \institution{Karlsruhe Institute of Technology}
 \streetaddress{Am Fasanengarten 5}
 \city{}
 \state{}
 \country{}
 \postcode{}
}
\email{thorsten.strufe@kit.edu}

\renewcommand{\shortauthors}{Todt et al.}

\begin{abstract}
A lack of suitable datasets has limited the research into the privacy risks of novel smart city sensors, such as thermal cameras, depth cameras, and lidar.
Given the number of unsubstantiated privacy claims and their potential widespread deployment into many people's everyday life, understanding the privacy risks of these sensors -- in isolation and in like-for-like comparisons -- is crucial.
With \emph{MultiGait}, we collected the first multi-sensor, multi-perspective, multi-session gait-focused dataset, for the corresponding, and additional more far-reaching investigations.
The dataset, validated with multiple state-of-the-art recognition systems, comprises various walking modes and annotated personal attributes for 199 individuals, to ensure the benefit for advanced studies including cross-sensor recognition and anonymization at the edge.
\emph{MultiGait} represents a foundation for rigorous privacy investigations, demonstrated through an extensive identity inference benchmark across eight sensors, four perspectives, and three recording sessions.
Our benchmark incidentally reveals that sensors often assumed to be privacy-friendly do still entail considerable identity inference risks, while the poor cross-session generalization of existing methods underscores an important research gap.
\end{abstract}



\received{20 February 2007}
\received[revised]{12 March 2009}
\received[accepted]{5 June 2009}

\maketitle

\section{Introduction}
Ubiquitous sensor systems promise to make smart cities an efficient and safe public space.
While their utility, such as the chance to observe potential crimes and culprits, is the primary goal of their deployment \cite{mohanty_everything_2016, lacinak_smart_2017, troisi_managing_2022}, there are also concerns:
An abundance of sensors means that any human participating in public life will be recorded by them.
These recordings contain rich biometric information, and even just gait can potentially be used to infer identities, activities, and sensitive attributes such as emotional or medical conditions \cite{wan_survey_2019, sepas-moghaddam_deep_2023, shen_comprehensive_2025, slemensek_human_2023, nixon_automatic_2006}.
This means that the sensors deployed in public spaces are powerful equipment for potential surveillance and raise significant privacy concerns.

These concerns are exacerbated when video cameras are replaced with other sensors, such as thermal or depth cameras, radar, or lidar, with the claim that those are more privacy-friendly -- particularly in comparison.
These unsubstantiated claims by both research \cite{gade2016thermal, collini_flexible_2024, baghezza_profile_2022, vales_iot_2024, gad_total_2025, yamaguchi_human_2018, zhao_mid_2019} and practice \cite{lynred_privacyLWIR_nodate, axiscomm_privacyLWIR_2022, snap4city_privacyLWIR_nodate, nypd_thermal_2025, uspolice_radar_2015} potentially foster a false sense of privacy and unintended levels of surveillance in public spaces.
To inform the privacy-utility trade-off of the deployment of different (combinations of) sensors, a better understanding of their capabilities is needed.
However, the absence of multi-sensor, multi-session datasets makes it difficult to rigorously evaluate and directly compare the privacy risks and utility trade-offs of different sensors.
Novel datasets can help to mitigate this by providing a principled foundation for rigorous benchmarking and systematic analysis, thereby informing evidence-based technology deployment.
This is a key requirement for privacy-respecting smart cities with privacy by design at the sensor through well-informed deployment and anonymization at the edge. %

The availability of extensive datasets of individuals captured with (visual-light) cameras has helped to establish their privacy risks at depth:
Facial images facilitate recognition of individuals with robust identification capabilities even at the scale of millions of humans~\cite{cao_vggface2_2018, deng_arcface_2019}.
Given full body videos, gait is claimed to be a highly distinctive biometric trait~\cite{hanisch_understanding_2023}, enabling identity, attribute and activity inference \cite{wan_survey_2019, sepas-moghaddam_deep_2023, shen_comprehensive_2025, slemensek_human_2023, nixon_automatic_2006}.
The thorough understanding of such privacy risks has resulted in research and practice to propose the use of alternative sensors, such as thermal and depth cameras, lidar, mmWave radar, and WiFi as a mitigation of privacy concerns~\cite{gade2016thermal, collini_flexible_2024, baghezza_profile_2022, vales_iot_2024, gad_total_2025, yamaguchi_human_2018, zhao_mid_2019, lynred_privacyLWIR_nodate, axiscomm_privacyLWIR_2022, snap4city_privacyLWIR_nodate, nypd_thermal_2025, uspolice_radar_2015}. %
There is, however, no rigorous research investigating their actual privacy risks.
A fundamental obstacle is the fact that existing datasets are limited in four ways (cmp. Table~\ref{tab:datasets} for an overview).
First, for some sensors only very small datasets exist \cite{tan_efficient_2006, xue_infrared_2010, decann_investigating_2013, zhao_mid_2019, meng_gait_2020, cao_lightweight_2021, song_casia-e_2023}. While they facilitated initial investigations into privacy risks, these are not sufficiently comprehensive due to their small size.
Second, with little exceptions, existing datasets only contain data from a single sensor type \cite{sarkar_humanid_2005, yu_framework_2006, tan_efficient_2006, okumura_performance_2010, makihara_ou-isir_2012, decann_investigating_2013, Xu_CVA2017, takemura_multi-view_2018, zhao_mid_2019, meng_gait_2020, cao_lightweight_2021, zhang_learning_2022, zheng_gait_2022, song_casia-e_2023}.
This significantly limits comparison of sensors, both in terms of privacy and regarding their benefit for arbitrary recognition tasks.
It has hence been hard to argue for or against certain sensors in favor over others for a given use case.
The lack of multi-sensor datasets represents an additional challenge in light of recent investigations that highlight more complex privacy risks, which emerge through cross-sensor identification \cite{guo_camera-lidar_2025, wang_cross-modality_2024}.
Third, a vast majority of datasets are only covering a single recording session \cite{yu_framework_2006, tan_efficient_2006, xue_infrared_2010, okumura_performance_2010, makihara_ou-isir_2012, iwama_ou-isir_2012, hutchison_re-identification_2012, decann_investigating_2013, andersson_person_2015, Xu_CVA2017, takemura_multi-view_2018, zhao_mid_2019, tavares_gridds_2019, meng_gait_2020, cao_lightweight_2021, zheng_gait_2022, li_-depth_2023, shen_lidargait_2023, song_casia-e_2023, todt_BFId_2025, zhu_gait_2025}. This represents a significant simplification from any real-world scenario and hinders both the development and evaluation of recognition systems.
Fourth, existing datasets are designed for either identity or activity inference, but not both \cite{okumura_performance_2010, makihara_ou-isir_2012, iwama_ou-isir_2012, hutchison_re-identification_2012, decann_investigating_2013, andersson_person_2015, Xu_CVA2017, takemura_multi-view_2018, zhao_mid_2019, tavares_gridds_2019, meng_gait_2020, cao_lightweight_2021, zheng_gait_2022, li_-depth_2023, zhu_gait_2025}.
The development and evaluation of anonymizations, which are critical tools to mitigate privacy risks, require both:
It is necessary to evaluate identity inference as a proxy of privacy risks, as well as an instantiation of utility (such as activity or attribute inference), to evaluate privacy-utility trade-offs.

To enable investigations into the research gaps and overcoming the described limitations of existing datasets, we offer a new multi-session, multi-sensor, and multi-perspective gait-focused dataset that we call \emph{MultiGait}.
We recorded 199 individuals simultaneously and synchronized with video, depth and thermal (near- \& longwave-infrared) cameras, mmWave radar, lidar and WiFi (CSI \& BFI) from four perspectives each, yielding $32$ different information sources per individual.
Each recording includes five different activities and five poses, and is annotated with personal attributes.
The dataset thereby enables identity, activity and attribute inference, all of which is extensively validated.
In addition, we recorded multiple sessions of 32\% of the subjects (14\% even three times). %
The entire dataset is available for academic use, facilitating further research in this area.

MultiGait enables investigations into a wide range of research questions that were previously impossible.
This includes rigorously evaluating each sensor's privacy risks, both individually and, in particular, in a like-for-like comparison to another.
In addition to single-session identification, as common in existing work, MultiGait also enables testing multi-session identification, which provides a more challenging and realistic setting.
To demonstrate, we conduct a biometric identification benchmark in which we use state-of-the-art recognition systems for all sensors and measure identity inference accuracy for all perspectives for both single and multi-session.
We show decisively that many previously-considered privacy-friendly sensors can actually induce a high identity inference risk.
The benchmark serves as a foundation for future research in this area, particularly considering the research gap that it demonstrates with regards to multi-session identification accuracy -- a gap that existing datasets failed to reveal.

To summarize, our contributions are as follows:
\begin{itemize}
    \item A gait-focused dataset of 199 subjects, captured by video, depth and thermal cameras (NIR \& LWIR), lidar, radar and WiFi (CSI \& BFI) -- each from multiple perspectives.
    \item A validation of this dataset, showing the viability of activity and attribute inferences.
    \item A benchmark comparing state-of-the-art identity inferences across different sensors, perspectives and sessions.
\end{itemize}

\section{Background}%
Data that contains information about an individual's behavioral or physiological attributes is called biometric data \cite{dantcheva_what_2016}.
It can be used to infer the identity of an individual, their attributes (e.g., age, sex, height or weight) or current activity \cite{dantcheva_what_2016}. %
Biometric traits include the face and gait -- the way we walk -- which has been shown to be particularly distinctive \cite{hanisch_understanding_2023}.
Considering the widespread data collection in (social) media, (smart) cities and more, privacy issues arise, for example, when biometric data is collected or processed without consent.

The inference potential of recordings from visual-light RGB video recordings (hereafter referred to as \textit{video cameras}) has been investigated extensively.
Both face and gait have been shown to enable identification with high accuracies using deep learning recognition systems \cite{cao_vggface2_2018, deng_arcface_2019, wan_survey_2019, sepas-moghaddam_deep_2023, shen_comprehensive_2025}.
We would like to note that though systems might claim gait-based recognition, it is not always clear to which extent other biometric (particularly physiological) traits might play a role, especially in end-to-end learning based systems.
Investigations into this are currently lacking due to the absence of relevant datasets that extend beyond (some) physiological attributes for example through multi-session.

Due to the privacy risks resulting from the high identification potential, proposals have been made to replace video cameras with alternative sensors that supposedly can achieve similar utility while reducing privacy risks.
Commonly mentioned sensors in this area include depth \cite{stone_evaluation_2011, planinc_introducing_2013} and thermal cameras \cite{lintvedt_thermal_2023, collini_flexible_2024, baghezza_profile_2022}, mmWave radar \cite{vales_iot_2024, zhao_mid_2019}, lidar \cite{yamaguchi_human_2018, gad_total_2025} and joint communication and sensing (JCAS) including WiFi sensing \cite{zeng_wiwho_2016, ma_wifi_2020, li_deep_2022}. %
At the same time, recognition systems that generally enable inferences via any of these sensors have been proposed (see \cref{sec:recsystems}).
Their exact inference potential, however, still remains unclear due to the limitations of current datasets, as seen in the following section.%

\section{Related Work}
\begin{table*}
  \caption{Overview over existing gait-focused datasets (chronologically). \mbox{\textsuperscript{a}type of environment:} \mbox{os = outside + supervised}, \mbox{ou = outside + unsupervised}, \mbox{i+o = inside and outside, supervised}; \textsuperscript{b}only 12 subjects; \textsuperscript{c}inconsistent between subjects (uncontrolled); \textsuperscript{d}manually annotated, no ground truth}
  \label{tab:datasets}
  \centering
  \setlength{\tabcolsep}{4.6pt}
  \begin{tabular}{@{}lcccccc|ccccccccc@{}}
    \toprule
    Dataset & Subjects & persp. & sessions & env.\textsuperscript{a} & act. & att. & video & depth & NIR & LWIR & lidar & radar & CSI & BFI & audio \\
    \midrule
    humanID \cite{sarkar_humanid_2005} & 122 & 2 & 2 & os                          & 3 & 7 & \checkmark\\
    CASIA B \cite{yu_framework_2006}     & 124 & 11 & 1 & lab                        & 2 & & \checkmark\\
    CASIA C \cite{tan_efficient_2006}    & 153 & 1 & 1 & lab                       & 4 & & & & & \checkmark\\
    Tianjin \cite{xue_infrared_2010}                 & 23 & 1 & 1 & lab         & 4 & 2 & \checkmark & & & \checkmark\\
    OU-ISIR Large \cite{okumura_performance_2010}          & 1,035 & 2 & 1 & lab  & 1 & 2 & \checkmark\\
    OU-ISIR C \cite{makihara_ou-isir_2012}   & 200 & 25 & 1 & lab                        & 1 & 2 & \checkmark\\
    OU-ISIR LP \cite{iwama_ou-isir_2012}     & 4,007 & 2 & 1 & lab                       & 1 & 2 & \checkmark\\
    PAVIS \cite{hutchison_re-identification_2012}  & 79 & 1 & 1 & lab      & 1 & & \checkmark & \checkmark\\
    WOSAG \cite{decann_investigating_2013}            & 155 & 4 & 1 & os             & 1 & 3 & & & \checkmark\\
    TUM-GAID \cite{hofmann_tum_2014}    & 305 & 1 & 2 & lab & 3 & 4 & \checkmark & \checkmark & & & & & & & \checkmark\\
    UFPEL \cite{andersson_person_2015}        & 140 & 1 & 1 & lab                 & 1 & & \checkmark & \checkmark\\
    OU-ISIR LP-Age \cite{Xu_CVA2017}          & 63,846 & 1 & 1 & lab                      & 1 & 2 & \checkmark\\
    OU-ISIR MVLP \cite{takemura_multi-view_2018}        & 10,307 & 7 & 1 & lab     & 1 & 2 & \checkmark\\
    mID \cite{zhao_mid_2019}                & 12 & 1 & 1 & lab      & 1 & 4 & & & & & & \checkmark\\
    mmGait \cite{meng_gait_2020}      & 95 & 4 & 1 & lab        & 1 & 4 & & & & & & \checkmark\\
    LW-WiID \cite{cao_lightweight_2021} & 50 & 1 & 1 & lab & 1 & & & & & & & & \checkmark\\
    FVG \cite{zhang_learning_2022}                    & 226 & 3 & 2\textsuperscript{b} & os     & 5 & & \checkmark\\
    Gait3D \cite{zheng_gait_2022}      & 4,000 & 39\textsuperscript{c} & 1 & ou    & 1\textsuperscript{c} & & \checkmark\\
    SUSTech1K \cite{shen_lidargait_2023}   & 1050 & 1 & 1 & os                         & 2 & & \checkmark & & & & \checkmark\\
    CASIA E \cite{song_casia-e_2023}         & 1,014 & 13 & 1 & os                       & 3 & 4 & \checkmark\\
    CASIA E LWIR \cite{song_casia-e_2023}    & 270 & 1 & 1 & os                          & 1 & & & & & \checkmark\\
    BFId \cite{todt_BFId_2025} & 197 & 4 & 1 & lab & 4 & & & & & & & & \checkmark & \checkmark \\
    GREW \cite{zhu_gait_2025}        & 26,345 & 882\textsuperscript{c} & 1 & ou  & 1\textsuperscript{c} & 2\textsuperscript{d} & \checkmark\\
    \midrule
    \textbf{MultiGait (ours)} & 199 & 4 & 3 & lab & 4 & 8 & \checkmark & \checkmark & \checkmark & \checkmark & \checkmark & \checkmark & \checkmark & \checkmark & \\
    \bottomrule
  \end{tabular}
\end{table*}

Several datasets containing some biometrics have been recorded in the past.
\cref{tab:datasets} presents an overview of existing gait-focused datasets that are commonly used in research. %
Due to the large number of small single-sensor single-perspective datasets, it is hard to present a comprehensive list, and we rather focus on datasets that have multiple sensors, perspectives, and/or sessions.

Datasets containing concurrent recordings of multiple sensors, for instance for comparison, are especially rare.
Some include both video and depth videos (due to Microsoft Kinects recording both), such as \cite{hutchison_re-identification_2012, andersson_person_2015}.
For any other combination of two sensors only a handful of datasets exist, such as Tianjin \cite{xue_infrared_2010} (video and LWIR) and SUSTech1K \cite{shen_lidargait_2023} (video and lidar). %
The only dataset with three sensors is TUM-GAID \cite{hofmann_tum_2014}, which contains video, depth and audio (all recorded via a single Microsoft Kinect).
This makes it challenging to effectively and fairly compare both the utility and privacy of different sensors or their combinations, and to investigate cross-sensor recognition, sensor fusion, and similar tasks.

Another significant limitation of existing datasets is the lack of multiple sessions.
Only TUM-GAID \cite{hofmann_tum_2014}, FVG \cite{zhang_learning_2022} and humanID \cite{sarkar_humanid_2005} record individuals across multiple sessions.
All other datasets dedicate recordings from the same session for both training and testing.
This is unrealistic, given that many factors, such as clothing or lighting, likely change between sessions.
To mimic real deployments, the split between training and testing should be based on sessions.

While some datasets contain recordings from multiple perspectives, the majority does not.
In some,  the angle towards the walking path is varied.
Filming is commonly kept parallel to the ground, but
rarely are cameras angled from above, like a ceiling-mounted surveillance camera in real environments.
Finally, some datasets \cite{zheng_gait_2022, zhu_gait_2025} impress with large sizes and real-world scenarios, but their uncontrolled setting means potentially imprecise manual annotations for ground truths.
Uncontrolled variations within the dataset also mean that the impact of the variations cannot be assessed.
This however is crucial when attempting to use this knowledge to plan future deployment of sensors.
These approaches also raise ethical concerns: because they simply record public spaces and only inform via signs of recordings for research, it is questionable to which extent this constitutes voluntary and informed participation.%

\section{MultiGait Dataset}
\begin{figure*}[tb]
    \centering
    \begin{subfigure}{\textwidth}
        \vspace{.5em}
        \centering
        \setlength{\tabcolsep}{1pt}\begin{tabular}{cccccc}
            \includegraphics[height=6em]{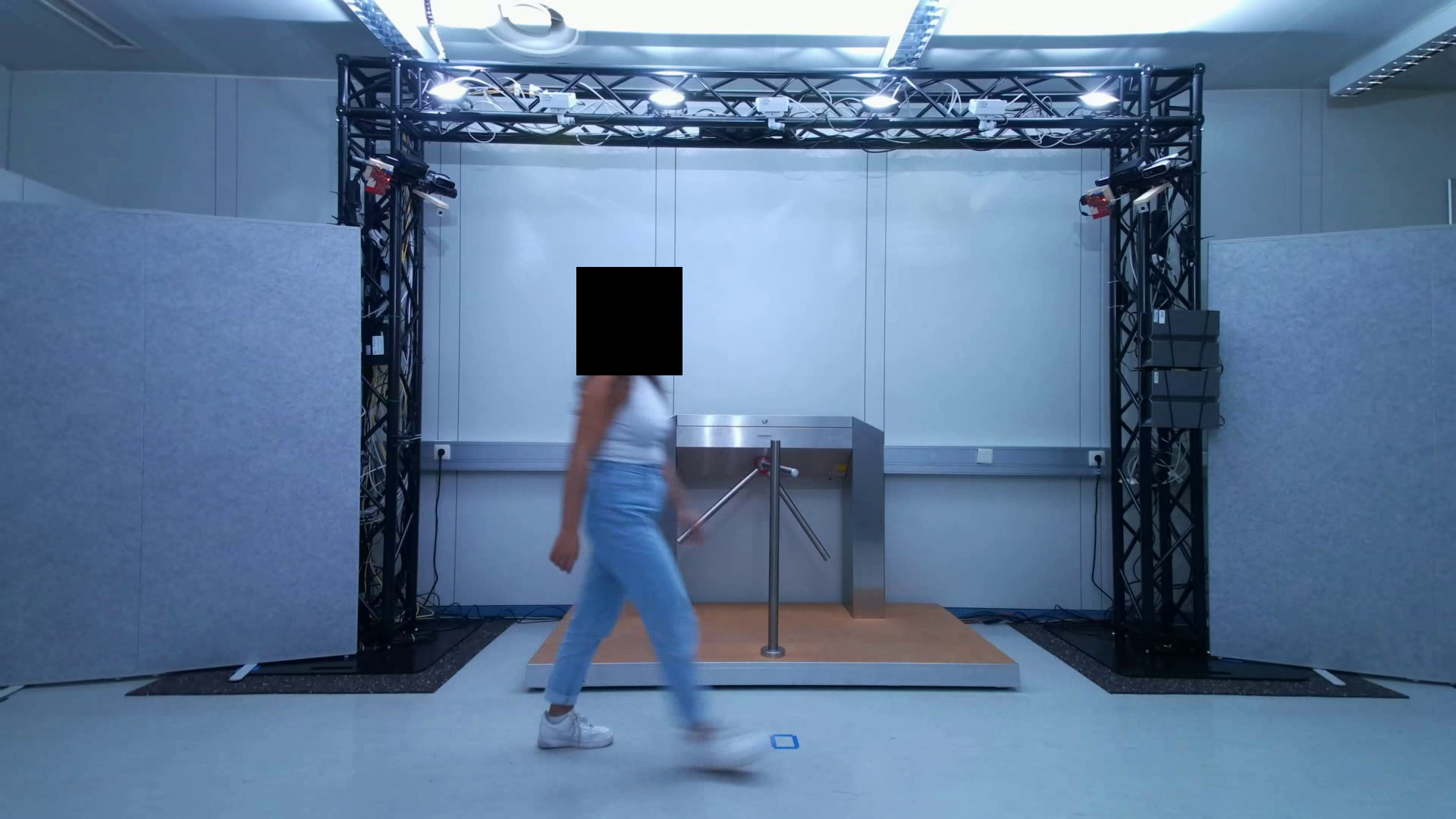} & \includegraphics[height=6em]{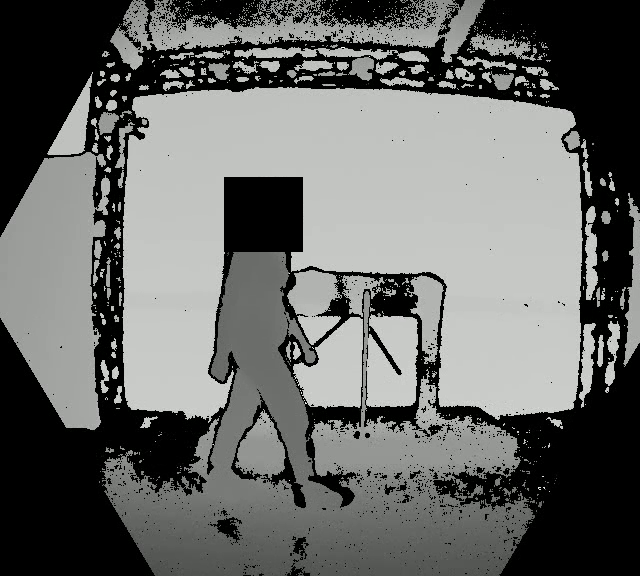} & \includegraphics[height=6em]{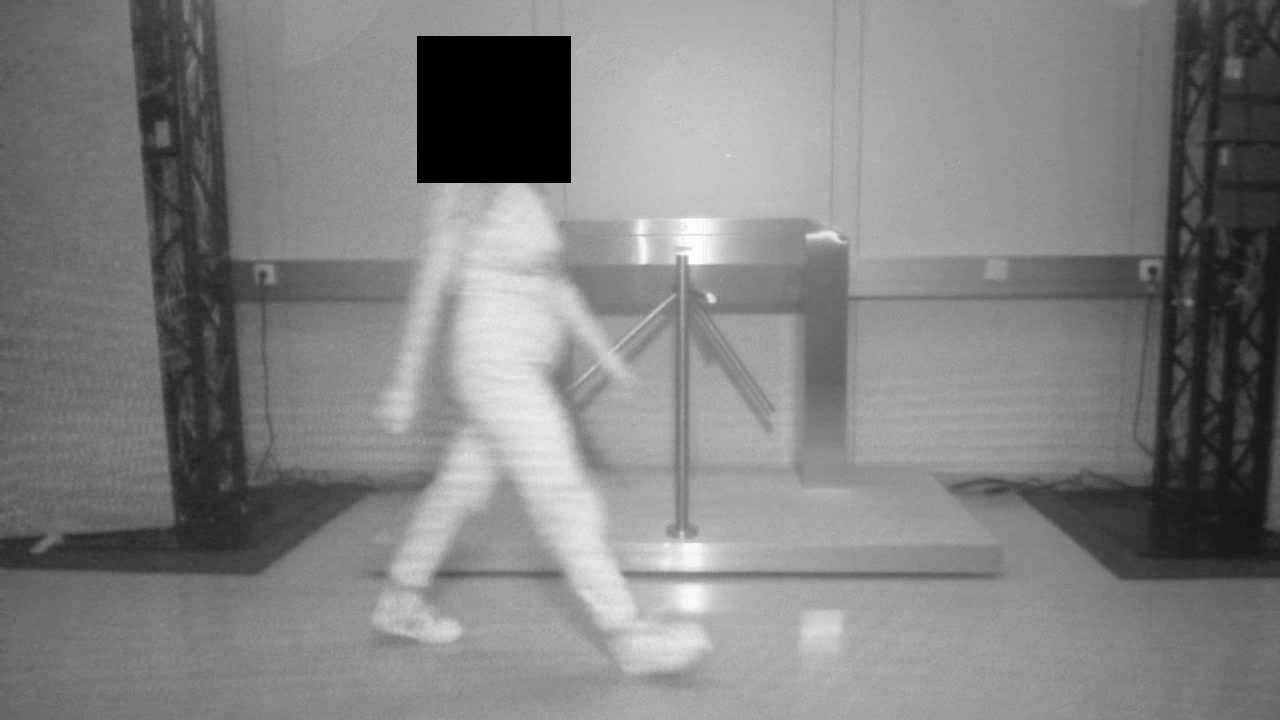} & \includegraphics[height=6em]{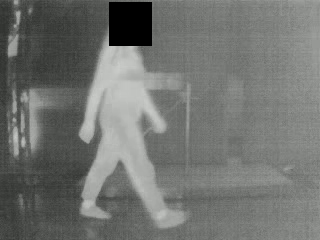} & \includegraphics[height=6em]{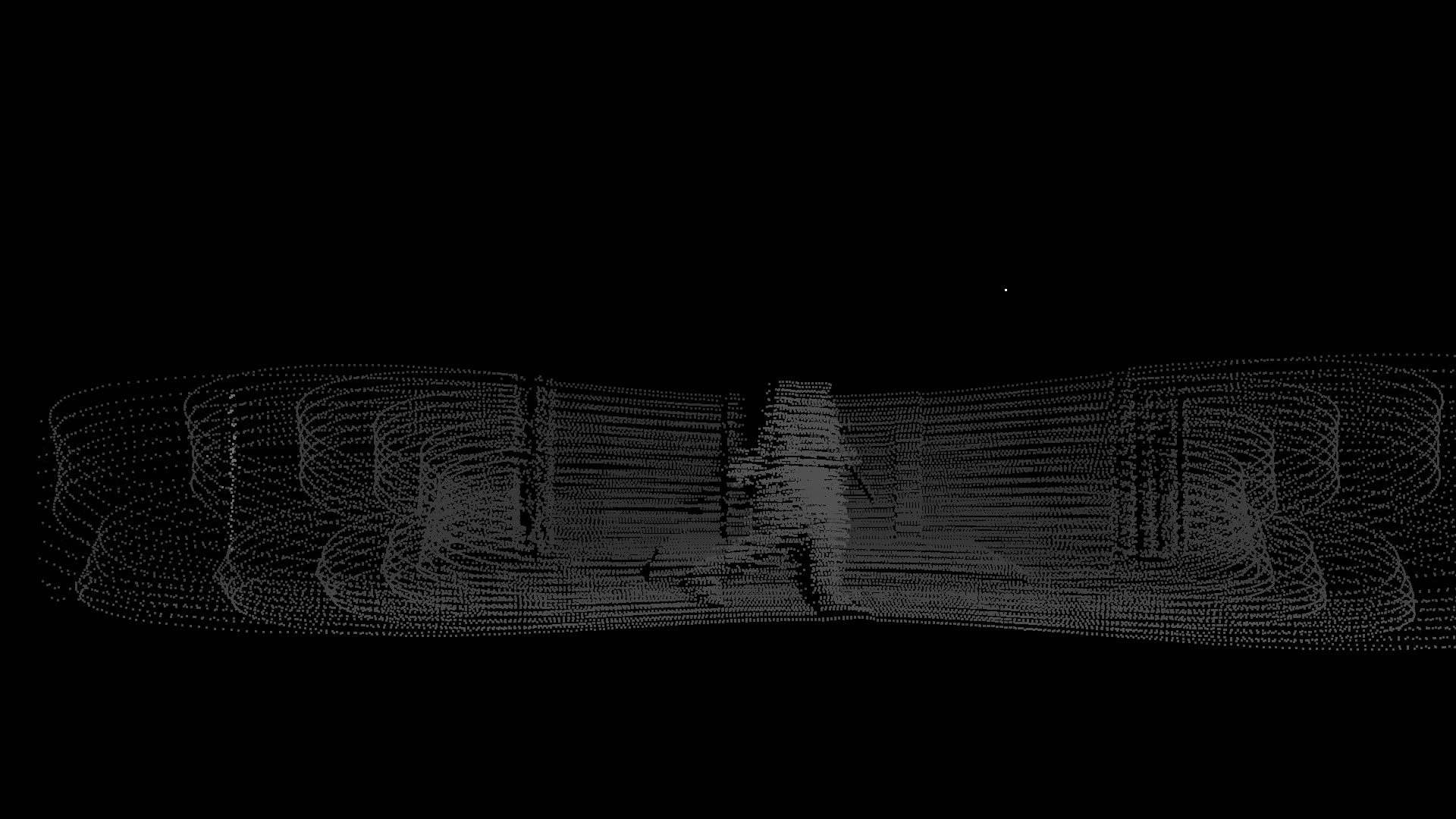} & \includegraphics[height=6em]{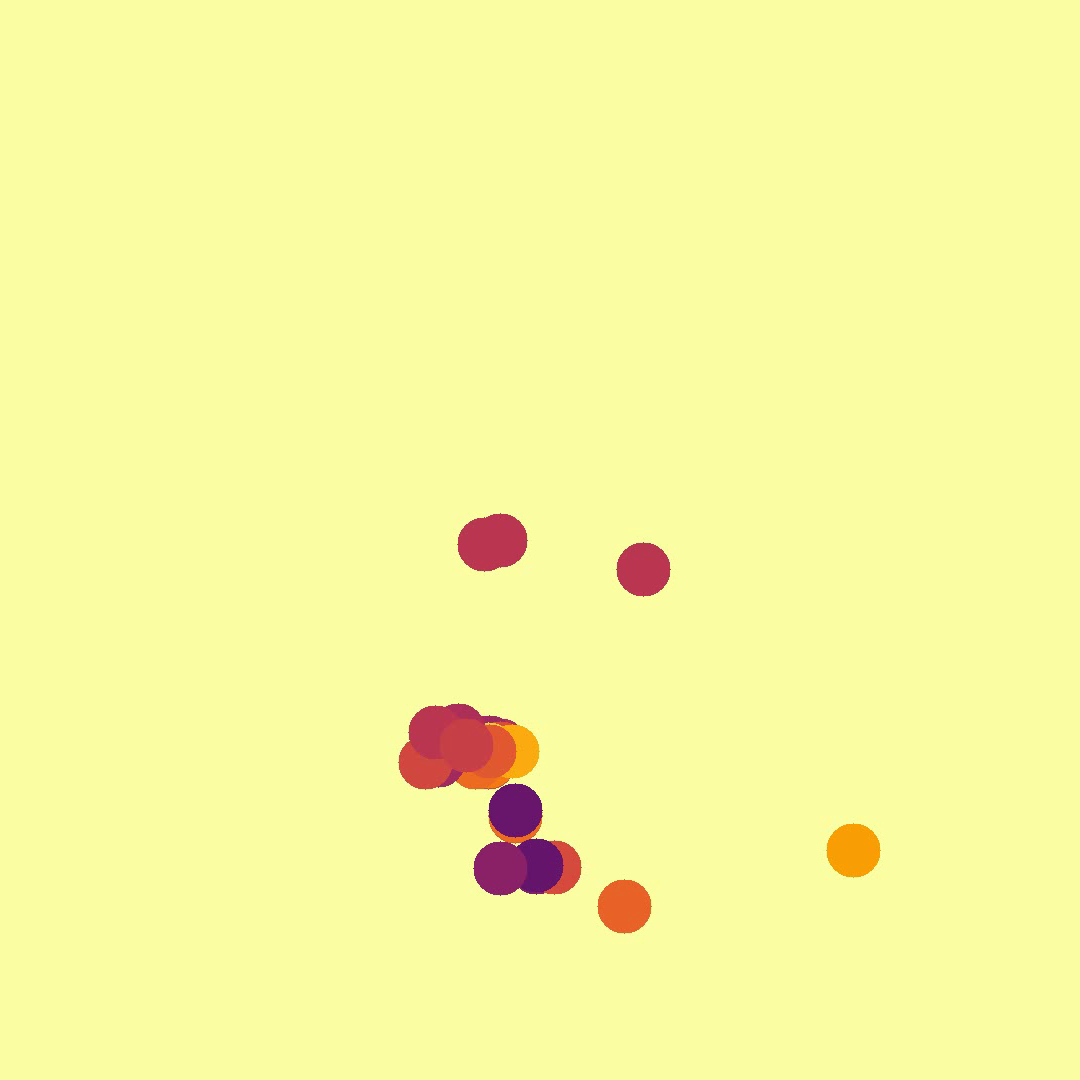}\\
        \end{tabular}
        \caption{Sample images from the different sensors in MultiGait. Left to right: video, depth, NIR, LWIR, lidar \& radar. For lidar and radar 2d projections of the higher dimensional point clouds are shown. No visualizations for CSI and BFI.}
    \end{subfigure}
    \begin{subfigure}{\textwidth}
        \centering
        \setlength{\tabcolsep}{1pt}\begin{tabular}{cccc}
            \includegraphics[height=6em]{figures/samples/base.mp4} & \includegraphics[height=6em]{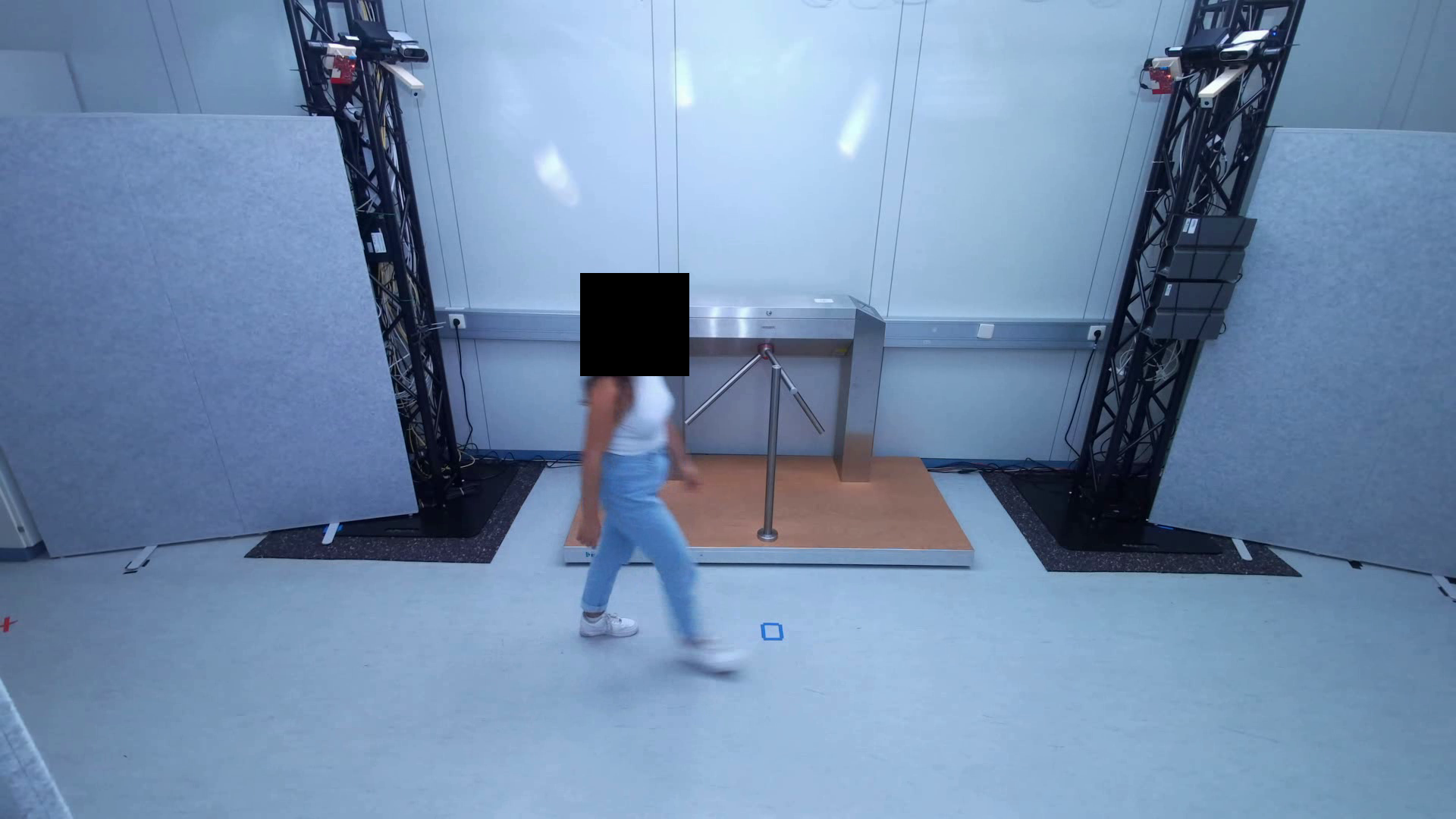} & \includegraphics[height=6em]{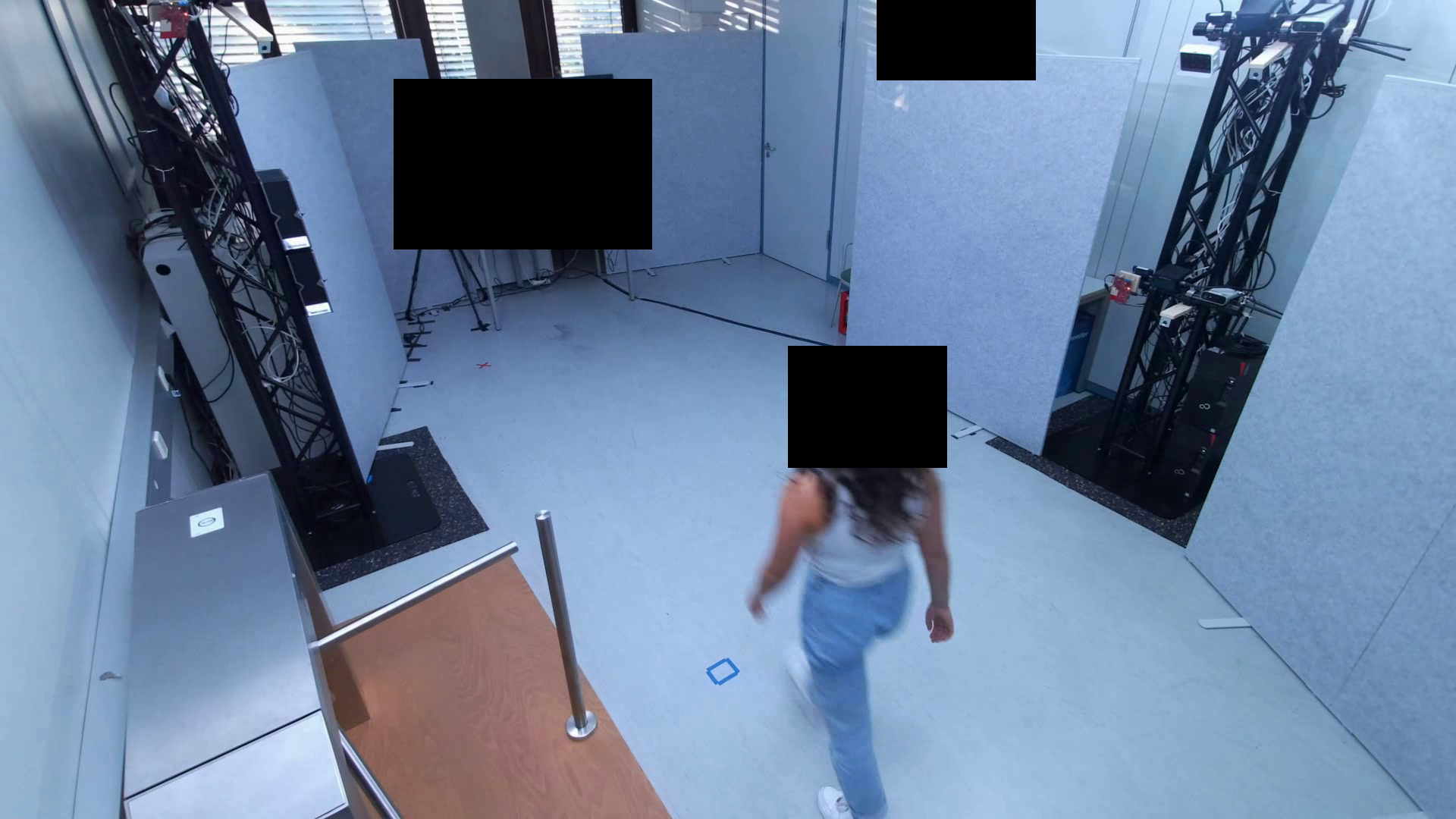} & \includegraphics[height=6em]{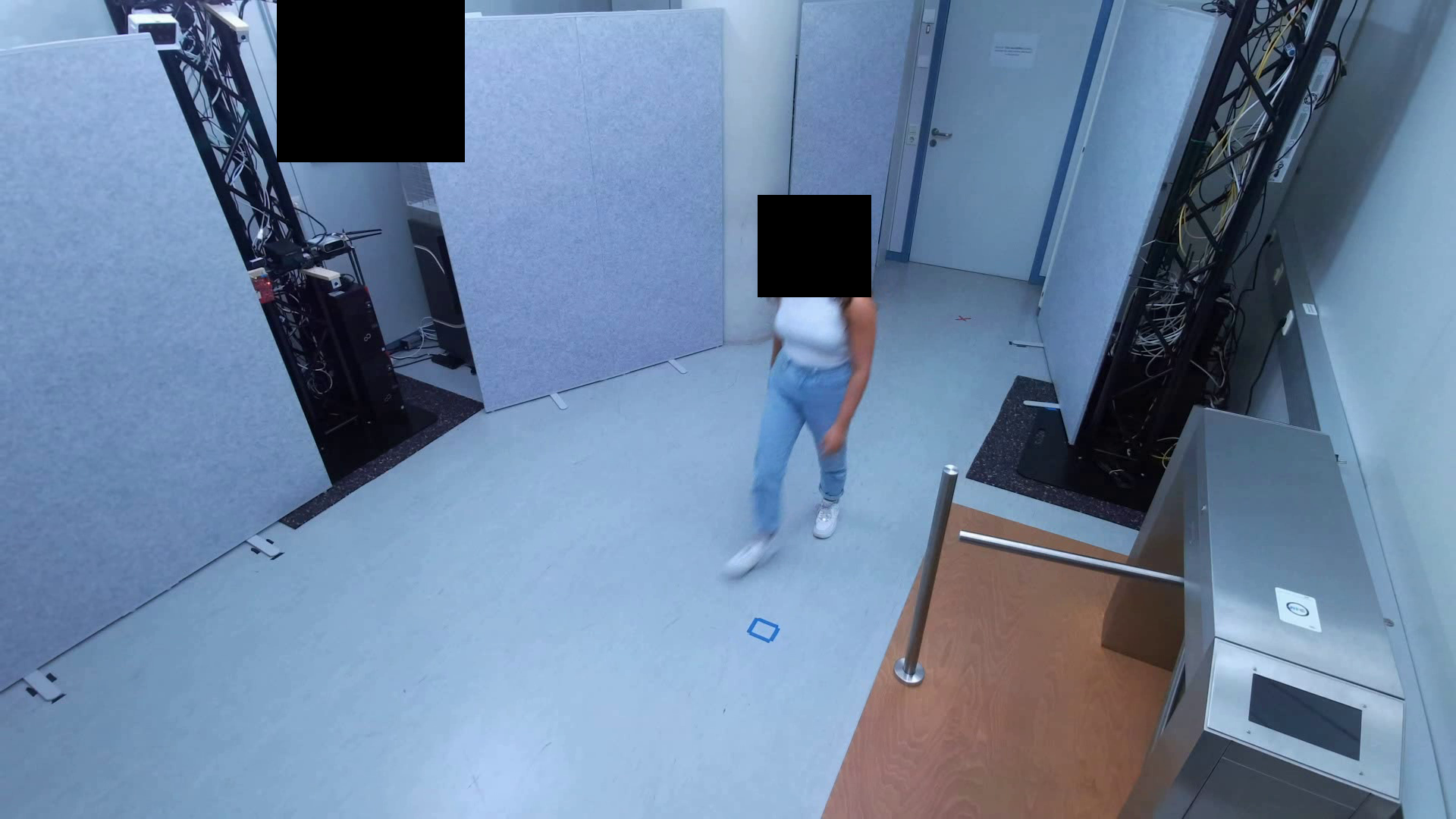}\\
        \end{tabular}
        \caption{Sample images from all perspectives in MultiGait. Left to right: center-low, center-high, left, right.}
    \end{subfigure}
    \begin{subfigure}{\textwidth}
        \centering
        \setlength{\tabcolsep}{1pt}\begin{tabular}{ccc}
            \includegraphics[height=6em]{figures/samples/base.mp4} & \includegraphics[height=6em]{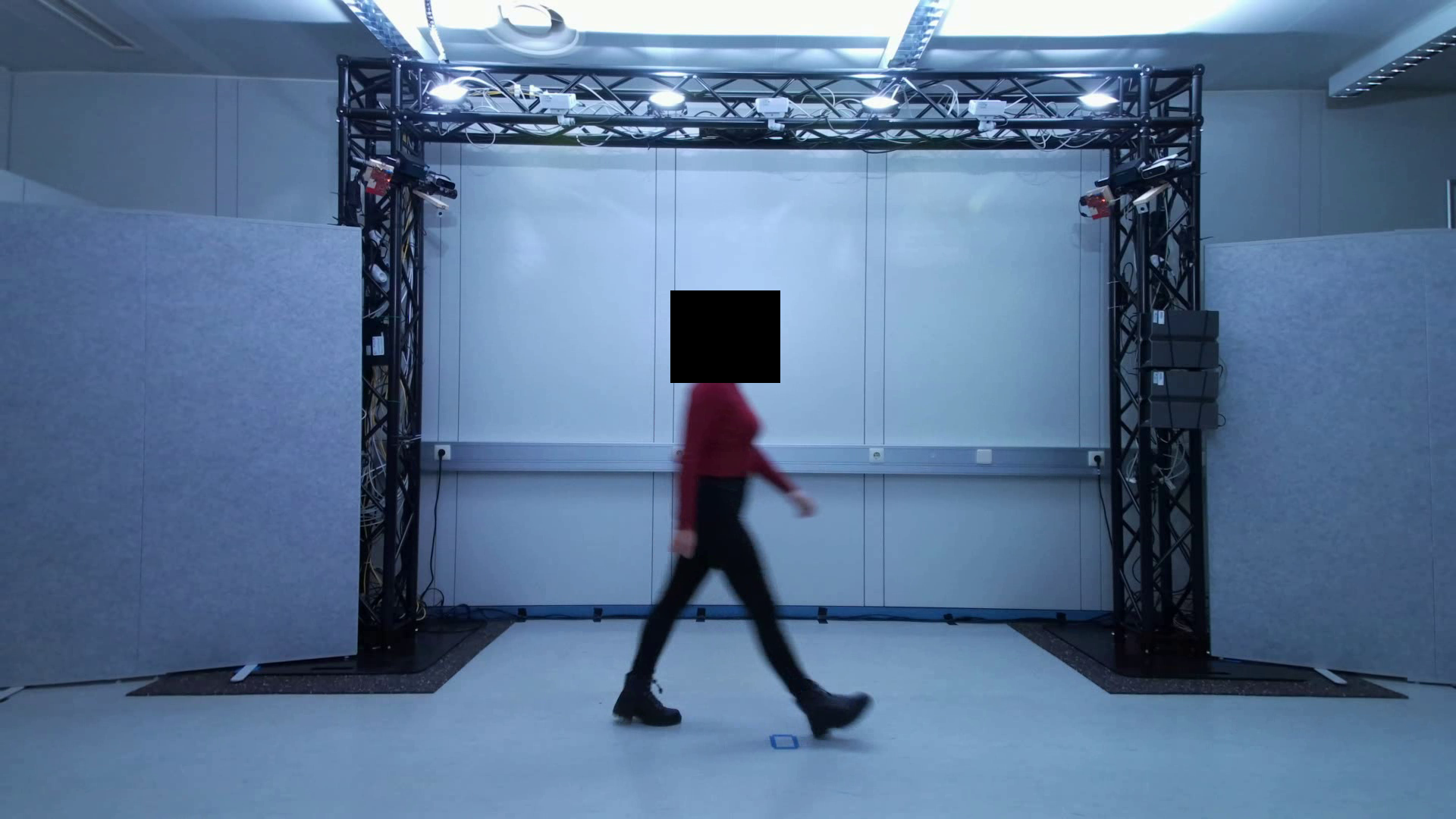} & \includegraphics[height=6em]{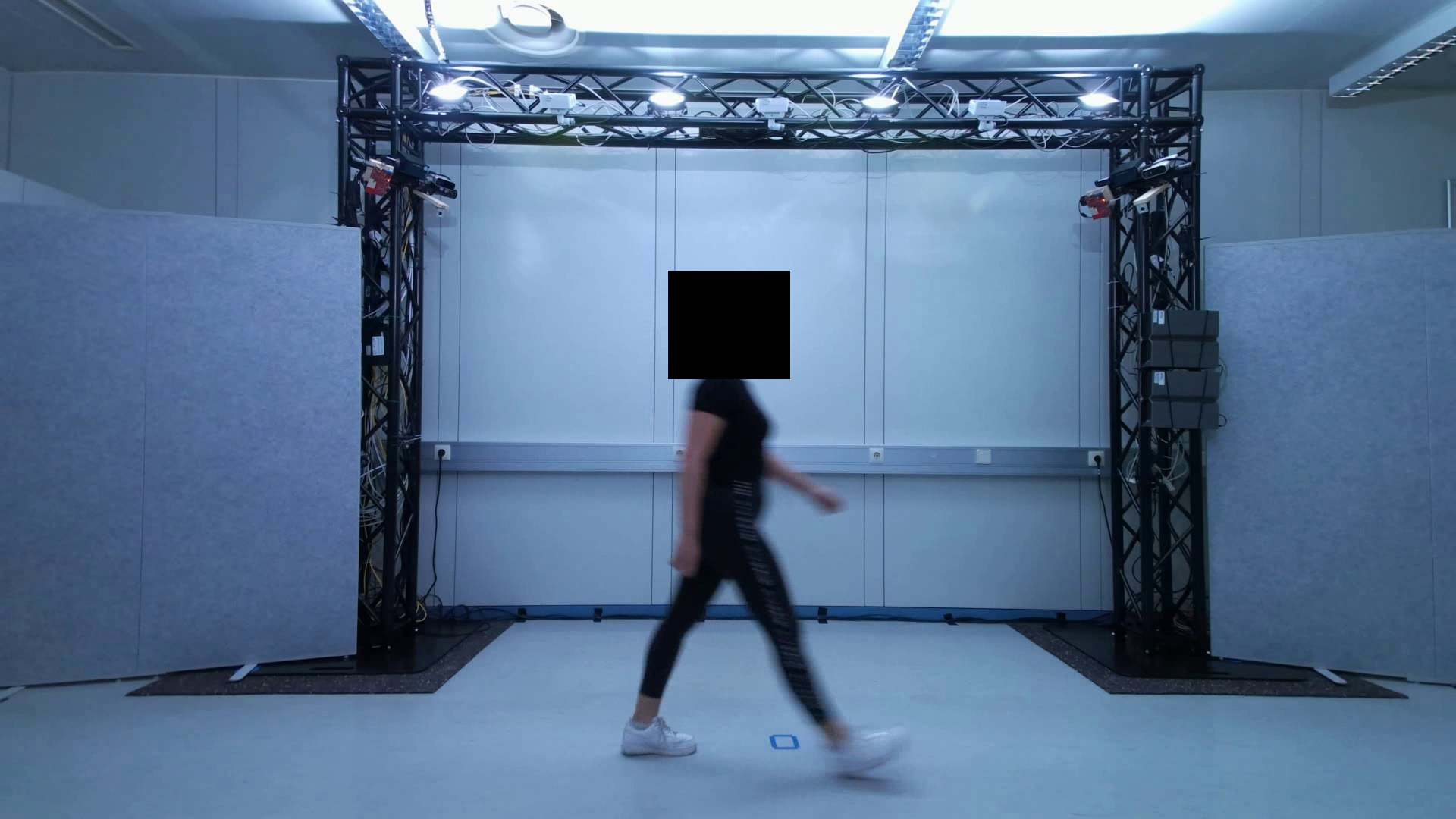}\\
        \end{tabular}
        \caption{Sample images of a participant from their three sessions in MultiGait.}
    \end{subfigure}
    \begin{subfigure}{\textwidth}
        \centering
        \setlength{\tabcolsep}{1pt}\begin{tabular}{ccccc}
            \includegraphics[height=6em]{figures/samples/base.mp4} & \includegraphics[height=6em]{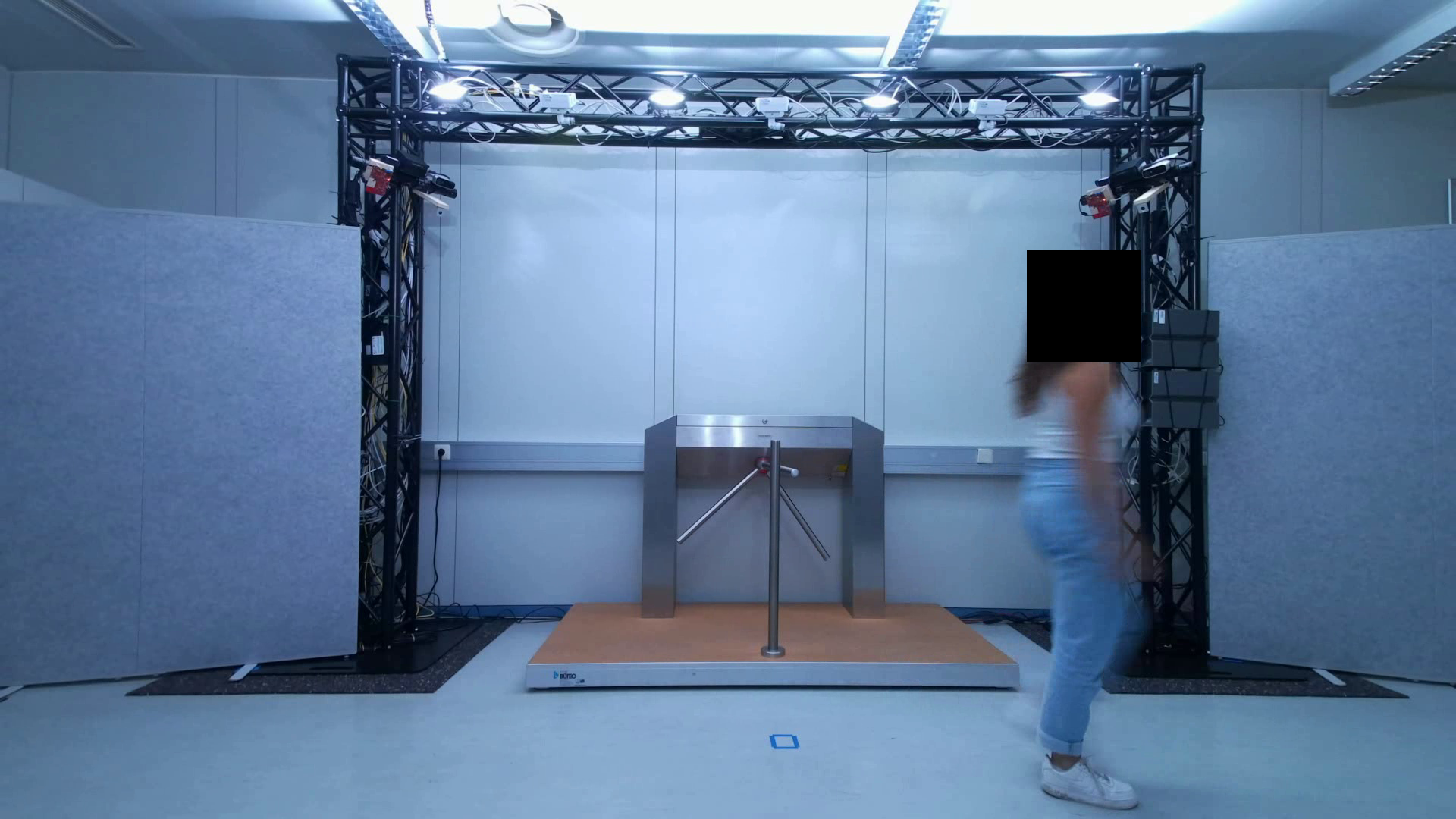} & \includegraphics[height=6em]{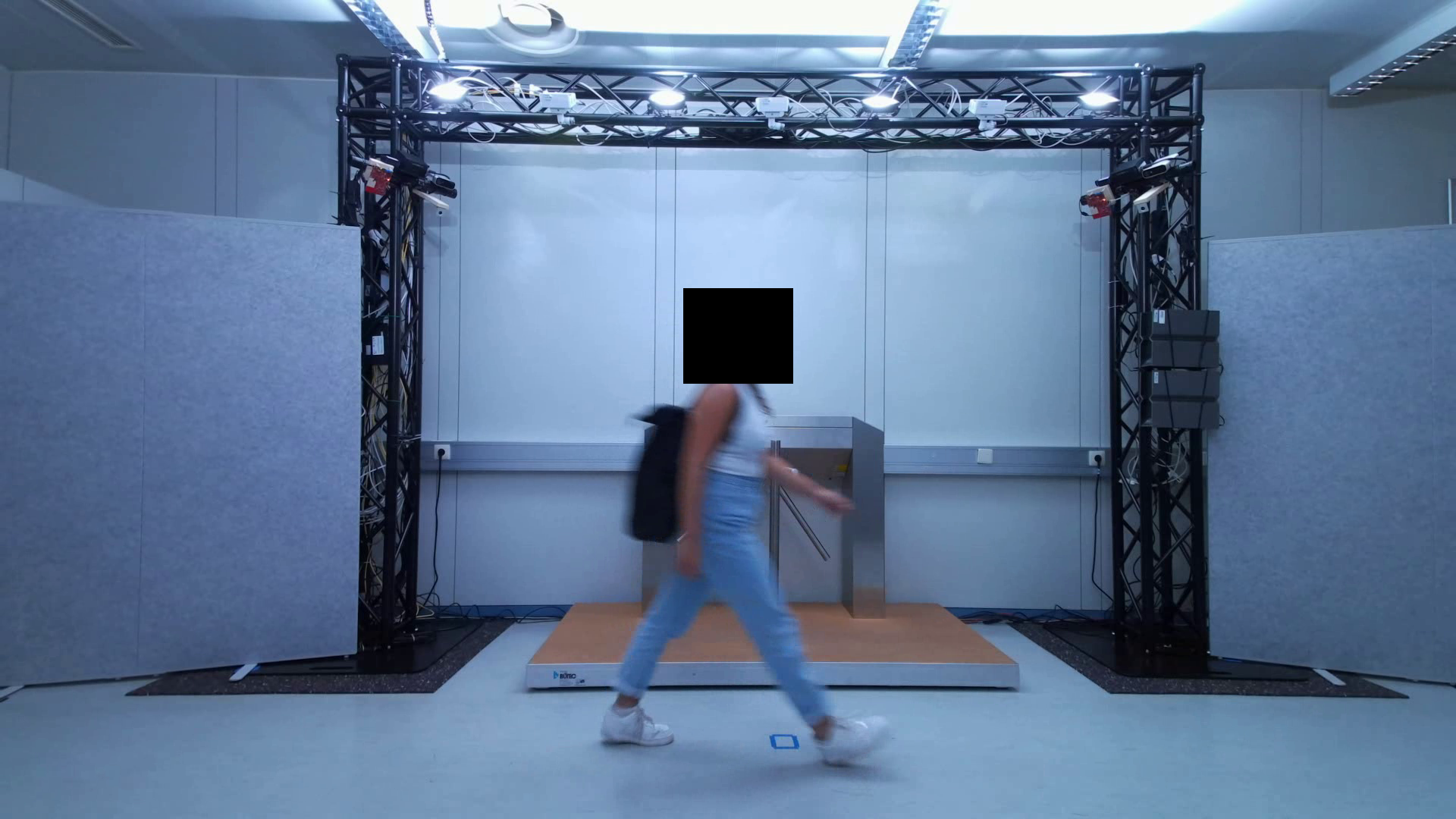} & \includegraphics[height=6em]{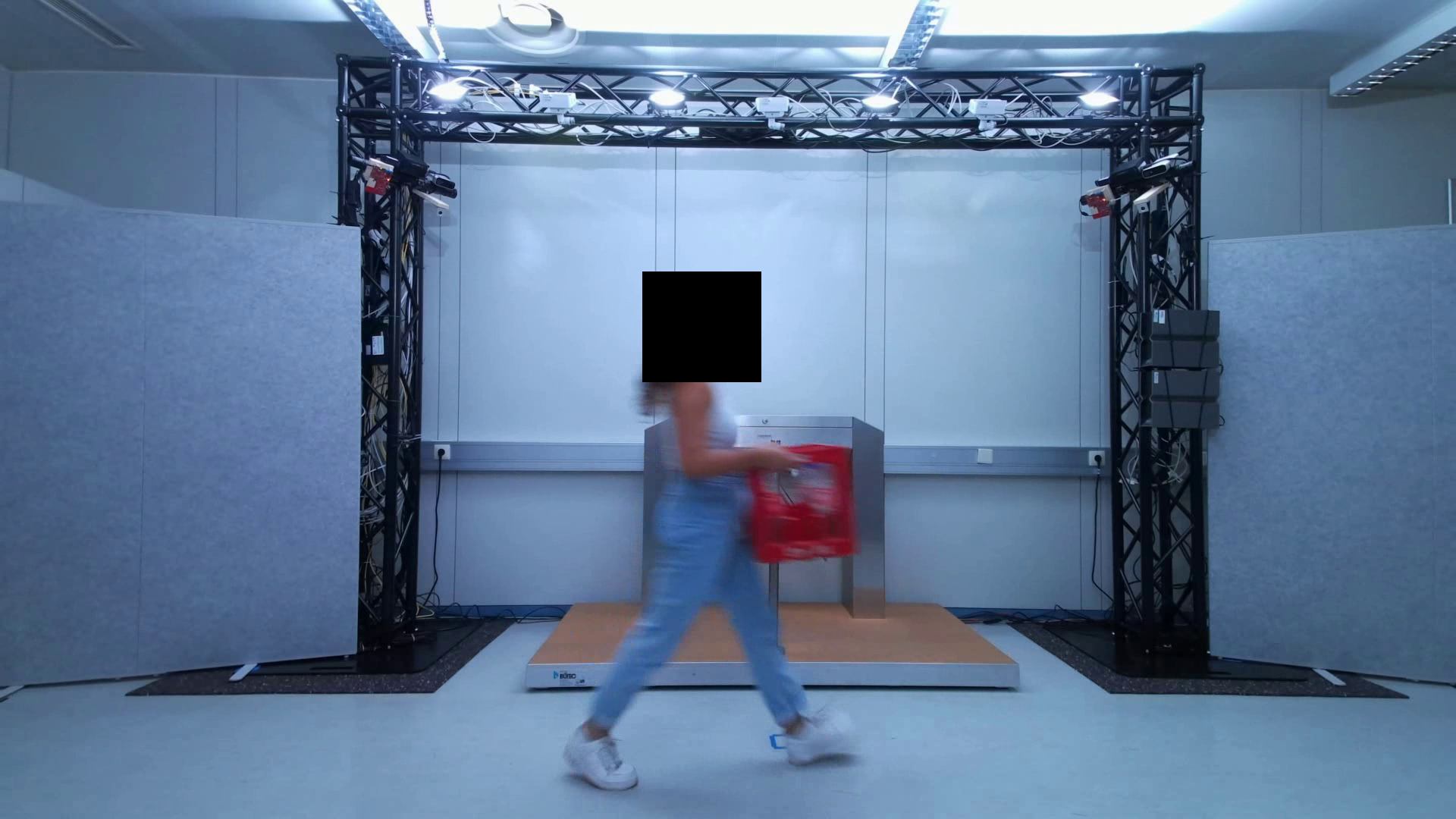} & \includegraphics[height=6em]{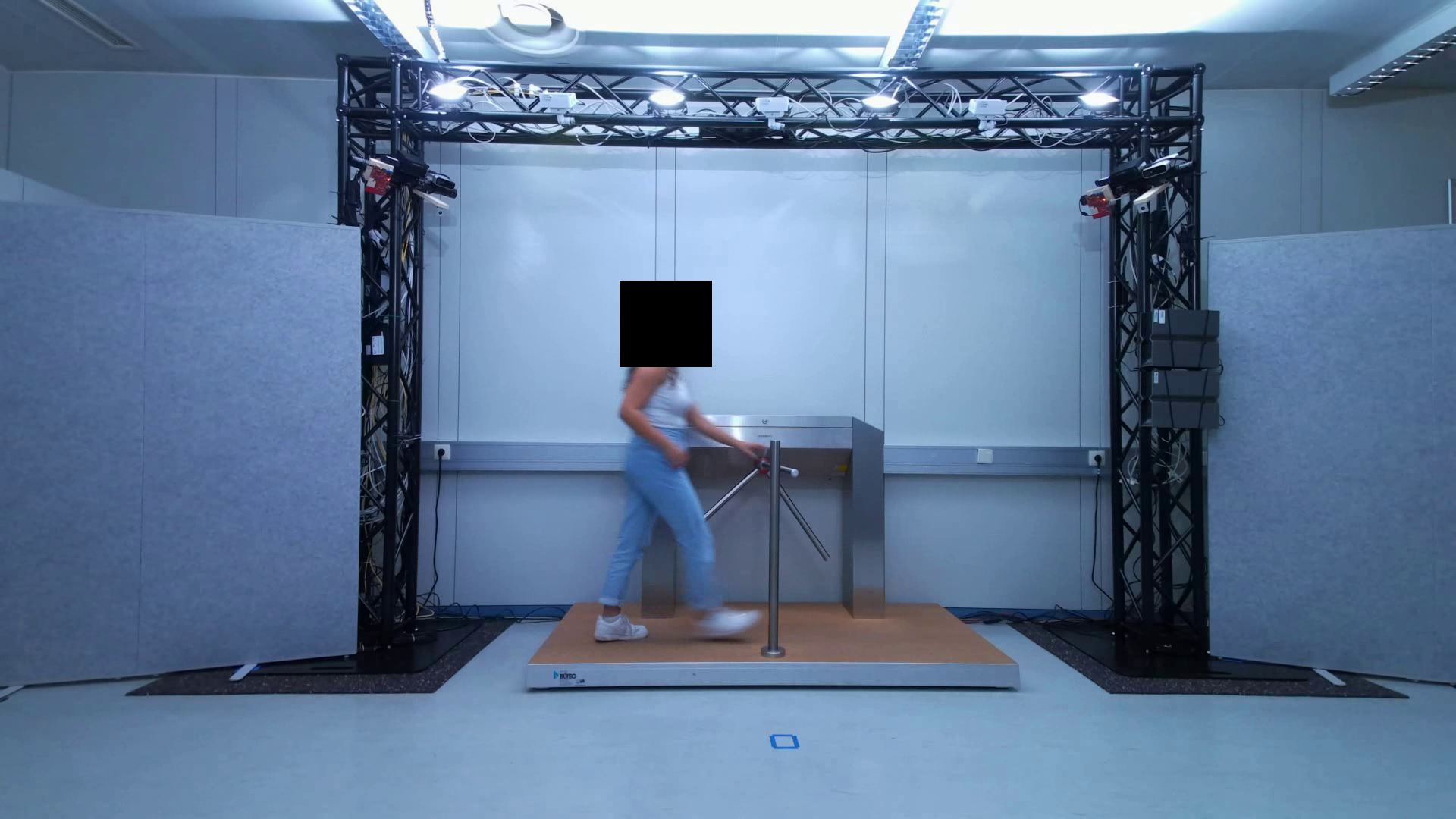}\\
        \end{tabular}
        \caption{Sample images from all activities performed in MultiGait. Left to right: normal, fast, with a backpack, with a bottle crate, and through a turnstile.}
    \end{subfigure}
    \begin{subfigure}{\textwidth}
        \centering
        \setlength{\tabcolsep}{1pt}\begin{tabular}{ccccc}
            \includegraphics[height=6em]{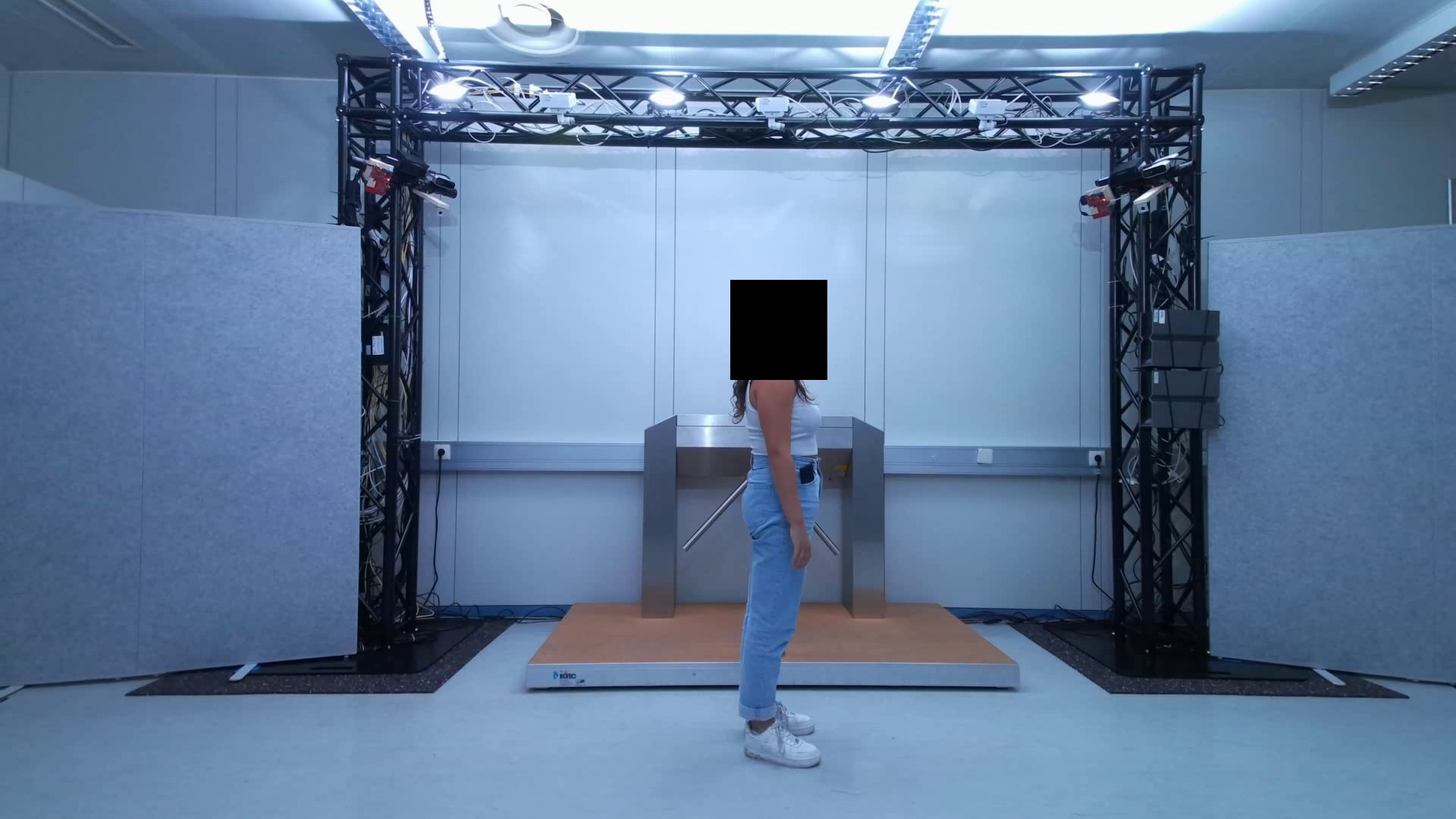} & \includegraphics[height=6em]{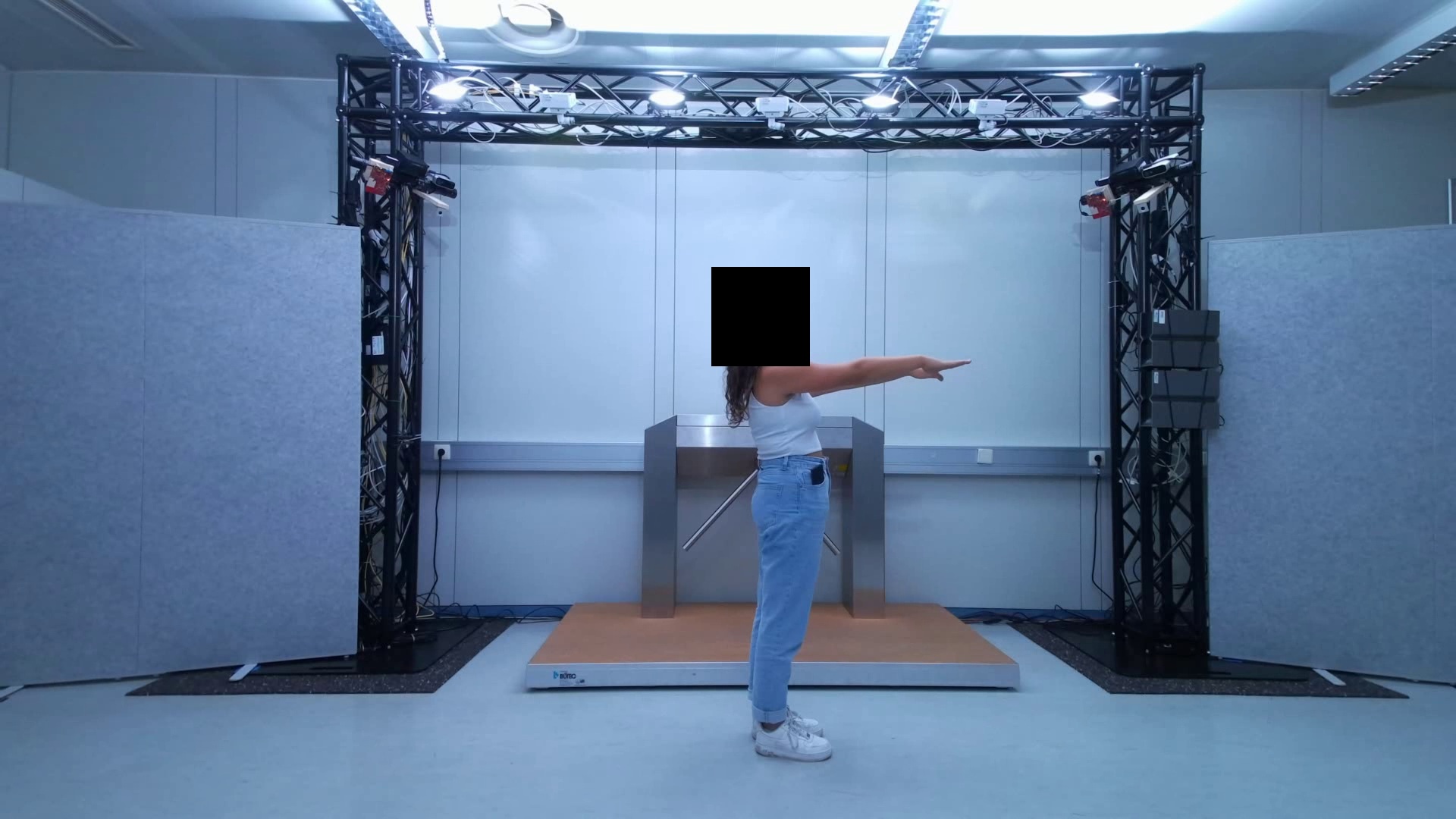} & \includegraphics[height=6em]{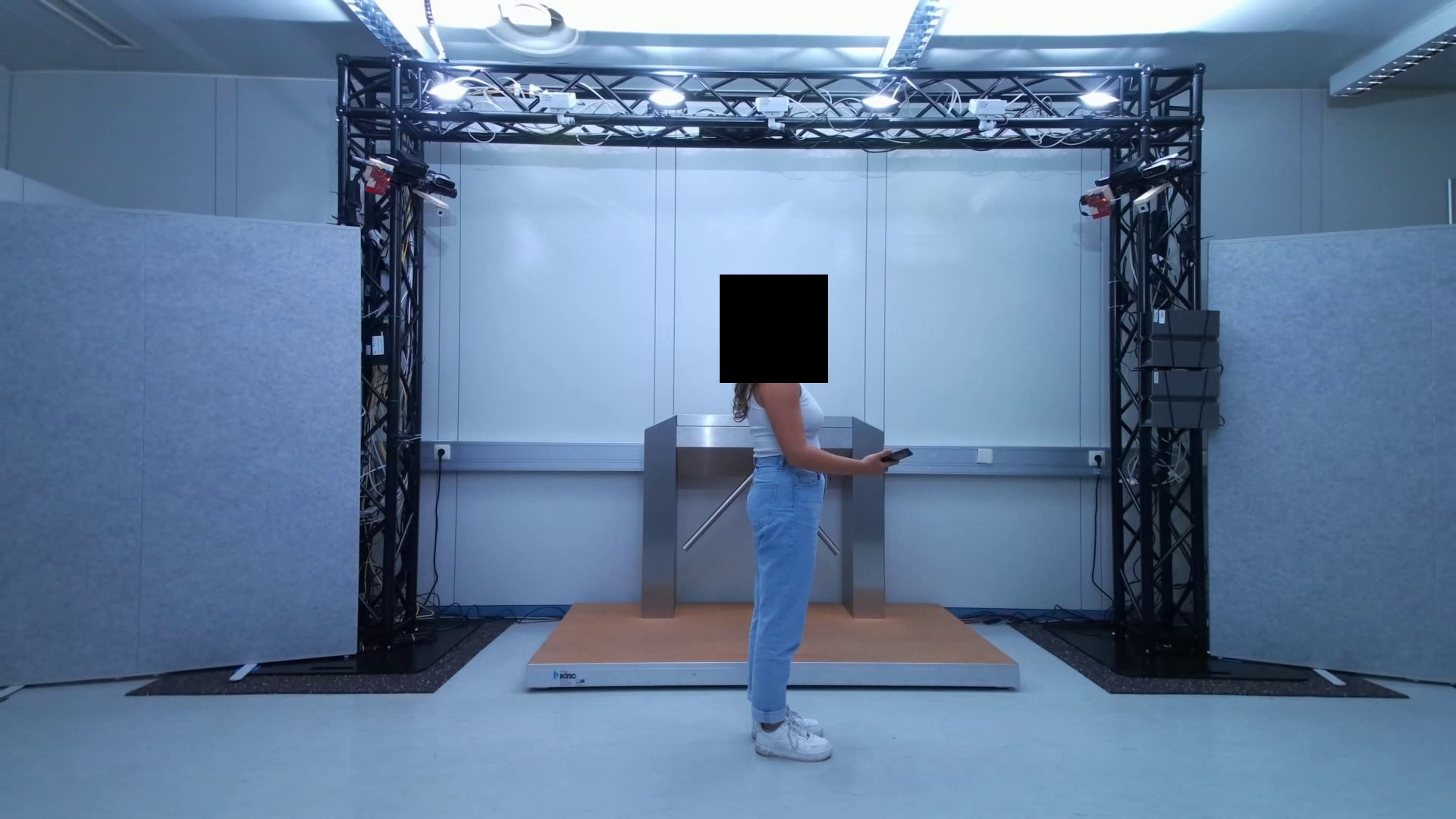} & \includegraphics[height=6em]{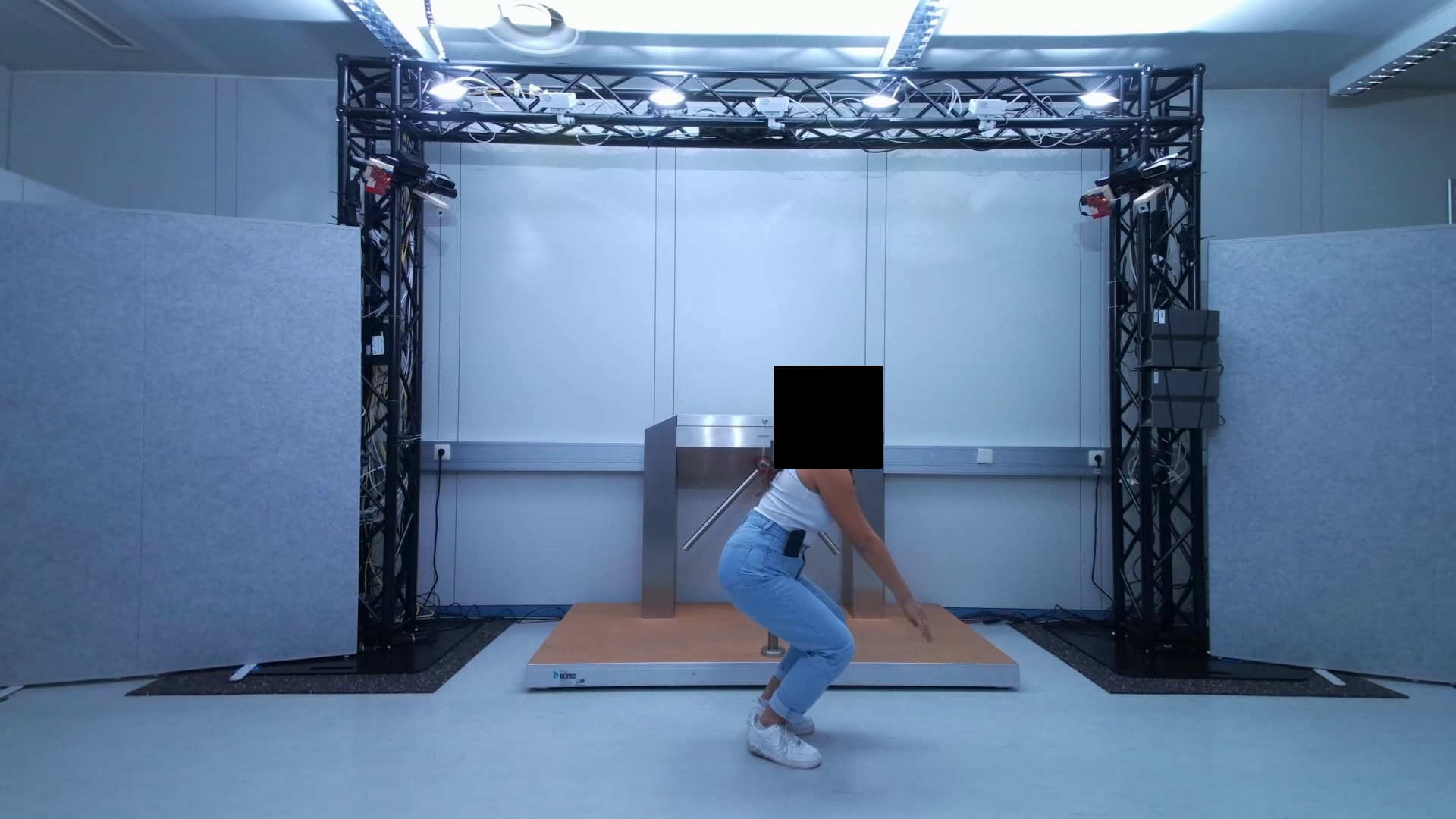} & \includegraphics[height=6em]{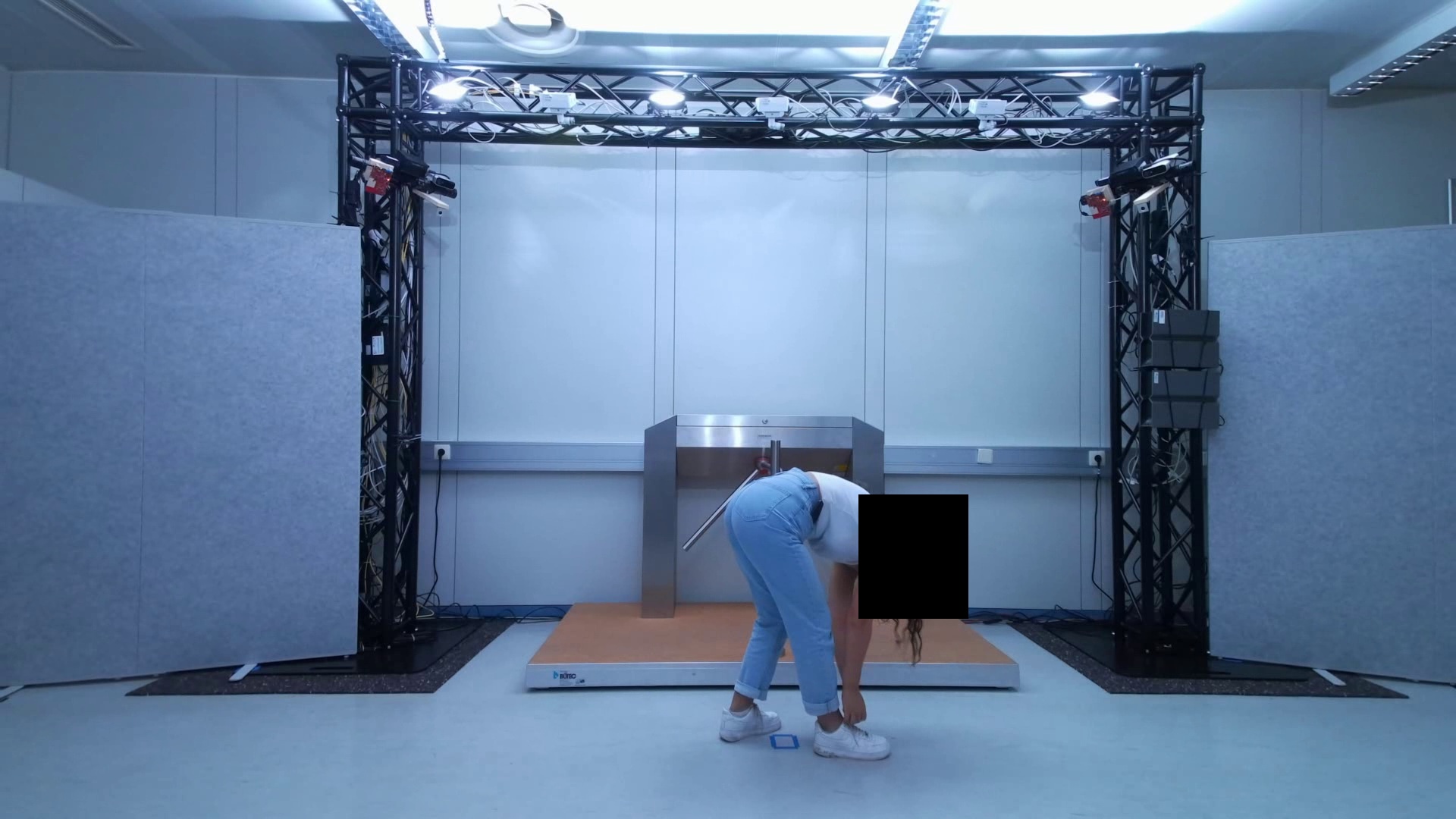} \\
        \end{tabular}
        \caption{Sample images from poses performed in MultiGait. Left to right: standing, standing with arms ahead, holding a smartphone, crouching, and pretending to tie their shoes.}
    \end{subfigure}
    \caption{Sample images over all dimensions of MultiGait. Note that we do not vary one-factor-at-a-time, but rather MultiGait is full-factorial. Also note that faces are redacted to preserve double-blindness during peer-review, but will be unredacted for the camera-ready version.}
    \label{fig:multigait}
\end{figure*}

To overcome the limitations of existing work established above, we created a novel gait-focused dataset that is multi-sensor, multi-perspective, and multi-session, and therefore name it \emph{MultiGait}.
Note, that we do not modify one factor at a time, but rather record 120 walk samples from all our 199 participants with every sensor, from every perspective, for every session in a full-factorial manner, yielding 32 information sources for each individual and session.
An overview over the data included in MultiGait is shown in \cref{fig:multigait}.
MultiGait is available to researchers.

\subsection{Sensors}
Our dataset contains the recordings of eight different sensor types, see \cref{tab:sensors} for an overview.
We recorded with video, depth, near-infrared and long-wave infrared thermal cameras, lidar, radar, and two WiFi sensing sources (channel state information and beamforming feedback information).
To enable fine grained comparisons, our setup includes multiple seemingly similar sensing technologies.
This enables future practitioners to use our dataset to investigate which (subset of) sensors are suitable for which tasks, evaluating both their utility and privacy.
We grouped all eight sensors together onto boards, see \cref{fig:sensors}.
We deployed four sensor boards in our lab.
The recordings of all sensors are synchronized to millisecond accuracy.

Similar to extant work, we used Microsoft Azure Kinects as a \textbf{video camera} baseline for comparison.
For consistent lighting, the area was lit indirectly with six 6500K flood lights.
The Kinects also contain a \textbf{depth camera}, a commonly proposed sensor for ``privacy-friendly'' surveillance \cite{kepski_fall_2014, rougier_fall_2011, villamizar_watchnet_2018, 10.1007/978-3-031-08645-8_62}. %

While related work considers "thermal" cameras \cite{collini_flexible_2024, tan_efficient_2006}, this terminology is imprecise as it refers to a range of infrared frequencies with different properties.
To both make this explicit and to allow for a comparison of different frequencies, our setup included sensing in the \textbf{near-infrared (NIR)} band and the \textbf{long-wave infrared (LWIR)} band.
For NIR, similar to existing work \cite{espinosa-duro_new_2013}, we modified logitech C920 HD Pro webcams by removing their NIR filter (which blocks NIR frequencies) and replacing them with visual-light filters which restrict light under 775nm wave length.
As common for applications of this frequency band (``night-vision''), we indirectly illuminated our lab using five spotlights emitting light with 850nm wave length.
For LWIR, we used seekthermal S314SPX. %

\textbf{Lidar} and \textbf{mmWave radar} are two frequently proposed smart city sensors which are commonly used for (semi-)autonomous driving.
We used Livox HAP TX as lidar sensors for their high resolution and Texas Instruments AWR2944 as mmWave radars based on the use of similar sensors in extant work \cite{gonzalez_inferring_2025, cheng_person_2022}.

For WiFi sensing, we recorded both \textbf{channel state information (CSI)} and \textbf{beamforming feedback information (BFI)}.
Wireless Sensing (recently standardized in IEEE 802.11bf \cite{ieee80211bf}) and Integrated Sensing and Communication (ICAS) in general are commonly proposed sensing systems for smart cities as they can take advantage of already deployed ubiquitous infrastructure \cite{10922050, murroni20236g}.
For each, a WiFi network in the 6 GHz band is setup in which benign clients communicate while additional nodes collect CSI and BFI, similar to related work \cite{todt_BFId_2025}, see the following section.

\begin{table}[tb]
    \centering
    \begin{tabular}{l|ccc}
         Modality & Hardware & Resolution & Framerate \\
         \toprule
         video & Microsoft Azure Kinect & 1920x1080 & 30fps\\
         depth & Microsoft Azure Kinect & 640x576 & 30fps\\
         NIR & logitech C920 HD Pro & 1280x720 & 10fps\\
         LWIR & seekthermal S314SPX & 320x240 & 23fps\\
         lidar & Livox HAP TX & \multicolumn{2}{c}{\textasciitilde 45,000 pts/s}\\
         radar & TI AWR2944 & \multicolumn{2}{c}{\textasciitilde 250 pts/s}\\
         CSI & intel AX210 & \multicolumn{2}{c}{\textasciitilde 285 pkgs/s}\\
         BFI & intel AX210 & \multicolumn{2}{c}{\textasciitilde 10 pkgs/s}\\
    \end{tabular}
    \caption{Overview over the sensors used in our setup}
    \label{tab:sensors}
\end{table}

\subsection{Physical Setup}
\begin{figure*}[tb]
    \centering
    \begin{subfigure}{0.38\linewidth}
        \vspace{.5em}
        \centering
        \includegraphics[width=0.98\linewidth]{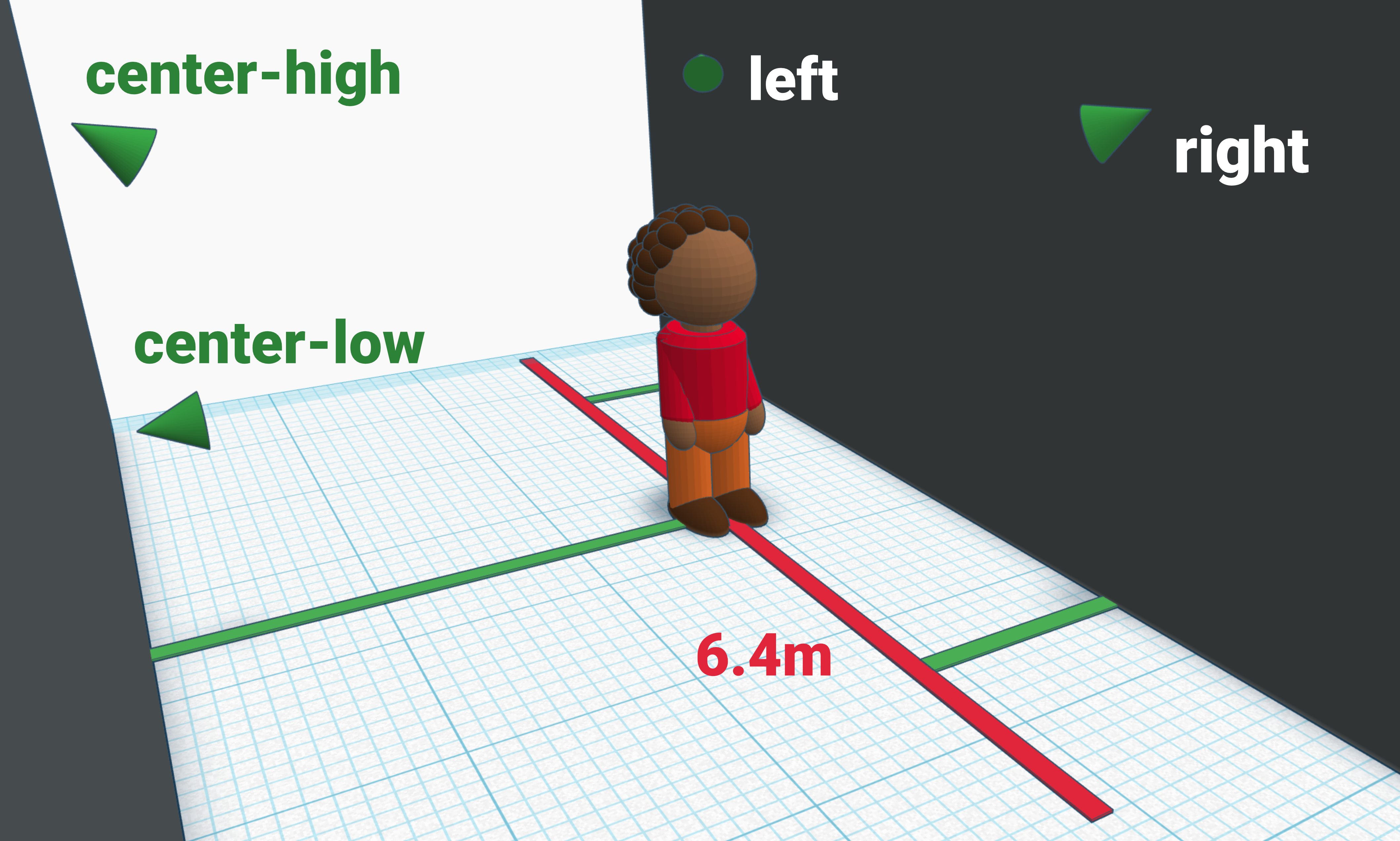}
        \caption{Physical layout of our setup.}
        \label{fig:layout}
    \end{subfigure}
    \begin{subfigure}{0.58\linewidth}
        \centering
        \includegraphics[width=0.98\linewidth]{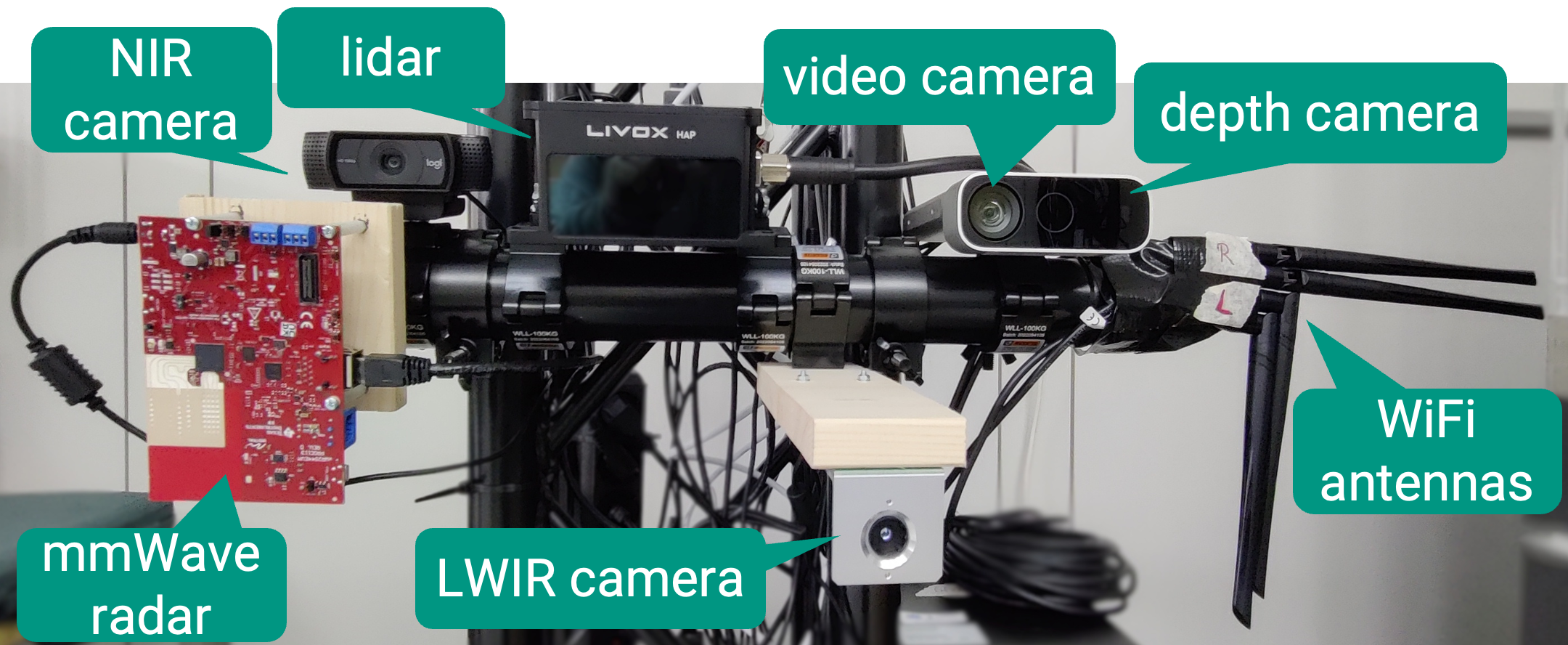}
        \caption{Setup of the sensors used.}
        \label{fig:sensors}
    \end{subfigure}
    \caption{Physical setup of the sensors for the recording of MultiGait.}
\end{figure*}
For a robust and reproducible evaluation of the inference potential of sensing technologies, we chose a lab setup over an uncontrollable outside scenario.
While the latter may be interesting for some of the chosen sensors, many sensors lack a proper evaluation even in lab conditions.
Further, as current datasets are generally limited to single-session, a lab setting allows us to investigate multi-session inference in a controlled scenario.
This characteristic also enables us to investigate the effect of any variation along our dataset's dimensions, which in turn enables a better-informed deployment of (combinations of) sensors in the future.
To enable a comparison of perspectives, each sensor type was present in four different locations, see overview in \cref{fig:layout}.

While other datasets with multiple perspectives vary only the angle between walking path and line-of-sight, we argue that a more reasonable consideration is to test perspectives from above, mimicking surveillance cameras mounted from the ceiling.
Thus, we employ three additional perspectives center-high (directly above center-low), left and right which are mounted 2.2m above ground.
As a baseline perspective, we chose the center-low perspective, which is perpendicular to the walking line, recorded from around hip height.
We chose it because it is a common perspective in related work as it captures gait information best.

As opposed to the other sensors, WiFi sensing requires communication between two devices.
Both CSI-based and BFI-based sensing relies on objects altering the electro-magnetic wave that is traveling from the sender to the receiver in normal WiFi communication.
Its setup is therefore slightly more involved.
For both CSI and BFI, the setup contains an access point and four clients with two antennas each.

For BFI, all four clients communicate with their access point normally while an additional, not-connected, node records all beamforming reports.
For CSI, the access point communicates with an additional traffic-generation node, while all four perspectives record CSI at their locations.
For both BFI and CSI, the communication patterns were determined empirically to achieve the highest sample rates.

Due to WiFi sensing requiring communication between two parties, the location of the sensors are slightly different:
In the location of the right perspective, though at a height of 90cm, we mounted the two WiFi access points, one for CSI and one for BFI.
The WiFi-right perspective as reported in our results is actually located around 2m further out right on the axis created by the left and right perspective. %
This WiFi right perspective as well as the WiFi left perspective are both located at a height of 90cm. %
The setup uses the lowest two non-overlapping 160 MHz channels in the 6GHz band with TP-Link Archer BE800 as access points and Intel AX210 WiFi NICs as clients.

\subsection{Recording}

We designed a user study to create a dataset that not only allows evaluating identity inferences across different factors, but also to perform attribute and activity inference.
This is done both to enable the evaluation of the inference potential of all the recorded sensors, and to allow for evaluations of utility-preserving anonymization methods using our dataset.
As anonymization methods generally need to make a trade-off between privacy and utility, both must be evaluated.
With our dataset, privacy -- or the absence thereof -- may be demonstrated through the inability to infer identities, while utility may be evaluated through the continued ability to infer certain attributes or activities.

Our participants were asked to perform five different activities: walking normally, walking fast (instruction: \textit{as if you were late to a meeting}), walking with a backpack, walking while carrying a bottle crate, and walking through a turnstile.
All activities were repeated ten times (back and forth), with the exception of normal walking which was recorded twenty times to allow for additional training data for identification experiments.
This part of the experiment took the participants 25 minutes on average. %

As we expect gait-based identification, at least with some sensors, to already work well, we also recorded poses to allow for more difficult inference tests in the future.
Each participant recorded each of the following poses five times in a randomized order: standing, standing with arms ahead, holding a smartphone, crouching, and pretending to tie their shoes.
To enable attribute inference, participants were asked to fill out a demographics survey that includes age, gender, hair and skin color, nationality (as a proxy for origin/ethnicity) and shoe size.
Additionally, we measured the height and weight of all participants.

\paragraph{Study Participants}
Participants were recruited from a local student panel using ORSEE \cite{greiner_subject_2015} which reduces biases in the recruitment process by standardizing and depersonalizing it.
We required all participants to be over the age of 18 and to be able to walk without aid.
Further, we required participants to understand German in order to fill out the demographics survey, and understand participant and data protection information.
Finally, to increase the quality of recorded data, participants were instructed to avoid loose clothing (e.g. dresses or oversized fits) and not to wear shoes with heels.

\begin{figure}[tb]
    \centering
    \includegraphics[width=0.47\textwidth]{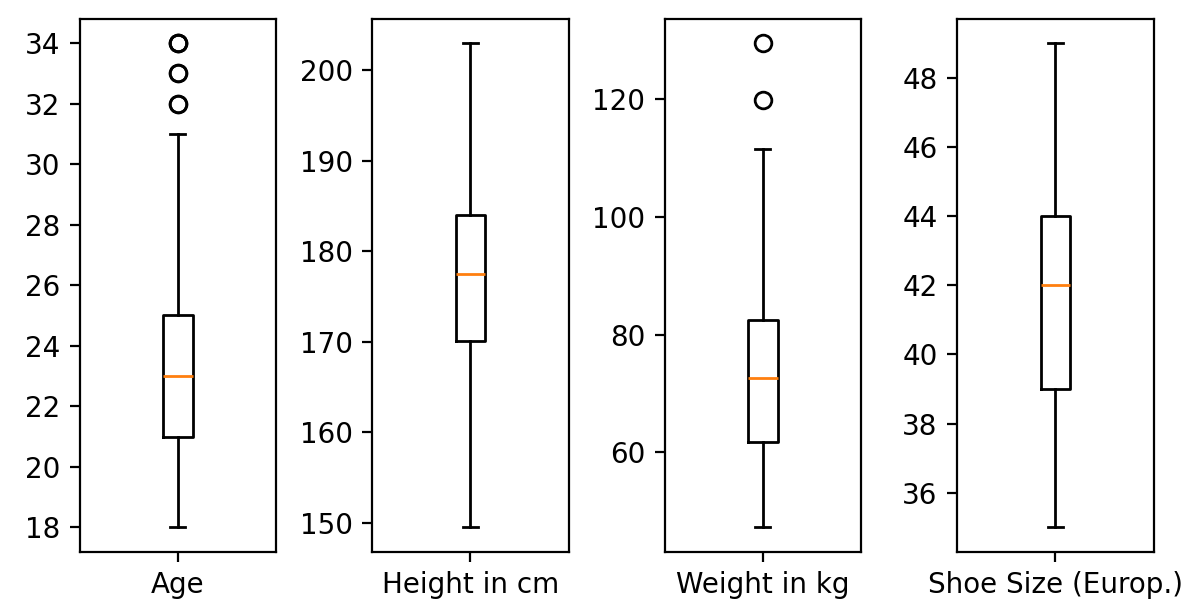}
    \caption{Numerical demographics of our participants as boxplots.}
    \label{fig:demo-numerical}
\end{figure}

Of our participants, 58.8\% identified as male and 40\% identified as female.
Our participants had an average age of $23.4 \pm 3.4$ years, height of $177 \pm 9.7$ cm and weight of $73.8 \pm 14.5$ kg, see also \cref{fig:demo-numerical}.
Further details on the categorical demographic information can be found in \cref{sec:demographics} in the Appendix.

\paragraph{Time of recording}
The dataset was recorded in three distinct sessions.
The first session was recorded over two weeks in November 2024.
All 199 participants were recorded for the first time during this session and demographic information was collected.
We invited all 199 individuals back for both the second and third session in February and March 2025, respectively.
As such, not all participants were present for each of the additional sessions. %
In total, we recorded 135 participants once, 36 participants twice and 28 participants three times, for our total of 199 participants.

\paragraph{Ethical Considerations}
For our user study, we obtained ethics approval from our university's ethics commission (IRB) and coordinated the data processing with our university's data protection office.
Participants were reminded that they could refuse to answer any question and withdraw from the study at any point.
For the first session, participation took approximately 60 minutes and was compensated with 15 Euro.
For the second and third sessions, participation took approximately 30 minutes and was compensated with 10 Euro.
The data collection information form explicitly highlighted that biometric data, as part of the special categories of personal data defined by Art. 9 of the GDPR would be recorded and processed. %

\paragraph{Open Science Considerations}%
In addition to the general agreement to participate in the study, participants could voluntarily agree to allow the sharing of their recorded data with other researchers.
Of our 199 participants, 183 (92\%) agreed to this.
While, for the sake of reproducible and open science, we would like to publish MultiGait without any limitations, we consider this a reasonable trade-off.
This is because of the sensitivity of the recorded data, its misuse potential and the uncertainty which inferences are possible with the data.

\paragraph{Dataset Release Policy}
We will publish the raw recordings from all sensors, all perspectives and all sessions from the 183 participants that agreed to this publication with this paper.
As set out in our data protection statement, our dataset is available for research purposes.
Therefore, interested researchers will need to complete an access request form (which restricts re-distribution and use for non-commercial purposes).
The full form can be found in \cref{sec:datasetrelease} of our Appendix.
We consider this an acceptable trade-off in light of the sensitivity of the data, the expected ethical conduct in research involving human participants, and in-line with release policies of similar datasets. %

\subsection{Post-Processing}
We performed general processing on the recorded data.
The primary part was cutting the recordings from each sensor for each participant into the individual walking samples, i.e. one sample for each time the participant walked the length of our setup.
We create exact timings using the video recording of the center-low perspective.
There, we compared each frame of the recording to an image of the empty room.
Similarity is highest when the participant was out of frame at each of the two ends of the walking path, so this is used as the timestamps for cutting all sensors' recordings.
For instances where too many or too few samples were detected, we adjusted timings manually.

As we have converted webcams to NIR cameras, we process their recordings by selecting only the value (V) channel from a HSV representation of each pixel's color.

For CSI, we processed the recordings by removing all reports with a different amount of subcarriers than 53 which was the most common amount in our recordings.
We keep both the phases and the magnitudes for all datapoints.

For BFI, we split our recordings based on the different perspectives that the reports were sent from.
Then, we parse them according to the IEEE 802.11 standard, resulting in 10 angles for each of the 74 channels.

\begin{table*}
  \caption{Measured accuracies for all sensors, and all perspectives in our activity recognition experiment.}
  \label{tab:activities}
  \centering
  \small
  \begin{tabular}{@{}llcccc@{}}
    \toprule
    Sensor & Rec. System & Center-Low & Center-High & Left & Right \\
    \midrule
    video & GaitBase \cite{fan_opengait_2023} & $93.2\% \pm 0.9$ & $92.9\% \pm 0.7$ & $94.0\% \pm 0.5$ & $91.0\% \pm 0.4$	\\
    depth & GaitBase \cite{fan_opengait_2023} & $96.2\% \pm 0.1$ & $96.0\% \pm 0.2$ & $93.1\% \pm 0.2$ & $96.5\% \pm 0.2$	\\
    NIR   & GaitBase \cite{fan_opengait_2023} & $97.6\% \pm 0.0$ & $97.2\% \pm 0.1$ & $96.4\% \pm 0.1$ & $95.6\% \pm 0.2$	\\
    LWIR  & GaitBase \cite{fan_opengait_2023} & $91.6\% \pm 0.1$ & $89.4\% \pm 0.3$ & $91.7\% \pm 0.1$ & $90.9\% \pm 0.3$	\\
    lidar & LidarGait \cite{shen_lidargait_2023} & $98.1\% \pm 0.2$ & $97.8\% \pm 0.0$ & $97.7\% \pm 0.0$ & $97.9\% \pm 0.1$	\\
    radar & SRPNet \cite{cheng_person_2022} & $31.1\% \pm 0.9$ & --- & $57.1\% \pm 0.6$ & $64.8\% \pm 0.7$	\\
    CSI   & BFId \cite{todt_BFId_2025} & $41.1\% \pm 11.2$ & $39.9\% \pm 7.6$ & $46.1\% \pm 13.7$ & $37.8\% \pm 6.1$	\\
    BFI & BFId \cite{todt_BFId_2025} & $38.6\% \pm 1.1$ & $36.9\% \pm 1.9$ & $34.3\% \pm 1.2$ & $36.3\% \pm 1.4$	\\
    \bottomrule
  \end{tabular}
\end{table*}

\begin{table*}
  \caption{Measured accuracies for selected sensors, and all perspectives in our activity recognition experiment using within-identity splitting.}
  \label{tab:activitiesrelwork}
  \centering
  \small
  \begin{tabular}{@{}llcccc@{}}
    \toprule
    Sensor & Rec. System & Center-Low & Center-High & Left & Right \\
    \midrule
    radar & SRPNet \cite{cheng_person_2022} & $47.9\% \pm 0.6$ & --- & $58.4\% \pm 1.1$ & $67.6\% \pm 0.6$	\\
    CSI &  BFId \cite{todt_BFId_2025} & $43.2\% \pm 11.3$ & $42.7\% \pm 9.3$ & $46.4\% \pm 10.5$ & $51.2\% \pm 9.2$	\\
    BFI & BFId \cite{todt_BFId_2025} & $64.7\% \pm 1.2$ & $67.2\% \pm 0.6$ & $68.3\% \pm 0.7$ & $81.5\% \pm 1.0$ \\
    \bottomrule
  \end{tabular}
\end{table*}

\subsection{Validation}
We validated MultiGait by performing attribute, activity and identity inferences.
See the following benchmark section for details on identity inferences and the following two subsections for details on attribute and activity inference.
In general we find that all inferences, that we would expect to be possible from related work, to also be possible on our dataset.

\subsubsection{Activity Inference}

In our dataset, every participant performs five different activities: walking normally, walking fast (as if they were late to a meeting), with a backpack, carrying a bottle crate and through a turnstile.
The goal of activity inference is to distinguish these activities.
We therefore trained machine learning models %
with the activities as labels and tested to which extent the classification of unseen samples is correct.

While for many of the sensors present in MultiGait, there exist purpose-built activity recognition systems, we used the same systems as for our identity inference benchmark. %
This therefore presents a simplified scenario with un-optimized recognition systems, underscoring the robustness of activity recognition on our dataset.
We used the best performing identity inference recognition system for each sensor and trained them using the different activities as labels.
For this, we used the same hyperparameters that we found for these sensor-system combinations for identity inference, as these have already shown high efficacy. %

While we recorded double the amount of samples for the normal walking style, in order to have more samples for identification, we used only half of them for our activity inference, so that the dataset is balanced.
As our pre-processing scheme is not optimized for the turnstile activity, we do not use it for this validation.
We are particularly interested to which extent activity recognition can generalize across identities.
We therefore split our dataset based on identities, i.e. 80\% of identities (with all their samples for all activities) were used as training data while the remaining 20\% of identities were used for testing.

The results for this experiment can be found in \cref{tab:activities}.
We find that we can, as expected, robustly infer activities on all imaging-based sensors, as well as lidar, with accuracies in excess of 90\%.
While we still significantly outperform chance-level (25\%) for radar, CSI and BFI, our results do not match existing activity inference literature on these sensors.
We investigate this disparity in the following.

We hypothesize that this disparity is due to the model not generalizing the activities beyond specific individuals. Thus, us splitting the dataset in such a way that any single identity is either in the training or the test dataset, results in significantly lower accuracy.
At the same time, many other datasets for activity inference only have very few individuals, because they rather focus on a large number of activities.
Therefore, the same identity is required to be in both the training and test dataset, which could help inference performance.

To test this hypothesis, we conducted additional experiments for radar, CSI and BFI where we infer activities, but rather than splitting by identity, we split our data within identities.
This means that for every identity, for every activity 80\% of samples are used for training and the remaining 20\% for testing.

The result can be found in \cref{tab:activitiesrelwork}.
We find that this does increase activity inference accuracy, supporting our hypothesis.
For CSI and BFI, accuracy is significantly increased and is now closer to related work. While we still do not match accuracy entirely, we attribute the remaining difference to related work using specifically optimized hyperparameters, higher sample rates and different activities.
For radar, the difference is relatively small, though it was also already higher than CSI and BFI.

\subsubsection{Attribute Inference}
\begin{figure*}[tb]
    \centering
    \includegraphics[width=0.98\textwidth]{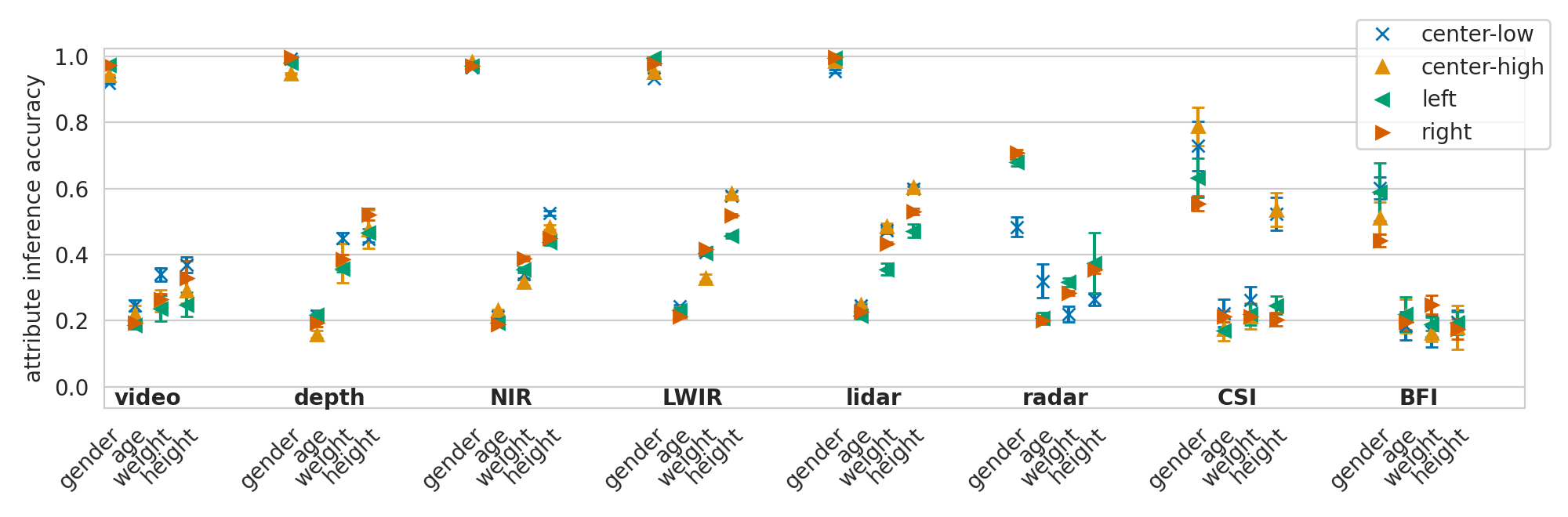}
    \caption{Measured accuracies for all sensors, all attributes and all perspectives in our attribute inference experiment.}
    \label{fig:attr}
\end{figure*}

When recording our dataset, we also collected personal attributes of each subject.
For attribute inference, the goal is to infer these attributes from the recordings.
For our validation, we used age, gender\footnote{We would like to point out that we use the term 'gender recognition' here because the labels that were used for training and testing are the gender identity that participants input in our study. At the same time, we acknowledge that features learned by our system likely refer to the biological sex of individuals which can inherently result in misclassification.}, height, and weight as attributes for inference, though MultiGait does contain more, as these are the most commonly used in related work. %

Similarly to our activity recognition experiment, we re-used our identity inference recognition systems and their hyperparameters.
For gender, we simply used the categories as labels.
For each numerical attribute, we created five non-overlapping categories, such that they are as balanced as possible.
Note that due to the attributes not being unique, this still results in slightly un-balanced categories.
We therefore report balanced accuracy in this experiment.
We ignored identities that did not provide an answer for the specific attribute or that input a custom response.

Our results can be found in \cref{fig:attr}, see \cref{tab:attributes} in the Appendix for details.
We find that gender can be robustly inferred with all imaging-based sensors and lidar.
Inference of age rarely exceeds the chance level accuracy of 20\%. We consider this a result of our biased dataset that contains only a limited age range of 18-34, see \cref{sec:demographics}. This therefore means that categories are very similar which makes this inference difficult.
While height and weight inference exceed chance level, it is lower than we expected. We consider this to be a result of our pre-processing that normalizes height and width of the extracted silhouettes. As such, information is lost that could be used by a more purpose-built system. %

\section{Biometric Identity Inference Benchmark}

MultiGait allows us to investigate the identity inference potential of smart city sensors -- and therefore their potential privacy risks -- in more depth compared to extant work.
Particularly, since the vast majority of previous datasets considers single sensors in isolation, it remains unclear how the identification potential under otherwise unchanged circumstances varies between sensors. %
Additionally, existing datasets only contain information from individuals from a single session, so the impact of, for example, changes in lighting, clothing, or cadence remains underexplored.
Finally, we consider the impact of different perspectives to be under-explored: Existing work either considers very specific changes in viewing angle (while keeping the relation of the sensor to the ground unchanged) or collects it in an uncontrolled manner.
To investigate these aspects, we selected multiple state-of-the-art identification systems for each of our sensors.
Subsequently, we conducted two experiments, one for single-session identity inference, and one for multi-session identity inference. Both experiments cover every perspective and sensor-system-combination.
We explicitly include a single-session experiment to highlight the discrepancy between single- and multi-session.

\subsection{Recognition System Selection}\label{sec:recsystems}
As it is unclear which existing recognition system would yield the highest identification accuracy, we selected multiple state-of-the-art systems for each type of sensor.
We considered systems from recent or seminal publications, systems that were frequently used as baselines, and systems occurring in benchmark papers.
For our experiment, we curated the set of systems to a) have been previously demonstrated to deliver high identification accuracies on large datasets, b) feature a diverse group of fundamental designs instead of incremental improvements of each other, and c) be easy to reproduce (available implementation) or replicate (comprehensive description of functionality).

For our imaging-based sensors, we chose GaitBase \cite{fan_opengait_2023}, DeepGaitv2 \cite{fan_opengait_2025}, GaitPart \cite{fan_gaitpart_2020}, GaitSet \cite{chao_gaitset_2019}, GaitGL \cite{lin_gait_2021} and GEINet \cite{shiraga_geinet_2016}. %
For lidar, we chose LidarGait \cite{shen_lidargait_2023}, LidarGait++ \cite{shen_lidargait_2025}, and GEINet \cite{shiraga_geinet_2016}.
For radar, we chose mmGaitNet \cite{meng_gait_2020}, SRPNet \cite{cheng_person_2022} and mID \cite{zhao_mid_2019}.
For CSI, we chose BFId \cite{todt_BFId_2025}, LW-WiID \cite{cao_lightweight_2021}, CAUTION \cite{wang_caution_2022} and FreeSense \cite{xin_freesense_2016}.
Finally, for BFI, we are limited to BFId \cite{todt_BFId_2025} as the only existing BFI-based identification system.

\subsection{Implementation}
We used the SEBA framework \cite{todt2024sebastrongevaluationbiometric} to implement our experiments.
We used existing open-source implementations of the approaches whenever possible, including OpenGait \cite{fan_opengait_2023} for many video and lidar recognition systems.
The code will be made publicly available with the publication of this paper.

\paragraph{Modifications to existing Systems}
Some recognition systems and their implementations required modifications to fit the setup in our biometric identification benchmark.
All of them are documented in the following section.

For GaitBase \cite{fan_opengait_2023}, DeepGaitv2 \cite{fan_opengait_2025}, GaitSet \cite{chao_gaitset_2019}, GaitPart \cite{fan_gaitpart_2020}, and GaitGL \cite{lin_gait_2021} we used the implementations in the OpenGait framework \cite{fan_opengait_2025}.
Only GEINet \cite{shiraga_geinet_2016} was implemented by us without functional changes from the original paper.

For radar, we implemented all methods ourselves using the information from the original papers.
However, due to changes in our setup compared to theirs and occasional missing information, some changes and assumptions had to be made.
Given that recordings in MultiGait only have one person at a time, the multi-person clustering stage following DBScan for mmGaitNet \cite{meng_gait_2020}, SRPNet \cite{cheng_person_2022} and mID \cite{zhao_mid_2019} was omitted.
For the same reason, we did not use the Hungarian algorithm.
The original implementation of mmGaitNet integrated the points of two recording devices, while we only use one perspective at a time. %
Furthermore, we substituted the signal strength in the input data with the relative timestamps of the points.
For SRPNet, it was necessary to replace the softmax activation function in the attention module with a sigmoid function.
This change was made to enable meaningful and learnable gating of the feature vector.
For mID, due to our significantly larger dataset, we were forced to disable augmentation during training, as to not exceed resource constraints.

For lidar, we used the implementations of LidarGait \cite{shen_lidargait_2023} and LidarGait++ \cite{shen_lidargait_2025} in OpenGait \cite{fan_opengait_2025}.
We used the same GEINet \cite{shiraga_geinet_2016} implementation as for the imaging-based sensors after using the same pre-processing as LidarGait.
For all three methods, the background was removed by eliminating all points outside of a bounding box that we manually determined to include all points from walking individuals, but not the floor or background walls.
LidarGait's pre-processing was modified slightly since our lidar sensor is not spherical.
To create the silhouettes, we simply rendered all points of a frame as a depth map from the view point of the center-low perspective.
The frame length was determined empirically based on the time before sensed points duplicated.

For all CSI and BFI recognition systems, we used the implementations from \cite{todt_BFId_2025}. %

\paragraph{Sensor-Specific Pre-Processing}
Most sensors' recordings required pre-processing. %
For example, all imaging-based sensors required silhouette extraction for all chosen recognition systems.
We tested a variety of common approaches in the literature for each sensor and empirically determined the best.
For video, we use the DeepLabv3 deep learning segmentation model \cite{10.1007/978-3-030-01234-2_49}.
For depth, we simply subtract an image of the empty room for every frame.
For NIR and LWIR, we use the MOG2 background segmentation algorithm \cite{10.5555/1018428.1020644}, as implemented in OpenCV \cite{bradski_opencv_2000}.
In all cases, we afterwards applied thresholding and denoising.
If the recognition system's description specified any (additional) pre-processing, we followed their instructions, particularly for radar, lidar, CSI and BFI.

\subsection{Experiment Setup}
For our single-session experiments, we randomly selected a single session for each participant.
Then, we followed the state-of-the-art by splitting the dataset 70/10/20 into training, validation and test datasets within identities, i.e. for each identity 70\% of samples were used for training, 10\% for validation and the remaining 20\% for testing.

For every sensor-recognition system combination we conducted a hyperparameter optimization using the center-low perspective on the training and validation datasets of the single-session experiments.
Optimization was done with the Tree-structured Parzen Estimator of the Optuna library \cite{optuna_2019}.
Where available, we tested the hyperparameters suggested by the authors of the recognition system first.
We optimized until we either found an accuracy above 99\% or to a maximum of 50 iterations.
We then used these hyperparameters for all experiments of this sensor-system combination, including other perspectives. %
Due to the required resources, we consider hyperparameter optimizing for every individual experiment to be out-of-scope. %

For our identification experiments, we used the hyperparameters with the highest validation accuracy and the conjunction of the training and validation dataset for training while evaluating with the testing dataset.
To account for variability in model training, we repeated every experiment five times, as common in related work. %
Unless otherwise stated, we report the mean and standard deviation of the identification accuracy.

For our multi-session experiments, we tested additional hyperparameter optimizations on five random sensor-system combinations, but with the exception of radar (where validation accuracy still remained below 10\%), we found no improvements.
Therefore, to save resources, we used the same hyperparameters for our multi-session experiment as for our single-session experiment.
We dedicated the second session of our recordings as our testing dataset, while all recordings from the other weeks are used for training.
In other words, the recognition systems did not observe any sample during training that was recorded in the same recording session as the testing data points.
Thus, no session related information of the testing session could have been learned.
This also means that our training data contains more individuals than our testing data, as MultiGait only contains two or more sessions for 64 out of the total 199 individuals.

\subsection{Results}

\paragraph{Single-session}
\begin{figure*}[tb]
    \centering
    \includegraphics[width=0.98\textwidth]{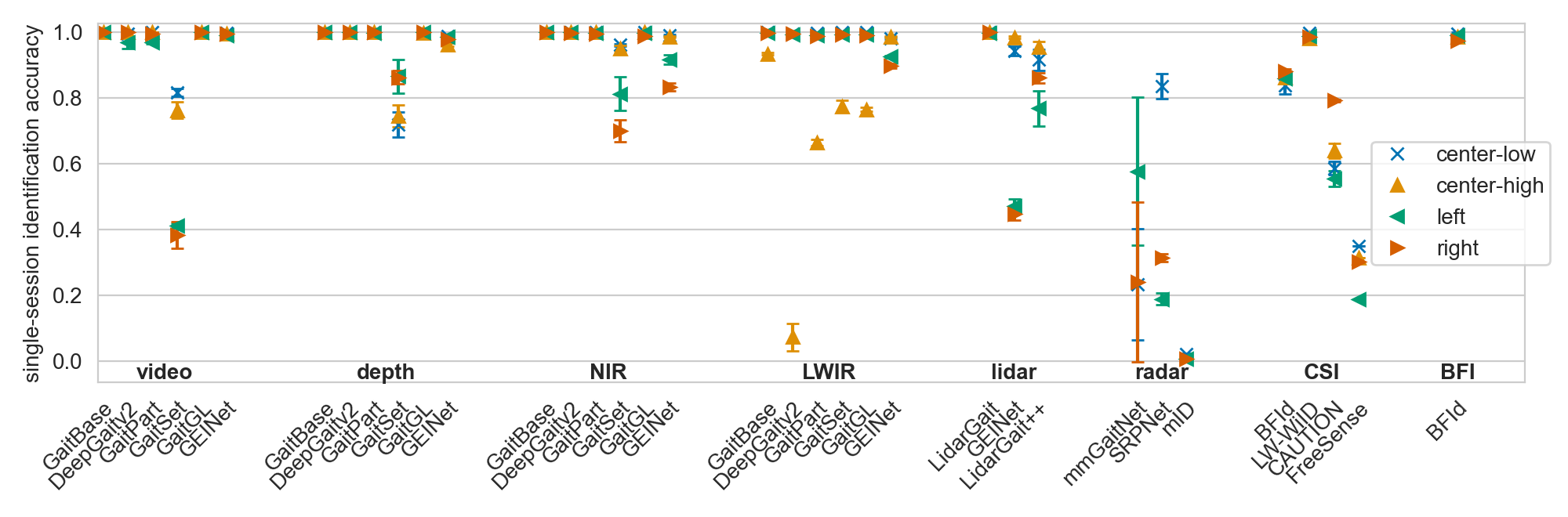}
    \caption{Measured accuracies for all sensors, all recognition systems and all perspectives in our single-session experiment.}
    \label{fig:ss}
\end{figure*}

The results of our single-session experiment can be found in \cref{fig:ss}. Please refer to \cref{tab:ss} in the Appendix for the full details.

Overall, our results show that identity inference is possible with all of the sensing technologies that are present in MultiGait.
All sensors show high identification potential with at least one recognition system's accuracy per sensor exceeding 80\% and even exceeding 99\% for all sensors excluding radar.

We generally find very little difference between perspectives. %
A notable exception is the center-high perspective of LWIR: Most recognition systems achieve 20\%-points lower accuracy than on the other perspectives. DeepGaitv2 even only achieves 7.3\% identification accuracy. %
We find this to be due to inconsistencies in silhouette extraction.

We find very little difference between all imaging based sensors with almost all measurements exceeding 98\% accuracy. %
This suggests that once silhouettes have been extracted from the respective recordings, the actual sensing technology has little impact on recognition accuracy. %
An exception is GaitSet, where many experiments perform significantly worse compared to other recognition systems.
This could be due to GaitSet being designed for cross-view scenarios, which we do not test here.

For non-imaging based sensors, some recognition systems, such as mmGaitNet (where we also see very high variance), mID, and FreeSense do not achieve high identification accuracies on our dataset, particularly in comparison to their claims on other datasets. %
In these cases, we performed additional experiments to validate our implementation and experiment design, see the last paragraph of this section.

\paragraph{Multi-Session}
\begin{figure*}[tb]
    \centering
    \includegraphics[width=0.98\textwidth]{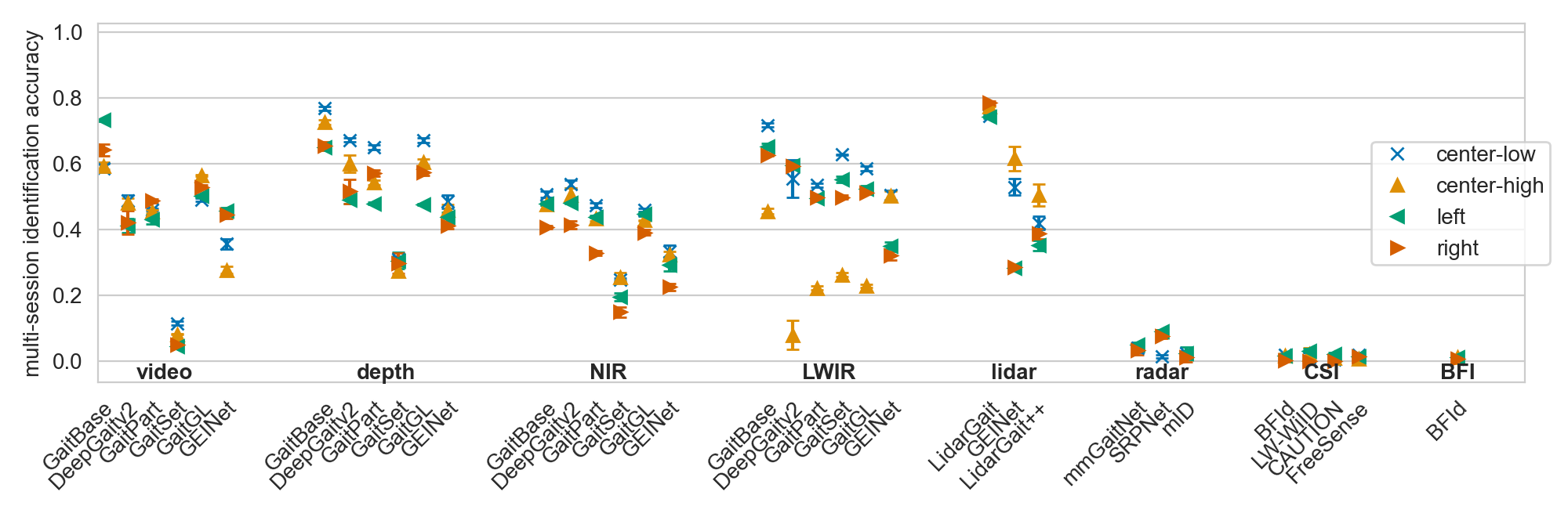}
    \caption{Measured accuracies for all sensors, all recognition systems and all perspectives in our multi-session experiment.}
    \label{fig:ms}
\end{figure*}
The results for our multi-session experiments can be found in \cref{fig:ms}. Please refer to \cref{tab:ms} in the Appendix for the full details.

Contrary to the single session results, we find that existing recognition systems do not yet generalize well to multiple sessions.
We find no accuracy higher than 80\% showing that reliable identification across sessions is not yet possible.

While recognition systems for imaging-based sensors and lidar are at least able to correctly identify in the majority of cases, systems using radar and WiFi artifacts do not significantly exceed chance-level accuracy. %
Most of the radar and WiFi recognition systems are softmax classifiers, which by design learn data-to-label relationships, compared to representation-based classifiers (predominantly used by imaging-based recognition systems) that learn data-to-data relationships.
Therefore, we hypothesize that the softmax-based classifiers do not properly disaggregate session-information from biometric-information and thus fail to generalize to unseen sessions. %

We see some of the same trends as we did for single-session with GaitSet performing worse than other imaging-based systems and low variances across perspectives. %

Interestingly, non-video imaging-based sensors (i.e.~thermal cameras, depth cameras, and lidar) do \emph{not} lack behind video cameras in identification accuracy.
While one might expect video recordings to have the highest quality and therefore identification accuracy, this is not the case.
Particularly for the center-low and center-high perspectives, video cameras are actually out-performed by depth by often more than 10 percentage-points and and by LWIR on the center-low perspective.
This could be explained by the fact that current gait-focused recognition systems only use a subset of the information present in the recordings: the extracted silhouettes. %
This shows that alternative sensors are not necessarily more privacy-friendly by design than video cameras.

\paragraph{Investigation of Disparities to Related Work in Single-Session Identification}
From our single-session experiment, we consider the performance of mmGaitNet \cite{meng_gait_2020} on our radar data which we find to both significantly lower and with a higher variability than its original paper.
We also consider the performance of mID \cite{zhao_mid_2019} on radar data to be below what is expected from its original paper.
Finally, we consider FreeSense \cite{xin_freesense_2016} on CSI-based WiFi sensing to be underperforming.

For each of these cases, we test their identification performance in a modified single-session experiment where we randomly select a subset of identities from MultiGait to match the number of individuals tested in their original papers.
This tests whether the low results are due to the large size of MultiGait.

\begin{table*}
  \caption{Measured accuracies for selected sensors and recognition systems for all perspectives in comparison to the original results.}
  \label{tab:ssrelwork}
  \centering
  \small
  \begin{tabular}{@{}llcccccc@{}}
    \toprule
    Sensor & Rec. System & No. identities & Original claim & Center-Low & Center-High & Left & Right \\
    \midrule
    radar & mmGaitNet \cite{meng_gait_2020} & 10 & 90\% & $81.0\% \pm 3.0$ & --- & $59.5\% \pm 3.1$ & $72.3\% \pm 2.1$	\\
                & & 20 & 80\% & $73.4\% \pm 0.7$ & --- & $50.6\% \pm 1.8$ & $52.3\% \pm 4.1$\\
          & mID \cite{zhao_mid_2019} & 12 & 89\% & $20.4\% \pm 1.0$ & --- & $8.3\% \pm 0.0$ & $8.3\% \pm 0.0$	\\
    \midrule
    CSI & FreeSense \cite{xin_freesense_2016} & 6 & 88.9\% & $87.5\%$ & $70.8\%$ & $63.3\%$ & $91.7\%$\\
    \bottomrule
  \end{tabular}
\end{table*}

The results can be found in \cref{tab:ssrelwork}.
With the exception of mID, we find that with the smaller datasets we can match the originally claimed accuracies.
This implies that we correctly replicated these recognition systems, however, they simply do not scale to the size of MultiGait.
For mID, we find that we can only outperform random guessing in the center-low perspective in which we run our hyperparameter optimization.
This suggests that the efficacy of this recognition system is strongly dependent on its hyperparameters and we may not yet have optimal ones.

\section{Discussion}

The benchmark yields two different key findings:
First, we show in our single-session experiment that all tested sensing technologies have high potential for identity inference.
This refutes many previous claims that some of the tested sensing technologies are privacy-preserving and instead calls for work on mitigation approaches, such as anonymizations.

Second, we show in our multi-session experiment that existing gait-focused recognition systems are not yet able to reliably generalize across sessions.
It is unclear to which extent this is a limitation of current recognition systems or whether the biometric data that the sensors record simply do not allow for such inferences because they are not robust or stable enough.
Considering the lack of relevant datasets until now, we expect future improvements to the recognition systems. %
Particularly, the only recently proposed systems for mmWave radar, CSI and BFI appear to be unsuited for this task, highlighting this as a current research gap. %
The apparent difference in identification accuracy between single and multi-session setups also further highlights the importance of multi-session datasets, such as MultiGait.

\subsection{Implications}
MultiGait has significant implications for research and practice alike.
First, it allows for a wealth of novel experiments and insights.
Apart from enabling training and evaluation of recognition systems for multi-session, it also is the largest available dataset for some of the included sensors (e.g., radar and NIR). %
By including various personal attributes for each participant and multiple activities, it enables attribute and activity inference experiments.
This is especially important for developing anonymization methods, which ideally need one dataset with enough subjects to test identity inference (as a proxy for anonymity) and multiple activities to test activity inference (as a proxy for utility).

Second, the synchronized, simultaneous recording of multiple sensors and multiple perspectives enables development of multi- and cross-sensor, as well as multi-perspective, recognition systems.
As current recognition systems mostly rely on a single information source (e.g., only video cameras from one perspective), it  remains largely unknown how their combination affects the privacy-utility trade-off.
Thus, MultiGait provides a significant contribution to investigations on the robust modeling of human mobility.

Third, as we have shown that all tested sensing technologies have high identification potential, their deployment in smart cities can introduce significant privacy risks.
As such, the deployment of these sensors should be subject to the same public scrutiny as video cameras and their data must be treated under the same legal framework as any other personal, biometric data. %

\subsection{Limitations}
While MultiGait is designed to be a comprehensive dataset that addresses many of the limitations of existing datasets, we acknowledge some shortcomings.
In particular, some video gait-focused datasets include more subjects.
However, given the sheer effort it takes to record datasets that contain the wealth of information as ours (multi-sensor, multi-session, number of activities and repetitions,~\dots) while also attaining the proper informed consent from participants, creating larger datasets is outside of the scope of academic research.
We purposefully restricted our dataset to a lab environment to ensure more reliable, reproducible sensor comparisons and a best-case scenario for multi-session experiments, even though an uncontrolled outdoor setting would be more realistic.
Further, the evaluation of anonymization methods does not require datasets with a large number of individuals, as recent work on their evaluation methodology has shown \cite{hanisch_false_2024}.

MultiGait also is not representative of the entire population.
It is limited to able-bodied individuals and biased in terms of ethnicity and age.
While such a homogeneous population sample makes identification theoretically even more challenging and thus possibly underestimates the identification risk, the generalization of the results always must be contextualized with these biases.

As is common, our benchmark is not able to test every single available recognition system, but rather representative state-of-the-art approaches had to be selected.
While this presents a limitation to our benchmark, we consider the impact negligible, mitigated by selecting a variety of diverse approaches.

\subsection{Future Work}
Based on our two key findings, we highlight two areas for future work.
First, considering the high identification potential of all sensing technologies in this paper, we consider the development of privacy-enhancing technologies for them critical.
Second, considering the research gap highlighted by our multi-session experiments, we suggest work on improved recognition systems such as developing representation-based learning approaches for radar, CSI and BFI.

In addition to these areas of future work based on our benchmark, the MultiGait dataset enables a range of future research directions.
On the technical side, work on cross-sensor recognition systems and multi-modal fusion is enabled by MultiGait containing a large variety of sensing technologies.
On a socio-technical side, we enable a more well-informed design and deployment process for sensor systems through a clearer privacy-utility trade-off. This means that (combinations of) sensors, their resolution, their field-of-view, and similar can be chosen in such a way that the specific utility goal of the sensor system can be achieved while providing utmost privacy for example in combination with anonymization method, potentially located at the edge. This can result in privacy by design at the sensor.

\section{Conclusion}
In this paper, we introduced MultiGait, a dataset containing the recordings of eight sensing technologies across multiple perspectives and sessions in a synchronized, full-factorial design.
MultiGait enabled us to conduct the first comparative benchmark of the privacy risks of smart city sensing technologies.
It showed that many sensors -- though claimed to be privacy-friendly -- actually have a high identification risk, but at the same time, current recognition systems fail to generalize to multiple sessions, highlighting a research gap. %
Furthermore, MultiGait enables future rigorous assessment of robustness, modeling of human motion and activities, and temporal generalization under controlled yet realistic conditions.
We release MultiGait as a community resource to drive reproducible research, fair benchmarking, and the next generation of multi-modal and privacy-aware sensing systems.

\bibliographystyle{ACM-Reference-Format}
\bibliography{main}


\begin{thebibliography}{81}


\ifx \showCODEN    \undefined \def \showCODEN     #1{\unskip}     \fi
\ifx \showISBNx    \undefined \def \showISBNx     #1{\unskip}     \fi
\ifx \showISBNxiii \undefined \def \showISBNxiii  #1{\unskip}     \fi
\ifx \showISSN     \undefined \def \showISSN      #1{\unskip}     \fi
\ifx \showLCCN     \undefined \def \showLCCN      #1{\unskip}     \fi
\ifx \shownote     \undefined \def \shownote      #1{#1}          \fi
\ifx \showarticletitle \undefined \def \showarticletitle #1{#1}   \fi
\ifx \showURL      \undefined \def \showURL       {\relax}        \fi
\providecommand\bibfield[2]{#2}
\providecommand\bibinfo[2]{#2}
\providecommand\natexlab[1]{#1}
\providecommand\showeprint[2][]{arXiv:#2}

\bibitem[Akiba et~al\mbox{.}(2019)]%
        {optuna_2019}
\bibfield{author}{\bibinfo{person}{Takuya Akiba}, \bibinfo{person}{Shotaro Sano}, \bibinfo{person}{Toshihiko Yanase}, \bibinfo{person}{Takeru Ohta}, {and} \bibinfo{person}{Masanori Koyama}.} \bibinfo{year}{2019}\natexlab{}.
\newblock \showarticletitle{Optuna: A next-generation hyperparameter optimization framework}. In \bibinfo{booktitle}{\emph{Proceedings of the 25th ACM SIGKDD international conference on knowledge discovery \& data mining}}. \bibinfo{pages}{2623--2631}.
\newblock


\bibitem[Andersson and Araujo(2015)]%
        {andersson_person_2015}
\bibfield{author}{\bibinfo{person}{Virginia Andersson} {and} \bibinfo{person}{Ricardo Araujo}.} \bibinfo{year}{2015}\natexlab{}.
\newblock \showarticletitle{Person {{Identification Using Anthropometric}} and {{Gait Data}} from {{Kinect Sensor}}}.
\newblock \bibinfo{journal}{\emph{Proceedings of the AAAI Conference on Artificial Intelligence}} \bibinfo{volume}{29}, \bibinfo{number}{1} (\bibinfo{date}{Feb.} \bibinfo{year}{2015}).
\newblock
\showISSN{2374-3468, 2159-5399}
\href{https://doi.org/10.1609/aaai.v29i1.9212}{doi:\nolinkurl{10.1609/aaai.v29i1.9212}}


\bibitem[Baghezza et~al\mbox{.}(2022)]%
        {baghezza_profile_2022}
\bibfield{author}{\bibinfo{person}{Rani Baghezza}, \bibinfo{person}{K{\'e}vin Bouchard}, \bibinfo{person}{Abdenour Bouzouane}, {and} \bibinfo{person}{Charles {Gouin-Vallerand}}.} \bibinfo{year}{2022}\natexlab{}.
\newblock \showarticletitle{Profile {{Recognition}} for {{Accessibility}} and {{Inclusivity}} in {{Smart Cities Using}} a {{Thermal Imaging Sensor}} in an {{Embedded System}}}.
\newblock \bibinfo{journal}{\emph{IEEE Internet of Things Journal}} \bibinfo{volume}{9}, \bibinfo{number}{10} (\bibinfo{date}{May} \bibinfo{year}{2022}), \bibinfo{pages}{7491--7509}.
\newblock
\showISSN{2327-4662}
\href{https://doi.org/10.1109/JIOT.2021.3127137}{doi:\nolinkurl{10.1109/JIOT.2021.3127137}}


\bibitem[Barbosa et~al\mbox{.}(2012)]%
        {hutchison_re-identification_2012}
\bibfield{author}{\bibinfo{person}{Igor~Barros Barbosa}, \bibinfo{person}{Marco Cristani}, \bibinfo{person}{Alessio Del~Bue}, \bibinfo{person}{Loris Bazzani}, {and} \bibinfo{person}{Vittorio Murino}.} \bibinfo{year}{2012}\natexlab{}.
\newblock \showarticletitle{Re-Identification with {{RGB-D Sensors}}}.
\newblock In \bibinfo{booktitle}{\emph{Computer {{Vision}} -- {{ECCV}} 2012. {{Workshops}} and {{Demonstrations}}}}, \bibfield{editor}{\bibinfo{person}{David Hutchison}, \bibinfo{person}{Takeo Kanade}, \bibinfo{person}{Josef Kittler}, \bibinfo{person}{Jon~M. Kleinberg}, \bibinfo{person}{Friedemann Mattern}, \bibinfo{person}{John~C. Mitchell}, \bibinfo{person}{Moni Naor}, \bibinfo{person}{Oscar Nierstrasz}, \bibinfo{person}{C.~Pandu~Rangan}, \bibinfo{person}{Bernhard Steffen}, \bibinfo{person}{Madhu Sudan}, \bibinfo{person}{Demetri Terzopoulos}, \bibinfo{person}{Doug Tygar}, \bibinfo{person}{Moshe~Y. Vardi}, \bibinfo{person}{Gerhard Weikum}, \bibinfo{person}{Andrea Fusiello}, \bibinfo{person}{Vittorio Murino}, {and} \bibinfo{person}{Rita Cucchiara}} (Eds.). Vol.~\bibinfo{volume}{7583}. \bibinfo{publisher}{Springer Berlin Heidelberg}, \bibinfo{address}{Berlin, Heidelberg}, \bibinfo{pages}{433--442}.
\newblock
\showISBNx{978-3-642-33862-5 978-3-642-33863-2}
\href{https://doi.org/10.1007/978-3-642-33863-2\_43}{doi:\nolinkurl{10.1007/978-3-642-33863-2\_43}}


\bibitem[BBC(2015)]%
        {uspolice_radar_2015}
\bibfield{author}{\bibinfo{person}{BBC}.} \bibinfo{year}{2015}\natexlab{}.
\newblock \bibinfo{title}{Radar that 'sees' through walls raises privacy concerns}.
\newblock \bibinfo{howpublished}{\url{https://www.bbc.com/news/technology-30904218} [Accessed: 2025-11-13]}.
\newblock


\bibitem[Bradski(2000)]%
        {bradski_opencv_2000}
\bibfield{author}{\bibinfo{person}{Gary Bradski}.} \bibinfo{year}{2000}\natexlab{}.
\newblock \showarticletitle{The {{openCV}} Library.}
\newblock \bibinfo{journal}{\emph{Dr. Dobb's Journal: Software Tools for the Professional Programmer}} \bibinfo{volume}{25}, \bibinfo{number}{11} (\bibinfo{year}{2000}), \bibinfo{pages}{120--123}.
\newblock


\bibitem[Cao et~al\mbox{.}(2018)]%
        {cao_vggface2_2018}
\bibfield{author}{\bibinfo{person}{Qiong Cao}, \bibinfo{person}{Li Shen}, \bibinfo{person}{Weidi Xie}, \bibinfo{person}{Omkar~M. Parkhi}, {and} \bibinfo{person}{Andrew Zisserman}.} \bibinfo{year}{2018}\natexlab{}.
\newblock \bibinfo{title}{{{VGGFace2}}: {{A}} Dataset for Recognising Faces across Pose and Age}.
\newblock
\showeprint[arxiv]{1710.08092}~[cs]


\bibitem[Cao et~al\mbox{.}(2021)]%
        {cao_lightweight_2021}
\bibfield{author}{\bibinfo{person}{Yangjie Cao}, \bibinfo{person}{Zhiyi Zhou}, \bibinfo{person}{Chenxi Zhu}, \bibinfo{person}{Pengsong Duan}, \bibinfo{person}{Xianfu Chen}, {and} \bibinfo{person}{Jie Li}.} \bibinfo{year}{2021}\natexlab{}.
\newblock \showarticletitle{A {{Lightweight Deep Learning Algorithm}} for {{WiFi-Based Identity Recognition}}}.
\newblock \bibinfo{journal}{\emph{IEEE Internet of Things Journal}} \bibinfo{volume}{8}, \bibinfo{number}{24} (\bibinfo{date}{Dec.} \bibinfo{year}{2021}), \bibinfo{pages}{17449--17459}.
\newblock
\showISSN{2327-4662, 2372-2541}
\href{https://doi.org/10.1109/JIOT.2021.3078782}{doi:\nolinkurl{10.1109/JIOT.2021.3078782}}


\bibitem[Chao et~al\mbox{.}(2019)]%
        {chao_gaitset_2019}
\bibfield{author}{\bibinfo{person}{Hanqing Chao}, \bibinfo{person}{Yiwei He}, \bibinfo{person}{Junping Zhang}, {and} \bibinfo{person}{Jianfeng Feng}.} \bibinfo{year}{2019}\natexlab{}.
\newblock \showarticletitle{{{GaitSet}}: Regarding Gait as a Set for Cross-View Gait Recognition}. In \bibinfo{booktitle}{\emph{Proceedings of the {{Thirty-Third AAAI Conference}} on {{Artificial Intelligence}} and {{Thirty-First Innovative Applications}} of {{Artificial Intelligence Conference}} and {{Ninth AAAI Symposium}} on {{Educational Advances}} in {{Artificial Intelligence}}}} \emph{(\bibinfo{series}{{{AAAI}}'19/{{IAAI}}'19/{{EAAI}}'19}, Vol.~\bibinfo{volume}{33})}. \bibinfo{publisher}{AAAI Press}, \bibinfo{address}{Honolulu, Hawaii, USA}, \bibinfo{pages}{8126--8133}.
\newblock
\showISBNx{978-1-57735-809-1}
\href{https://doi.org/10.1609/aaai.v33i01.33018126}{doi:\nolinkurl{10.1609/aaai.v33i01.33018126}}


\bibitem[Chen et~al\mbox{.}(2018)]%
        {10.1007/978-3-030-01234-2_49}
\bibfield{author}{\bibinfo{person}{Liang-Chieh Chen}, \bibinfo{person}{Yukun Zhu}, \bibinfo{person}{George Papandreou}, \bibinfo{person}{Florian Schroff}, {and} \bibinfo{person}{Hartwig Adam}.} \bibinfo{year}{2018}\natexlab{}.
\newblock \showarticletitle{Encoder-Decoder with Atrous Separable Convolution for Semantic Image Segmentation}. In \bibinfo{booktitle}{\emph{Computer Vision -- ECCV 2018}}, \bibfield{editor}{\bibinfo{person}{Vittorio Ferrari}, \bibinfo{person}{Martial Hebert}, \bibinfo{person}{Cristian Sminchisescu}, {and} \bibinfo{person}{Yair Weiss}} (Eds.). \bibinfo{publisher}{Springer International Publishing}, \bibinfo{address}{Cham}, \bibinfo{pages}{833--851}.
\newblock
\showISBNx{978-3-030-01234-2}


\bibitem[Cheng and Liu(2022)]%
        {cheng_person_2022}
\bibfield{author}{\bibinfo{person}{Yuwei Cheng} {and} \bibinfo{person}{Yimin Liu}.} \bibinfo{year}{2022}\natexlab{}.
\newblock \showarticletitle{Person {{Reidentification Based}} on {{Automotive Radar Point Clouds}}}.
\newblock \bibinfo{journal}{\emph{IEEE Transactions on Geoscience and Remote Sensing}}  \bibinfo{volume}{60} (\bibinfo{year}{2022}), \bibinfo{pages}{1--13}.
\newblock
\showISSN{1558-0644}
\href{https://doi.org/10.1109/TGRS.2021.3073664}{doi:\nolinkurl{10.1109/TGRS.2021.3073664}}


\bibitem[Collini et~al\mbox{.}(2024)]%
        {collini_flexible_2024}
\bibfield{author}{\bibinfo{person}{Enrico Collini}, \bibinfo{person}{Luciano Alessandro~Ipsaro Palesi}, \bibinfo{person}{Paolo Nesi}, \bibinfo{person}{Gianni Pantaleo}, {and} \bibinfo{person}{William Zhao}.} \bibinfo{year}{2024}\natexlab{}.
\newblock \showarticletitle{Flexible Thermal Camera Solution for {{Smart}} City People Detection and Counting}.
\newblock \bibinfo{journal}{\emph{Multimedia Tools and Applications}} \bibinfo{volume}{83}, \bibinfo{number}{7} (\bibinfo{date}{Feb.} \bibinfo{year}{2024}), \bibinfo{pages}{20457--20485}.
\newblock
\showISSN{1573-7721}
\href{https://doi.org/10.1007/s11042-023-16374-x}{doi:\nolinkurl{10.1007/s11042-023-16374-x}}


\bibitem[Communications(2022)]%
        {axiscomm_privacyLWIR_2022}
\bibfield{author}{\bibinfo{person}{Axis Communications}.} \bibinfo{year}{2022}\natexlab{}.
\newblock \bibinfo{title}{Thermal Cameras}.
\newblock \bibinfo{howpublished}{\url{https://www.axis.com/dam/public/98/87/26/brochure--axis-thermal-cameras-en-US-360992.pdf} [Accessed: 2025-11-13]}.
\newblock


\bibitem[Dantcheva et~al\mbox{.}(2016)]%
        {dantcheva_what_2016}
\bibfield{author}{\bibinfo{person}{Antitza Dantcheva}, \bibinfo{person}{Petros Elia}, {and} \bibinfo{person}{Arun Ross}.} \bibinfo{year}{2016}\natexlab{}.
\newblock \showarticletitle{What {{Else Does Your Biometric Data Reveal}}? {{A Survey}} on {{Soft Biometrics}}}.
\newblock \bibinfo{journal}{\emph{IEEE Transactions on Information Forensics and Security}} \bibinfo{volume}{11}, \bibinfo{number}{3} (\bibinfo{date}{March} \bibinfo{year}{2016}), \bibinfo{pages}{441--467}.
\newblock
\showISSN{1556-6013, 1556-6021}
\href{https://doi.org/10.1109/TIFS.2015.2480381}{doi:\nolinkurl{10.1109/TIFS.2015.2480381}}


\bibitem[DeCann et~al\mbox{.}(2013)]%
        {decann_investigating_2013}
\bibfield{author}{\bibinfo{person}{Brian DeCann}, \bibinfo{person}{Arun Ross}, {and} \bibinfo{person}{Jeremy Dawson}.} \bibinfo{year}{2013}\natexlab{}.
\newblock \showarticletitle{Investigating Gait Recognition in the Short-Wave Infrared ({{SWIR}}) Spectrum: Dataset and Challenges}. In \bibinfo{booktitle}{\emph{{{SPIE Defense}}, {{Security}}, and {{Sensing}}}}, \bibfield{editor}{\bibinfo{person}{Ioannis Kakadiaris}, \bibinfo{person}{Walter~J. Scheirer}, {and} \bibinfo{person}{Laurence~G. Hassebrook}} (Eds.). \bibinfo{address}{Baltimore, Maryland, USA}, \bibinfo{pages}{87120J}.
\newblock
\href{https://doi.org/10.1117/12.2018145}{doi:\nolinkurl{10.1117/12.2018145}}


\bibitem[Deng et~al\mbox{.}(2019)]%
        {deng_arcface_2019}
\bibfield{author}{\bibinfo{person}{Jiankang Deng}, \bibinfo{person}{Jia Guo}, \bibinfo{person}{Niannan Xue}, {and} \bibinfo{person}{Stefanos Zafeiriou}.} \bibinfo{year}{2019}\natexlab{}.
\newblock \bibinfo{title}{Arcface: Additive angular margin loss for deep face recognition}.
\newblock \bibinfo{numpages}{4690--4699}~pages.
\newblock


\bibitem[{Espinosa-Dur{\'o}} et~al\mbox{.}(2013)]%
        {espinosa-duro_new_2013}
\bibfield{author}{\bibinfo{person}{Virginia {Espinosa-Dur{\'o}}}, \bibinfo{person}{Marcos {Faundez-Zanuy}}, {and} \bibinfo{person}{Ji{\v r}{\'i} Mekyska}.} \bibinfo{year}{2013}\natexlab{}.
\newblock \showarticletitle{A {{New Face Database Simultaneously Acquired}} in {{Visible}}, {{Near-Infrared}} and {{Thermal Spectrums}}}.
\newblock \bibinfo{journal}{\emph{Cognitive Computation}} \bibinfo{volume}{5}, \bibinfo{number}{1} (\bibinfo{date}{March} \bibinfo{year}{2013}), \bibinfo{pages}{119--135}.
\newblock
\showISSN{1866-9964}
\href{https://doi.org/10.1007/s12559-012-9163-2}{doi:\nolinkurl{10.1007/s12559-012-9163-2}}


\bibitem[Fan et~al\mbox{.}(2025)]%
        {fan_opengait_2025}
\bibfield{author}{\bibinfo{person}{Chao Fan}, \bibinfo{person}{Saihui Hou}, \bibinfo{person}{Junhao Liang}, \bibinfo{person}{Chuanfu Shen}, \bibinfo{person}{Jingzhe Ma}, \bibinfo{person}{Dongyang Jin}, \bibinfo{person}{Yongzhen Huang}, {and} \bibinfo{person}{Shiqi Yu}.} \bibinfo{year}{2025}\natexlab{}.
\newblock \showarticletitle{{{OpenGait}}: {{A Comprehensive Benchmark Study}} for {{Gait Recognition Towards Better Practicality}}}.
\newblock \bibinfo{journal}{\emph{IEEE Transactions on Pattern Analysis and Machine Intelligence}} (\bibinfo{year}{2025}), \bibinfo{pages}{1--18}.
\newblock
\showISSN{1939-3539}
\href{https://doi.org/10.1109/TPAMI.2025.3576283}{doi:\nolinkurl{10.1109/TPAMI.2025.3576283}}


\bibitem[Fan et~al\mbox{.}(2023)]%
        {fan_opengait_2023}
\bibfield{author}{\bibinfo{person}{Chao Fan}, \bibinfo{person}{Junhao Liang}, \bibinfo{person}{Chuanfu Shen}, \bibinfo{person}{Saihui Hou}, \bibinfo{person}{Yongzhen Huang}, {and} \bibinfo{person}{Shiqi Yu}.} \bibinfo{year}{2023}\natexlab{}.
\newblock \showarticletitle{{{OpenGait}}: {{Revisiting Gait Recognition Toward Better Practicality}}}. In \bibinfo{booktitle}{\emph{2023 {{IEEE}}/{{CVF Conference}} on {{Computer Vision}} and {{Pattern Recognition}} ({{CVPR}})}}. \bibinfo{publisher}{IEEE}, \bibinfo{address}{Vancouver, BC, Canada}, \bibinfo{pages}{9707--9716}.
\newblock
\showISBNx{979-8-3503-0129-8}
\href{https://doi.org/10.1109/CVPR52729.2023.00936}{doi:\nolinkurl{10.1109/CVPR52729.2023.00936}}


\bibitem[Fan et~al\mbox{.}(2020)]%
        {fan_gaitpart_2020}
\bibfield{author}{\bibinfo{person}{Chao Fan}, \bibinfo{person}{Yunjie Peng}, \bibinfo{person}{Chunshui Cao}, \bibinfo{person}{Xu Liu}, \bibinfo{person}{Saihui Hou}, \bibinfo{person}{Jiannan Chi}, \bibinfo{person}{Yongzhen Huang}, \bibinfo{person}{Qing Li}, {and} \bibinfo{person}{Zhiqiang He}.} \bibinfo{year}{2020}\natexlab{}.
\newblock \showarticletitle{{{GaitPart}}: {{Temporal Part-Based Model}} for {{Gait Recognition}}}. In \bibinfo{booktitle}{\emph{2020 {{IEEE}}/{{CVF Conference}} on {{Computer Vision}} and {{Pattern Recognition}} ({{CVPR}})}}. \bibinfo{publisher}{IEEE}, \bibinfo{address}{Seattle, WA, USA}, \bibinfo{pages}{14213--14221}.
\newblock
\showISBNx{978-1-7281-7168-5}
\href{https://doi.org/10.1109/CVPR42600.2020.01423}{doi:\nolinkurl{10.1109/CVPR42600.2020.01423}}


\bibitem[Gad and Sadek(2025)]%
        {gad_total_2025}
\bibfield{author}{\bibinfo{person}{Peter Gad} {and} \bibinfo{person}{Mohamed Sadek}.} \bibinfo{year}{2025}\natexlab{}.
\newblock \showarticletitle{Total {{Road Control}}: {{A LiDAR-Powered Vision}} for~{{Smart}} and~{{Safe City Transformations}}}.
\newblock In \bibinfo{booktitle}{\emph{Internet of {{Vehicles}} and {{Computer Vision Solutions}} for {{Smart City Transformations}}}}, \bibfield{editor}{\bibinfo{person}{Anuj Abraham}, \bibinfo{person}{Shitala Prasad}, \bibinfo{person}{Ahmed Alhammadi}, \bibinfo{person}{Thierry Lestable}, {and} \bibinfo{person}{Ferdaous Chaabane}} (Eds.). \bibinfo{publisher}{Springer Nature Switzerland}, \bibinfo{address}{Cham}, \bibinfo{pages}{291--307}.
\newblock
\showISBNx{978-3-031-72959-1}
\href{https://doi.org/10.1007/978-3-031-72959-1\_13}{doi:\nolinkurl{10.1007/978-3-031-72959-1\_13}}


\bibitem[Gade et~al\mbox{.}(2016)]%
        {gade2016thermal}
\bibfield{author}{\bibinfo{person}{Rikke Gade}, \bibinfo{person}{Thomas~B Moeslund}, \bibinfo{person}{S{\o}ren~Zebitz Nielsen}, \bibinfo{person}{Hans Skov-Petersen}, \bibinfo{person}{Hans~J{\o}rgen Andersen}, \bibinfo{person}{Kent Basselbjerg}, \bibinfo{person}{Hans~Thorhauge Dam}, \bibinfo{person}{Ole~B Jensen}, \bibinfo{person}{Anders J{\o}rgensen}, \bibinfo{person}{Harry Lahrmann}, {et~al\mbox{.}}} \bibinfo{year}{2016}\natexlab{}.
\newblock \showarticletitle{Thermal imaging systems for real-time applications in smart cities}.
\newblock \bibinfo{journal}{\emph{International Journal of Computer Applications in Technology}} \bibinfo{volume}{53}, \bibinfo{number}{4} (\bibinfo{year}{2016}), \bibinfo{pages}{291--308}.
\newblock


\bibitem[Gonzalez et~al\mbox{.}(2025)]%
        {gonzalez_inferring_2025}
\bibfield{author}{\bibinfo{person}{Cinthya Celina~Tamayo Gonzalez}, \bibinfo{person}{Simone Soderi}, \bibinfo{person}{Julian Todt}, \bibinfo{person}{Thorsten Strufe}, {and} \bibinfo{person}{Mauro Conti}.} \bibinfo{year}{2025}\natexlab{}.
\newblock \showarticletitle{Inferring {{Personal Attributes}} with a {{Mmwave Radar}}}. In \bibinfo{booktitle}{\emph{2025 {{IEEE Wireless Communications}} and {{Networking Conference}} ({{WCNC}})}}. \bibinfo{pages}{1--6}.
\newblock
\showISSN{1558-2612}
\href{https://doi.org/10.1109/WCNC61545.2025.10978264}{doi:\nolinkurl{10.1109/WCNC61545.2025.10978264}}


\bibitem[Greiner(2015)]%
        {greiner_subject_2015}
\bibfield{author}{\bibinfo{person}{Ben Greiner}.} \bibinfo{year}{2015}\natexlab{}.
\newblock \showarticletitle{Subject Pool Recruitment Procedures: Organizing Experiments with {{ORSEE}}}.
\newblock \bibinfo{journal}{\emph{Journal of the Economic Science Association}} \bibinfo{volume}{1}, \bibinfo{number}{1} (\bibinfo{date}{July} \bibinfo{year}{2015}), \bibinfo{pages}{114--125}.
\newblock
\showISSN{2199-6784}
\href{https://doi.org/10.1007/s40881-015-0004-4}{doi:\nolinkurl{10.1007/s40881-015-0004-4}}


\bibitem[Guo et~al\mbox{.}(2025a)]%
        {guo_camera-lidar_2025}
\bibfield{author}{\bibinfo{person}{Wenxuan Guo}, \bibinfo{person}{Yingping Liang}, \bibinfo{person}{Zhiyu Pan}, \bibinfo{person}{Ziheng Xi}, \bibinfo{person}{Jianjiang Feng}, {and} \bibinfo{person}{Jie Zhou}.} \bibinfo{year}{2025}\natexlab{a}.
\newblock \showarticletitle{Camera-{{LiDAR Cross-Modality Gait Recognition}}}. In \bibinfo{booktitle}{\emph{Computer {{Vision}} -- {{ECCV}} 2024}}, \bibfield{editor}{\bibinfo{person}{Ale{\v s} Leonardis}, \bibinfo{person}{Elisa Ricci}, \bibinfo{person}{Stefan Roth}, \bibinfo{person}{Olga Russakovsky}, \bibinfo{person}{Torsten Sattler}, {and} \bibinfo{person}{G{\"u}l Varol}} (Eds.). \bibinfo{publisher}{Springer Nature Switzerland}, \bibinfo{address}{Cham}, \bibinfo{pages}{439--455}.
\newblock
\showISBNx{978-3-031-72754-2}
\href{https://doi.org/10.1007/978-3-031-72754-2\_25}{doi:\nolinkurl{10.1007/978-3-031-72754-2\_25}}


\bibitem[Guo et~al\mbox{.}(2025b)]%
        {zhu_gait_2025}
\bibfield{author}{\bibinfo{person}{Xianda Guo}, \bibinfo{person}{Zheng Zhu}, \bibinfo{person}{Tian Yang}, \bibinfo{person}{Beibei Lin}, \bibinfo{person}{Junjie Huang}, \bibinfo{person}{Jiankang Deng}, \bibinfo{person}{Guan Huang}, \bibinfo{person}{Jie Zhou}, {and} \bibinfo{person}{Jiwen Lu}.} \bibinfo{year}{2025}\natexlab{b}.
\newblock \showarticletitle{Gait Recognition in the Wild: A Large-Scale Benchmark and NAS-Based Baseline}.
\newblock \bibinfo{journal}{\emph{IEEE Transactions on Pattern Analysis and Machine Intelligence}} \bibinfo{volume}{47}, \bibinfo{number}{6} (\bibinfo{year}{2025}), \bibinfo{pages}{4535--4552}.
\newblock
\href{https://doi.org/10.1109/TPAMI.2025.3546482}{doi:\nolinkurl{10.1109/TPAMI.2025.3546482}}


\bibitem[Hanisch et~al\mbox{.}(2023)]%
        {hanisch_understanding_2023}
\bibfield{author}{\bibinfo{person}{Simon Hanisch}, \bibinfo{person}{Evelyn Muschter}, \bibinfo{person}{Admantini Hatzipanayioti}, \bibinfo{person}{Shu-Chen Li}, {and} \bibinfo{person}{Thorsten Strufe}.} \bibinfo{year}{2023}\natexlab{}.
\newblock \showarticletitle{Understanding {{Person Identification Through Gait}}}.
\newblock \bibinfo{journal}{\emph{Proceedings on Privacy Enhancing Technologies}} (\bibinfo{year}{2023}).
\newblock
\showISSN{2299-0984}


\bibitem[Hanisch et~al\mbox{.}(2024)]%
        {hanisch_false_2024}
\bibfield{author}{\bibinfo{person}{Simon Hanisch}, \bibinfo{person}{Julian Todt}, \bibinfo{person}{Jose Patino}, \bibinfo{person}{Nicholas Evans}, {and} \bibinfo{person}{Thorsten Strufe}.} \bibinfo{year}{2024}\natexlab{}.
\newblock \showarticletitle{A {{False Sense}} of {{Privacy}}: {{Towards}} a {{Reliable Evaluation Methodology}} for the {{Anonymization}} of {{Biometric Data}}}.
\newblock \bibinfo{journal}{\emph{Proceedings on Privacy Enhancing Technologies}} \bibinfo{volume}{2024}, \bibinfo{number}{1} (\bibinfo{date}{Jan.} \bibinfo{year}{2024}), \bibinfo{pages}{116--132}.
\newblock
\showISSN{2299-0984}
\href{https://doi.org/10.56553/popets-2024-0008}{doi:\nolinkurl{10.56553/popets-2024-0008}}


\bibitem[Hofmann et~al\mbox{.}(2014)]%
        {hofmann_tum_2014}
\bibfield{author}{\bibinfo{person}{Martin Hofmann}, \bibinfo{person}{J{\"u}rgen Geiger}, \bibinfo{person}{Sebastian Bachmann}, \bibinfo{person}{Bj{\"o}rn Schuller}, {and} \bibinfo{person}{Gerhard Rigoll}.} \bibinfo{year}{2014}\natexlab{}.
\newblock \showarticletitle{The {{TUM Gait}} from {{Audio}}, {{Image}} and {{Depth}} ({{GAID}}) Database: {{Multimodal}} Recognition of Subjects and Traits}.
\newblock \bibinfo{journal}{\emph{Journal of Visual Communication and Image Representation}} \bibinfo{volume}{25}, \bibinfo{number}{1} (\bibinfo{date}{Jan.} \bibinfo{year}{2014}), \bibinfo{pages}{195--206}.
\newblock
\showISSN{10473203}
\href{https://doi.org/10.1016/j.jvcir.2013.02.006}{doi:\nolinkurl{10.1016/j.jvcir.2013.02.006}}


\bibitem[IEEE(2025)]%
        {ieee80211bf}
\bibfield{author}{\bibinfo{person}{IEEE}.} \bibinfo{year}{2025}\natexlab{}.
\newblock \bibinfo{booktitle}{\emph{IEEE Standard for Information Technology -- Telecommunications and Information Exchange Between Systems Local and Metropolitan Area Networks -- Specific Requirements - Part 11: Wireless LAN Medium Access Control (MAC) and Physical Layer (PHY) Specifications - Amendment 4: Enhancements for Wireless LAN Sensing, IEEE Std 802.11bf}}.
\newblock \bibinfo{type}{{T}echnical {R}eport}. \bibinfo{institution}{IEEE}.
\newblock


\bibitem[Iwama et~al\mbox{.}(2012)]%
        {iwama_ou-isir_2012}
\bibfield{author}{\bibinfo{person}{H. Iwama}, \bibinfo{person}{M. Okumura}, \bibinfo{person}{Y. Makihara}, {and} \bibinfo{person}{Y. Yagi}.} \bibinfo{year}{2012}\natexlab{}.
\newblock \showarticletitle{The {{OU-ISIR Gait Database Comprising}} the {{Large Population Dataset}} and {{Performance Evaluation}} of {{Gait Recognition}}}.
\newblock \bibinfo{journal}{\emph{IEEE Transactions on Information Forensics and Security}} \bibinfo{volume}{7}, \bibinfo{number}{5} (\bibinfo{date}{Oct.} \bibinfo{year}{2012}), \bibinfo{pages}{1511--1521}.
\newblock
\showISSN{1556-6013, 1556-6021}
\href{https://doi.org/10.1109/TIFS.2012.2204253}{doi:\nolinkurl{10.1109/TIFS.2012.2204253}}


\bibitem[Kepski and Kwolek(2014)]%
        {kepski_fall_2014}
\bibfield{author}{\bibinfo{person}{Michal Kepski} {and} \bibinfo{person}{Bogdan Kwolek}.} \bibinfo{year}{2014}\natexlab{}.
\newblock \showarticletitle{Fall Detection Using Ceiling-Mounted {{3D}} Depth Camera}. In \bibinfo{booktitle}{\emph{2014 {{International Conference}} on {{Computer Vision Theory}} and {{Applications}} ({{VISAPP}})}}, Vol.~\bibinfo{volume}{2}. \bibinfo{pages}{640--647}.
\newblock


\bibitem[Lacin{\'a}k and Ristvej(2017)]%
        {lacinak_smart_2017}
\bibfield{author}{\bibinfo{person}{Maro{\v s} Lacin{\'a}k} {and} \bibinfo{person}{Jozef Ristvej}.} \bibinfo{year}{2017}\natexlab{}.
\newblock \showarticletitle{Smart {{City}}, {{Safety}} and {{Security}}}.
\newblock \bibinfo{journal}{\emph{Procedia Engineering}}  \bibinfo{volume}{192} (\bibinfo{year}{2017}), \bibinfo{pages}{522--527}.
\newblock
\showISSN{18777058}
\href{https://doi.org/10.1016/j.proeng.2017.06.090}{doi:\nolinkurl{10.1016/j.proeng.2017.06.090}}


\bibitem[Li et~al\mbox{.}(2022)]%
        {li_deep_2022}
\bibfield{author}{\bibinfo{person}{Chenning Li}, \bibinfo{person}{Zhichao Cao}, {and} \bibinfo{person}{Yunhao Liu}.} \bibinfo{year}{2022}\natexlab{}.
\newblock \showarticletitle{Deep {{AI Enabled Ubiquitous Wireless Sensing}}: {{A Survey}}}.
\newblock \bibinfo{journal}{\emph{Comput. Surveys}} \bibinfo{volume}{54}, \bibinfo{number}{2} (\bibinfo{date}{March} \bibinfo{year}{2022}), \bibinfo{pages}{1--35}.
\newblock
\showISSN{0360-0300, 1557-7341}
\href{https://doi.org/10.1145/3436729}{doi:\nolinkurl{10.1145/3436729}}


\bibitem[Li et~al\mbox{.}(2023)]%
        {li_-depth_2023}
\bibfield{author}{\bibinfo{person}{Weijia Li}, \bibinfo{person}{Saihui Hou}, \bibinfo{person}{Chunjie Zhang}, \bibinfo{person}{Chunshui Cao}, \bibinfo{person}{Xu Liu}, \bibinfo{person}{Yongzhen Huang}, {and} \bibinfo{person}{Yao Zhao}.} \bibinfo{year}{2023}\natexlab{}.
\newblock \showarticletitle{An {{In-Depth Exploration}} of {{Person Re-Identification}} and {{Gait Recognition}} in {{Cloth-Changing Conditions}}}. In \bibinfo{booktitle}{\emph{2023 {{IEEE}}/{{CVF Conference}} on {{Computer Vision}} and {{Pattern Recognition}} ({{CVPR}})}}. \bibinfo{publisher}{IEEE}, \bibinfo{address}{Vancouver, BC, Canada}, \bibinfo{pages}{13824--13833}.
\newblock
\showISBNx{979-8-3503-0129-8}
\href{https://doi.org/10.1109/CVPR52729.2023.01328}{doi:\nolinkurl{10.1109/CVPR52729.2023.01328}}


\bibitem[Lin et~al\mbox{.}(2021)]%
        {lin_gait_2021}
\bibfield{author}{\bibinfo{person}{Beibei Lin}, \bibinfo{person}{Shunli Zhang}, {and} \bibinfo{person}{Xin Yu}.} \bibinfo{year}{2021}\natexlab{}.
\newblock \showarticletitle{Gait {{Recognition}} via {{Effective Global-Local Feature Representation}} and {{Local Temporal Aggregation}}}. In \bibinfo{booktitle}{\emph{2021 {{IEEE}}/{{CVF International Conference}} on {{Computer Vision}} ({{ICCV}})}}. \bibinfo{publisher}{IEEE}, \bibinfo{address}{Montreal, QC, Canada}, \bibinfo{pages}{14628--14636}.
\newblock
\showISBNx{978-1-6654-2812-5}
\href{https://doi.org/10.1109/ICCV48922.2021.01438}{doi:\nolinkurl{10.1109/ICCV48922.2021.01438}}


\bibitem[Lintvedt(2023)]%
        {lintvedt_thermal_2023}
\bibfield{author}{\bibinfo{person}{Naomi Lintvedt}.} \bibinfo{year}{2023}\natexlab{}.
\newblock \showarticletitle{Thermal {{Imaging}} in {{Robotics}} as a {{Privacy-Enhancing}} or {{Privacy-Invasive Measure}}? {{Misconceptions}} of {{Privacy}} When {{Using Thermal Cameras}} in {{Robots}}}.
\newblock \bibinfo{journal}{\emph{Digital Society}} \bibinfo{volume}{2}, \bibinfo{number}{3} (\bibinfo{date}{Sept.} \bibinfo{year}{2023}), \bibinfo{pages}{33}.
\newblock
\showISSN{2731-4669}
\href{https://doi.org/10.1007/s44206-023-00060-4}{doi:\nolinkurl{10.1007/s44206-023-00060-4}}


\bibitem[Ma et~al\mbox{.}(2020)]%
        {ma_wifi_2020}
\bibfield{author}{\bibinfo{person}{Yongsen Ma}, \bibinfo{person}{Gang Zhou}, {and} \bibinfo{person}{Shuangquan Wang}.} \bibinfo{year}{2020}\natexlab{}.
\newblock \showarticletitle{{{WiFi Sensing}} with {{Channel State Information}}: {{A Survey}}}.
\newblock \bibinfo{journal}{\emph{Comput. Surveys}} \bibinfo{volume}{52}, \bibinfo{number}{3} (\bibinfo{date}{May} \bibinfo{year}{2020}), \bibinfo{pages}{1--36}.
\newblock
\showISSN{0360-0300, 1557-7341}
\href{https://doi.org/10.1145/3310194}{doi:\nolinkurl{10.1145/3310194}}


\bibitem[Makihara et~al\mbox{.}(2012)]%
        {makihara_ou-isir_2012}
\bibfield{author}{\bibinfo{person}{Yasushi Makihara}, \bibinfo{person}{Hidetoshi Mannami}, \bibinfo{person}{Akira Tsuji}, \bibinfo{person}{Md.~Altab Hossain}, \bibinfo{person}{Kazushige Sugiura}, \bibinfo{person}{Atsushi Mori}, {and} \bibinfo{person}{Yasushi Yagi}.} \bibinfo{year}{2012}\natexlab{}.
\newblock \showarticletitle{The {{OU-ISIR Gait Database Comprising}} the {{Treadmill Dataset}}}.
\newblock \bibinfo{journal}{\emph{IPSJ Transactions on Computer Vision and Applications}}  \bibinfo{volume}{4} (\bibinfo{year}{2012}), \bibinfo{pages}{53--62}.
\newblock
\showISSN{1882-6695}
\href{https://doi.org/10.2197/ipsjtcva.4.53}{doi:\nolinkurl{10.2197/ipsjtcva.4.53}}


\bibitem[Meng et~al\mbox{.}(2020)]%
        {meng_gait_2020}
\bibfield{author}{\bibinfo{person}{Zhen Meng}, \bibinfo{person}{Song Fu}, \bibinfo{person}{Jie Yan}, \bibinfo{person}{Hongyuan Liang}, \bibinfo{person}{Anfu Zhou}, \bibinfo{person}{Shilin Zhu}, \bibinfo{person}{Huadong Ma}, \bibinfo{person}{Jianhua Liu}, {and} \bibinfo{person}{Ning Yang}.} \bibinfo{year}{2020}\natexlab{}.
\newblock \showarticletitle{Gait {{Recognition}} for {{Co-Existing Multiple People Using Millimeter Wave Sensing}}}.
\newblock \bibinfo{journal}{\emph{Proceedings of the AAAI Conference on Artificial Intelligence}} \bibinfo{volume}{34}, \bibinfo{number}{01} (\bibinfo{date}{April} \bibinfo{year}{2020}), \bibinfo{pages}{849--856}.
\newblock
\showISSN{2374-3468}
\href{https://doi.org/10.1609/aaai.v34i01.5430}{doi:\nolinkurl{10.1609/aaai.v34i01.5430}}


\bibitem[Michel(2025)]%
        {nypd_thermal_2025}
\bibfield{author}{\bibinfo{person}{C.D. Michel}.} \bibinfo{year}{2025}\natexlab{}.
\newblock \bibinfo{title}{NYPD Developing Remote Thermal Concealed Gun Detectors}.
\newblock \bibinfo{howpublished}{\url{https://michellawyers.com/nypd-developing-remote-thermal-concealed-gun-detectors/} [Accessed: 2025-11-13]}.
\newblock


\bibitem[Mohanty et~al\mbox{.}(2016)]%
        {mohanty_everything_2016}
\bibfield{author}{\bibinfo{person}{Saraju~P. Mohanty}, \bibinfo{person}{Uma Choppali}, {and} \bibinfo{person}{Elias Kougianos}.} \bibinfo{year}{2016}\natexlab{}.
\newblock \showarticletitle{Everything You Wanted to Know about Smart Cities: {{The Internet}} of Things Is the Backbone}.
\newblock \bibinfo{journal}{\emph{IEEE Consumer Electronics Magazine}} \bibinfo{volume}{5}, \bibinfo{number}{3} (\bibinfo{date}{July} \bibinfo{year}{2016}), \bibinfo{pages}{60--70}.
\newblock
\showISSN{2162-2256}
\href{https://doi.org/10.1109/MCE.2016.2556879}{doi:\nolinkurl{10.1109/MCE.2016.2556879}}


\bibitem[Mucha and Kampel(2022)]%
        {10.1007/978-3-031-08645-8_62}
\bibfield{author}{\bibinfo{person}{Wiktor Mucha} {and} \bibinfo{person}{Martin Kampel}.} \bibinfo{year}{2022}\natexlab{}.
\newblock \showarticletitle{Addressing Privacy Concerns in Depth Sensors}. In \bibinfo{booktitle}{\emph{Computers Helping People with Special Needs}}, \bibfield{editor}{\bibinfo{person}{Klaus Miesenberger}, \bibinfo{person}{Georgios Kouroupetroglou}, \bibinfo{person}{Katerina Mavrou}, \bibinfo{person}{Roberto Manduchi}, \bibinfo{person}{Mario Covarrubias~Rodriguez}, {and} \bibinfo{person}{Petr Pen{\'a}z}} (Eds.). \bibinfo{publisher}{Springer International Publishing}, \bibinfo{address}{Cham}, \bibinfo{pages}{526--533}.
\newblock
\showISBNx{978-3-031-08645-8}


\bibitem[Murroni et~al\mbox{.}(2023)]%
        {murroni20236g}
\bibfield{author}{\bibinfo{person}{Maurizio Murroni}, \bibinfo{person}{Matteo Anedda}, \bibinfo{person}{Mauro Fadda}, \bibinfo{person}{Pietro Ruiu}, \bibinfo{person}{Vlad Popescu}, \bibinfo{person}{Corneliu Zaharia}, {and} \bibinfo{person}{Daniele Giusto}.} \bibinfo{year}{2023}\natexlab{}.
\newblock \showarticletitle{6G—Enabling the new smart city: A survey}.
\newblock \bibinfo{journal}{\emph{Sensors}} \bibinfo{volume}{23}, \bibinfo{number}{17} (\bibinfo{year}{2023}), \bibinfo{pages}{7528}.
\newblock


\bibitem[Nixon and Carter(2006)]%
        {nixon_automatic_2006}
\bibfield{author}{\bibinfo{person}{Mark~S. Nixon} {and} \bibinfo{person}{John~N. Carter}.} \bibinfo{year}{2006}\natexlab{}.
\newblock \showarticletitle{Automatic {{Recognition}} by {{Gait}}}.
\newblock \bibinfo{journal}{\emph{Proc. IEEE}} \bibinfo{volume}{94}, \bibinfo{number}{11} (\bibinfo{date}{Nov.} \bibinfo{year}{2006}), \bibinfo{pages}{2013--2024}.
\newblock
\showISSN{1558-2256}
\href{https://doi.org/10.1109/JPROC.2006.886018}{doi:\nolinkurl{10.1109/JPROC.2006.886018}}


\bibitem[Nunes et~al\mbox{.}(2019)]%
        {tavares_gridds_2019}
\bibfield{author}{\bibinfo{person}{Jo{\~a}o~Ferreira Nunes}, \bibinfo{person}{Pedro~Miguel Moreira}, {and} \bibinfo{person}{Jo{\~a}o Manuel R.~S. Tavares}.} \bibinfo{year}{2019}\natexlab{}.
\newblock \showarticletitle{{{GRIDDS}} - {{A Gait Recognition Image}} and {{Depth Dataset}}}.
\newblock In \bibinfo{booktitle}{\emph{{{VipIMAGE}} 2019}}, \bibfield{editor}{\bibinfo{person}{Jo{\~a}o Manuel R.~S. Tavares} {and} \bibinfo{person}{Renato~Manuel Natal~Jorge}} (Eds.). Vol.~\bibinfo{volume}{34}. \bibinfo{publisher}{Springer International Publishing}, \bibinfo{address}{Cham}, \bibinfo{pages}{343--352}.
\newblock
\showISBNx{978-3-030-32039-3 978-3-030-32040-9}
\href{https://doi.org/10.1007/978-3-030-32040-9\_36}{doi:\nolinkurl{10.1007/978-3-030-32040-9\_36}}


\bibitem[Okumura et~al\mbox{.}(2010)]%
        {okumura_performance_2010}
\bibfield{author}{\bibinfo{person}{Mayu Okumura}, \bibinfo{person}{Haruyuki Iwama}, \bibinfo{person}{Yasushi Makihara}, {and} \bibinfo{person}{Yasushi Yagi}.} \bibinfo{year}{2010}\natexlab{}.
\newblock \showarticletitle{Performance Evaluation of Vision-Based Gait Recognition Using a Very Large-Scale Gait Database}. In \bibinfo{booktitle}{\emph{2010 {{Fourth IEEE International Conference}} on {{Biometrics}}: {{Theory}}, {{Applications}} and {{Systems}} ({{BTAS}})}}. \bibinfo{publisher}{IEEE}, \bibinfo{address}{Washington, DC, USA}, \bibinfo{pages}{1--6}.
\newblock
\showISBNx{978-1-4244-7581-0}
\href{https://doi.org/10.1109/BTAS.2010.5634525}{doi:\nolinkurl{10.1109/BTAS.2010.5634525}}


\bibitem[Planinc and Kampel(2013)]%
        {planinc_introducing_2013}
\bibfield{author}{\bibinfo{person}{Rainer Planinc} {and} \bibinfo{person}{Martin Kampel}.} \bibinfo{year}{2013}\natexlab{}.
\newblock \showarticletitle{Introducing the Use of Depth Data for Fall Detection}.
\newblock \bibinfo{journal}{\emph{Personal and Ubiquitous Computing}} \bibinfo{volume}{17}, \bibinfo{number}{6} (\bibinfo{date}{Aug.} \bibinfo{year}{2013}), \bibinfo{pages}{1063--1072}.
\newblock
\showISSN{1617-4917}
\href{https://doi.org/10.1007/s00779-012-0552-z}{doi:\nolinkurl{10.1007/s00779-012-0552-z}}


\bibitem[Rougier et~al\mbox{.}(2011)]%
        {rougier_fall_2011}
\bibfield{author}{\bibinfo{person}{Caroline Rougier}, \bibinfo{person}{Edouard Auvinet}, \bibinfo{person}{Jacqueline Rousseau}, \bibinfo{person}{Max Mignotte}, {and} \bibinfo{person}{Jean Meunier}.} \bibinfo{year}{2011}\natexlab{}.
\newblock \showarticletitle{Fall {{Detection}} from {{Depth Map Video Sequences}}}. In \bibinfo{booktitle}{\emph{Toward {{Useful Services}} for {{Elderly}} and {{People}} with {{Disabilities}}}}, \bibfield{editor}{\bibinfo{person}{Bessam Abdulrazak}, \bibinfo{person}{Sylvain Giroux}, \bibinfo{person}{Bruno Bouchard}, \bibinfo{person}{H{\'e}l{\`e}ne Pigot}, {and} \bibinfo{person}{Mounir Mokhtari}} (Eds.). \bibinfo{publisher}{Springer}, \bibinfo{address}{Berlin, Heidelberg}, \bibinfo{pages}{121--128}.
\newblock
\showISBNx{978-3-642-21535-3}
\href{https://doi.org/10.1007/978-3-642-21535-3\_16}{doi:\nolinkurl{10.1007/978-3-642-21535-3\_16}}


\bibitem[Sarkar et~al\mbox{.}(2005)]%
        {sarkar_humanid_2005}
\bibfield{author}{\bibinfo{person}{S. Sarkar}, \bibinfo{person}{P.J. Phillips}, \bibinfo{person}{Z. Liu}, \bibinfo{person}{I.R. Vega}, \bibinfo{person}{P. Grother}, {and} \bibinfo{person}{K.W. Bowyer}.} \bibinfo{year}{2005}\natexlab{}.
\newblock \showarticletitle{The {{humanID}} Gait Challenge Problem: Data Sets, Performance, and Analysis}.
\newblock \bibinfo{journal}{\emph{IEEE Transactions on Pattern Analysis and Machine Intelligence}} \bibinfo{volume}{27}, \bibinfo{number}{2} (\bibinfo{date}{Feb.} \bibinfo{year}{2005}), \bibinfo{pages}{162--177}.
\newblock
\showISSN{1939-3539}
\href{https://doi.org/10.1109/TPAMI.2005.39}{doi:\nolinkurl{10.1109/TPAMI.2005.39}}


\bibitem[{Sepas-Moghaddam} and Etemad(2023)]%
        {sepas-moghaddam_deep_2023}
\bibfield{author}{\bibinfo{person}{Alireza {Sepas-Moghaddam}} {and} \bibinfo{person}{Ali Etemad}.} \bibinfo{year}{2023}\natexlab{}.
\newblock \showarticletitle{Deep {{Gait Recognition}}: {{A Survey}}}.
\newblock \bibinfo{journal}{\emph{IEEE Transactions on Pattern Analysis and Machine Intelligence}} \bibinfo{volume}{45}, \bibinfo{number}{1} (\bibinfo{date}{Jan.} \bibinfo{year}{2023}), \bibinfo{pages}{264--284}.
\newblock
\showISSN{1939-3539}
\href{https://doi.org/10.1109/TPAMI.2022.3151865}{doi:\nolinkurl{10.1109/TPAMI.2022.3151865}}


\bibitem[Shen et~al\mbox{.}(2023)]%
        {shen_lidargait_2023}
\bibfield{author}{\bibinfo{person}{Chuanfu Shen}, \bibinfo{person}{Fan Chao}, \bibinfo{person}{Wei Wu}, \bibinfo{person}{Rui Wang}, \bibinfo{person}{George~Q. Huang}, {and} \bibinfo{person}{Shiqi Yu}.} \bibinfo{year}{2023}\natexlab{}.
\newblock \showarticletitle{{{LidarGait}}: {{Benchmarking 3D Gait Recognition}} with {{Point Clouds}}}. In \bibinfo{booktitle}{\emph{2023 {{IEEE}}/{{CVF Conference}} on {{Computer Vision}} and {{Pattern Recognition}} ({{CVPR}})}}. \bibinfo{publisher}{IEEE}, \bibinfo{address}{Vancouver, BC, Canada}, \bibinfo{pages}{1054--1063}.
\newblock
\href{https://doi.org/10.1109/cvpr52729.2023.00108}{doi:\nolinkurl{10.1109/cvpr52729.2023.00108}}


\bibitem[Shen et~al\mbox{.}(2025a)]%
        {shen_lidargait_2025}
\bibfield{author}{\bibinfo{person}{Chuanfu Shen}, \bibinfo{person}{Rui Wang}, \bibinfo{person}{Lixin Duan}, {and} \bibinfo{person}{Shiqi Yu}.} \bibinfo{year}{2025}\natexlab{a}.
\newblock \showarticletitle{LidarGait++: Learning Local Features and Size Awareness from LiDAR Point Clouds for 3D Gait Recognition}. In \bibinfo{booktitle}{\emph{2025 IEEE/CVF Conference on Computer Vision and Pattern Recognition (CVPR)}}. \bibinfo{pages}{6627--6636}.
\newblock
\href{https://doi.org/10.1109/CVPR52734.2025.00621}{doi:\nolinkurl{10.1109/CVPR52734.2025.00621}}


\bibitem[Shen et~al\mbox{.}(2025b)]%
        {shen_comprehensive_2025}
\bibfield{author}{\bibinfo{person}{Chuanfu Shen}, \bibinfo{person}{Shiqi Yu}, \bibinfo{person}{Jilong Wang}, \bibinfo{person}{George~Q. Huang}, {and} \bibinfo{person}{Liang Wang}.} \bibinfo{year}{2025}\natexlab{b}.
\newblock \showarticletitle{A {{Comprehensive Survey}} on {{Deep Gait Recognition}}: {{Algorithms}}, {{Datasets}}, and {{Challenges}}}.
\newblock \bibinfo{journal}{\emph{IEEE Transactions on Biometrics, Behavior, and Identity Science}} \bibinfo{volume}{7}, \bibinfo{number}{2} (\bibinfo{date}{April} \bibinfo{year}{2025}), \bibinfo{pages}{270--292}.
\newblock
\showISSN{2637-6407}
\href{https://doi.org/10.1109/TBIOM.2024.3486345}{doi:\nolinkurl{10.1109/TBIOM.2024.3486345}}


\bibitem[Shiraga et~al\mbox{.}(2016)]%
        {shiraga_geinet_2016}
\bibfield{author}{\bibinfo{person}{Kohei Shiraga}, \bibinfo{person}{Yasushi Makihara}, \bibinfo{person}{Daigo Muramatsu}, \bibinfo{person}{Tomio Echigo}, {and} \bibinfo{person}{Yasushi Yagi}.} \bibinfo{year}{2016}\natexlab{}.
\newblock \showarticletitle{{{GEINet}}: {{View-invariant}} Gait Recognition Using a Convolutional Neural Network}. In \bibinfo{booktitle}{\emph{2016 {{International Conference}} on {{Biometrics}} ({{ICB}})}}. \bibinfo{pages}{1--8}.
\newblock
\href{https://doi.org/10.1109/ICB.2016.7550060}{doi:\nolinkurl{10.1109/ICB.2016.7550060}}


\bibitem[Slemen{\v s}ek et~al\mbox{.}(2023)]%
        {slemensek_human_2023}
\bibfield{author}{\bibinfo{person}{Jan Slemen{\v s}ek}, \bibinfo{person}{Iztok Fister}, \bibinfo{person}{Jelka Ger{\v s}ak}, \bibinfo{person}{Bo{\v z}idar Bratina}, \bibinfo{person}{Vesna~Marija van Midden}, \bibinfo{person}{Zvezdan Pirto{\v s}ek}, \bibinfo{person}{Riko {\v S}afari{\v c}}, \bibinfo{person}{Jan Slemen{\v s}ek}, \bibinfo{person}{Iztok Fister}, \bibinfo{person}{Jelka Ger{\v s}ak}, \bibinfo{person}{Bo{\v z}idar Bratina}, \bibinfo{person}{Vesna~Marija van Midden}, \bibinfo{person}{Zvezdan Pirto{\v s}ek}, {and} \bibinfo{person}{Riko {\v S}afari{\v c}}.} \bibinfo{year}{2023}\natexlab{}.
\newblock \showarticletitle{Human {{Gait Activity Recognition Machine Learning Methods}}}.
\newblock \bibinfo{journal}{\emph{Sensors}} \bibinfo{volume}{23}, \bibinfo{number}{2} (\bibinfo{date}{Jan.} \bibinfo{year}{2023}).
\newblock
\showISSN{1424-8220}
\href{https://doi.org/10.3390/s23020745}{doi:\nolinkurl{10.3390/s23020745}}


\bibitem[Snap4City({[n.\,d.]})]%
        {snap4city_privacyLWIR_nodate}
\bibfield{author}{\bibinfo{person}{Snap4City}.} \bibinfo{year}{[n.\,d.]}\natexlab{}.
\newblock \bibinfo{title}{{GDPR} Compliant People Detection and Counting using Thermal Cameras}.
\newblock \bibinfo{howpublished}{\url{https://www.snap4city.org/drupal/node/805} [Accessed: 2025-11-13]}.
\newblock


\bibitem[Song et~al\mbox{.}(2023)]%
        {song_casia-e_2023}
\bibfield{author}{\bibinfo{person}{Chunfeng Song}, \bibinfo{person}{Yongzhen Huang}, \bibinfo{person}{Weining Wang}, {and} \bibinfo{person}{Liang Wang}.} \bibinfo{year}{2023}\natexlab{}.
\newblock \showarticletitle{{{CASIA-E}}: {{A Large Comprehensive Dataset}} for {{Gait Recognition}}}.
\newblock \bibinfo{journal}{\emph{IEEE Transactions on Pattern Analysis and Machine Intelligence}} \bibinfo{volume}{45}, \bibinfo{number}{3} (\bibinfo{date}{March} \bibinfo{year}{2023}), \bibinfo{pages}{2801--2815}.
\newblock
\showISSN{1939-3539}
\href{https://doi.org/10.1109/TPAMI.2022.3183288}{doi:\nolinkurl{10.1109/TPAMI.2022.3183288}}


\bibitem[Stone and Skubic(2011)]%
        {stone_evaluation_2011}
\bibfield{author}{\bibinfo{person}{Erik Stone} {and} \bibinfo{person}{Marjorie Skubic}.} \bibinfo{year}{2011}\natexlab{}.
\newblock \showarticletitle{Evaluation of an Inexpensive Depth Camera for In-Home Gait Assessment}.
\newblock \bibinfo{journal}{\emph{Journal of Ambient Intelligence and Smart Environments}} \bibinfo{volume}{3}, \bibinfo{number}{4} (\bibinfo{year}{2011}), \bibinfo{pages}{349--361}.
\newblock
\showISSN{18761364}
\href{https://doi.org/10.3233/AIS-2011-0124}{doi:\nolinkurl{10.3233/AIS-2011-0124}}


\bibitem[Takemura et~al\mbox{.}(2018)]%
        {takemura_multi-view_2018}
\bibfield{author}{\bibinfo{person}{Noriko Takemura}, \bibinfo{person}{Yasushi Makihara}, \bibinfo{person}{Daigo Muramatsu}, \bibinfo{person}{Tomio Echigo}, {and} \bibinfo{person}{Yasushi Yagi}.} \bibinfo{year}{2018}\natexlab{}.
\newblock \showarticletitle{Multi-View Large Population Gait Dataset and Its Performance Evaluation for Cross-View Gait Recognition}.
\newblock \bibinfo{journal}{\emph{IPSJ Transactions on Computer Vision and Applications}} \bibinfo{volume}{10}, \bibinfo{number}{1} (\bibinfo{date}{Dec.} \bibinfo{year}{2018}), \bibinfo{pages}{4}.
\newblock
\showISSN{1882-6695}
\href{https://doi.org/10.1186/s41074-018-0039-6}{doi:\nolinkurl{10.1186/s41074-018-0039-6}}


\bibitem[Tan et~al\mbox{.}(2006)]%
        {tan_efficient_2006}
\bibfield{author}{\bibinfo{person}{Daoliang Tan}, \bibinfo{person}{Kaiqi Huang}, \bibinfo{person}{Shiqi Yu}, {and} \bibinfo{person}{Tieniu Tan}.} \bibinfo{year}{2006}\natexlab{}.
\newblock \showarticletitle{Efficient {{Night Gait Recognition Based}} on {{Template Matching}}}. In \bibinfo{booktitle}{\emph{Proceedings of the 18th {{International Conference}} on {{Pattern Recognition}} - {{Volume}} 03}} \emph{(\bibinfo{series}{{{ICPR}} '06})}. \bibinfo{publisher}{IEEE Computer Society}, \bibinfo{address}{USA}, \bibinfo{pages}{1000--1003}.
\newblock
\showISBNx{978-0-7695-2521-1}
\href{https://doi.org/10.1109/ICPR.2006.478}{doi:\nolinkurl{10.1109/ICPR.2006.478}}


\bibitem[Thadathil and Saeed(2025)]%
        {10922050}
\bibfield{author}{\bibinfo{person}{Mohammed~Risal Thadathil} {and} \bibinfo{person}{Nasir Saeed}.} \bibinfo{year}{2025}\natexlab{}.
\newblock \showarticletitle{Smart Traffic Intersections: Leveraging ISAC and Millimeter-Wave for Advanced Vehicle Platooning}. In \bibinfo{booktitle}{\emph{2025 Global Information Infrastructure and Networking Symposium (GIIS)}}. \bibinfo{pages}{1--6}.
\newblock
\href{https://doi.org/10.1109/GIIS64151.2025.10922050}{doi:\nolinkurl{10.1109/GIIS64151.2025.10922050}}


\bibitem[Todt et~al\mbox{.}(2024)]%
        {todt2024sebastrongevaluationbiometric}
\bibfield{author}{\bibinfo{person}{Julian Todt}, \bibinfo{person}{Simon Hanisch}, {and} \bibinfo{person}{Thorsten Strufe}.} \bibinfo{year}{2024}\natexlab{}.
\newblock \bibinfo{title}{{SEBA}: Strong Evaluation of Biometric Anonymizations}.
\newblock
\showeprint[arxiv]{2407.06648}~[cs.CR]
\urldef\tempurl%
\url{https://arxiv.org/abs/2407.06648}
\showURL{%
\tempurl}


\bibitem[Todt et~al\mbox{.}(2025)]%
        {todt_BFId_2025}
\bibfield{author}{\bibinfo{person}{Julian Todt}, \bibinfo{person}{Felix Morsbach}, {and} \bibinfo{person}{Thorsten Strufe}.} \bibinfo{year}{2025}\natexlab{}.
\newblock \showarticletitle{{{BFId}}: {{Identity}} Inference Attacks Utilizing Beamforming Feedback Information}. In \bibinfo{booktitle}{\emph{Proceedings of 32nd {{ACM SIGSAC}} Conference on Computer and Communications Security ({{CCS}} '25), Taipei, October 13--17, 2025} (\bibinfo{edition}{2025} ed.)}. \bibinfo{publisher}{Association for Computing Machinery (ACM)}.
\newblock
\showISBNx{979-8-4007-1525-9}
\href{https://doi.org/10.1145/3719027.3765062}{doi:\nolinkurl{10.1145/3719027.3765062}}


\bibitem[Troisi et~al\mbox{.}(2022)]%
        {troisi_managing_2022}
\bibfield{author}{\bibinfo{person}{Orlando Troisi}, \bibinfo{person}{Mohamad Kashef}, {and} \bibinfo{person}{Anna Visvizi}.} \bibinfo{year}{2022}\natexlab{}.
\newblock \showarticletitle{Managing {{Safety}} and {{Security}} in the {{Smart City}}: {{Covid-19}}, {{Emergencies}} and {{Smart Surveillance}}}.
\newblock In \bibinfo{booktitle}{\emph{Managing {{Smart Cities}}: {{Sustainability}} and {{Resilience Through Effective Management}}}}, \bibfield{editor}{\bibinfo{person}{Anna Visvizi} {and} \bibinfo{person}{Orlando Troisi}} (Eds.). \bibinfo{publisher}{Springer International Publishing}, \bibinfo{address}{Cham}, \bibinfo{pages}{73--88}.
\newblock
\showISBNx{978-3-030-93585-6}
\href{https://doi.org/10.1007/978-3-030-93585-6_5}{doi:\nolinkurl{10.1007/978-3-030-93585-6_5}}


\bibitem[USA({[n.\,d.]})]%
        {lynred_privacyLWIR_nodate}
\bibfield{author}{\bibinfo{person}{Lynred USA}.} \bibinfo{year}{[n.\,d.]}\natexlab{}.
\newblock \bibinfo{title}{VISIBLE vs. THERMAL DETECTION: Advantages and Disadvantages}.
\newblock \bibinfo{howpublished}{\url{https://www.lynred-usa.com/homepage/about-us/blog/visible-vs-thermal-detection-advantages-and-disadvantages.html} [Accessed: 2025-11-13]}.
\newblock


\bibitem[Vales et~al\mbox{.}(2024)]%
        {vales_iot_2024}
\bibfield{author}{\bibinfo{person}{Valent{\'i}n~Barral Vales}, \bibinfo{person}{Tom{\'a}s {Dom{\'i}nguez-Bola{\~n}o}}, \bibinfo{person}{Carlos~J. Escudero}, {and} \bibinfo{person}{Jos{\'e}~A. {Garc{\'i}a-Naya}}.} \bibinfo{year}{2024}\natexlab{}.
\newblock \showarticletitle{An {{IoT System}} for {{Smart Building Combining Multiple mmWave FMCW Radars Applied}} to {{People Counting}}}.
\newblock \bibinfo{journal}{\emph{IEEE Internet of Things Journal}} \bibinfo{volume}{11}, \bibinfo{number}{21} (\bibinfo{date}{Nov.} \bibinfo{year}{2024}), \bibinfo{pages}{35306--35316}.
\newblock
\showISSN{2327-4662}
\href{https://doi.org/10.1109/JIOT.2024.3434707}{doi:\nolinkurl{10.1109/JIOT.2024.3434707}}


\bibitem[Villamizar et~al\mbox{.}(2018)]%
        {villamizar_watchnet_2018}
\bibfield{author}{\bibinfo{person}{M. Villamizar}, \bibinfo{person}{A. {Mart{\'i}nez-Gonz{\'a}lez}}, \bibinfo{person}{O. Can{\'e}vet}, {and} \bibinfo{person}{J-M. Odobez}.} \bibinfo{year}{2018}\natexlab{}.
\newblock \showarticletitle{{{WatchNet}}: {{Efficient}} and {{Depth-based Network}} for {{People Detection}} in {{Video Surveillance Systems}}}. In \bibinfo{booktitle}{\emph{2018 15th {{IEEE International Conference}} on {{Advanced Video}} and {{Signal Based Surveillance}} ({{AVSS}})}}. \bibinfo{pages}{1--6}.
\newblock
\href{https://doi.org/10.1109/AVSS.2018.8639165}{doi:\nolinkurl{10.1109/AVSS.2018.8639165}}


\bibitem[Wan et~al\mbox{.}(2019)]%
        {wan_survey_2019}
\bibfield{author}{\bibinfo{person}{Changsheng Wan}, \bibinfo{person}{Li Wang}, {and} \bibinfo{person}{Vir~V. Phoha}.} \bibinfo{year}{2019}\natexlab{}.
\newblock \showarticletitle{A {{Survey}} on {{Gait Recognition}}}.
\newblock \bibinfo{journal}{\emph{Comput. Surveys}} \bibinfo{volume}{51}, \bibinfo{number}{5} (\bibinfo{date}{Sept.} \bibinfo{year}{2019}), \bibinfo{pages}{1--35}.
\newblock
\showISSN{0360-0300, 1557-7341}
\href{https://doi.org/10.1145/3230633}{doi:\nolinkurl{10.1145/3230633}}


\bibitem[Wang et~al\mbox{.}(2022)]%
        {wang_caution_2022}
\bibfield{author}{\bibinfo{person}{Dazhuo Wang}, \bibinfo{person}{Jianfei Yang}, \bibinfo{person}{Wei Cui}, \bibinfo{person}{Lihua Xie}, {and} \bibinfo{person}{Sumei Sun}.} \bibinfo{year}{2022}\natexlab{}.
\newblock \showarticletitle{{{CAUTION}}: {{A Robust WiFi-Based Human Authentication System}} via {{Few-Shot Open-Set Recognition}}}.
\newblock \bibinfo{journal}{\emph{IEEE Internet of Things Journal}} \bibinfo{volume}{9}, \bibinfo{number}{18} (\bibinfo{date}{Sept.} \bibinfo{year}{2022}), \bibinfo{pages}{17323--17333}.
\newblock
\showISSN{2327-4662}
\href{https://doi.org/10.1109/JIOT.2022.3156099}{doi:\nolinkurl{10.1109/JIOT.2022.3156099}}


\bibitem[Wang et~al\mbox{.}(2024)]%
        {wang_cross-modality_2024}
\bibfield{author}{\bibinfo{person}{Rui Wang}, \bibinfo{person}{Chuanfu Shen}, \bibinfo{person}{Manuel~J. {Marin-Jimenez}}, \bibinfo{person}{George~Q. Huang}, {and} \bibinfo{person}{Shiqi Yu}.} \bibinfo{year}{2024}\natexlab{}.
\newblock \showarticletitle{Cross-{{Modality Gait Recognition}}: {{Bridging LiDAR}} and {{Camera Modalities}} for {{Human Identification}}}. In \bibinfo{booktitle}{\emph{2024 {{IEEE International Joint Conference}} on {{Biometrics}} ({{IJCB}})}}. \bibinfo{pages}{1--11}.
\newblock
\showISSN{2474-9699}
\href{https://doi.org/10.1109/IJCB62174.2024.10744428}{doi:\nolinkurl{10.1109/IJCB62174.2024.10744428}}


\bibitem[Xin et~al\mbox{.}(2016)]%
        {xin_freesense_2016}
\bibfield{author}{\bibinfo{person}{Tong Xin}, \bibinfo{person}{Bin Guo}, \bibinfo{person}{Zhu Wang}, \bibinfo{person}{Mingyang Li}, \bibinfo{person}{Zhiwen Yu}, {and} \bibinfo{person}{Xingshe Zhou}.} \bibinfo{year}{2016}\natexlab{}.
\newblock \showarticletitle{{{FreeSense}}: {{Indoor Human Identification}} with {{Wi-Fi Signals}}}. In \bibinfo{booktitle}{\emph{2016 {{IEEE Global Communications Conference}} ({{GLOBECOM}})}}. \bibinfo{pages}{1--7}.
\newblock
\href{https://doi.org/10.1109/GLOCOM.2016.7841847}{doi:\nolinkurl{10.1109/GLOCOM.2016.7841847}}


\bibitem[Xu et~al\mbox{.}(2017)]%
        {Xu_CVA2017}
\bibfield{author}{\bibinfo{person}{Chi Xu}, \bibinfo{person}{Yasushi Makihara}, \bibinfo{person}{Gakuto Ogi}, \bibinfo{person}{Xiang Li}, \bibinfo{person}{Yasushi Yagi}, {and} \bibinfo{person}{Jianfeng Lu}.} \bibinfo{year}{2017}\natexlab{}.
\newblock \showarticletitle{The OU-ISIR Gait Database Comprising the Large Population Dataset with Age and Performance Evaluation of Age Estimation}.
\newblock \bibinfo{journal}{\emph{IPSJ Trans. on Computer Vision and Applications}} \bibinfo{volume}{9}, \bibinfo{number}{24} (\bibinfo{year}{2017}), \bibinfo{pages}{1--14}.
\newblock


\bibitem[Xue et~al\mbox{.}(2010)]%
        {xue_infrared_2010}
\bibfield{author}{\bibinfo{person}{Zhaojun Xue}, \bibinfo{person}{Dong Ming}, \bibinfo{person}{Wei Song}, \bibinfo{person}{Baikun Wan}, {and} \bibinfo{person}{Shijiu Jin}.} \bibinfo{year}{2010}\natexlab{}.
\newblock \showarticletitle{Infrared Gait Recognition Based on Wavelet Transform and Support Vector Machine}.
\newblock \bibinfo{journal}{\emph{Pattern Recognition}} \bibinfo{volume}{43}, \bibinfo{number}{8} (\bibinfo{date}{Aug.} \bibinfo{year}{2010}), \bibinfo{pages}{2904--2910}.
\newblock
\showISSN{0031-3203}
\href{https://doi.org/10.1016/j.patcog.2010.03.011}{doi:\nolinkurl{10.1016/j.patcog.2010.03.011}}


\bibitem[Yamaguchi et~al\mbox{.}(2018)]%
        {yamaguchi_human_2018}
\bibfield{author}{\bibinfo{person}{Hirozumi Yamaguchi}, \bibinfo{person}{Akihito Hiromori}, {and} \bibinfo{person}{Teruo Higashino}.} \bibinfo{year}{2018}\natexlab{}.
\newblock \showarticletitle{A Human Tracking and Sensing Platform for Enabling Smart City Applications}. In \bibinfo{booktitle}{\emph{Proceedings of the {{Workshop Program}} of the 19th {{International Conference}} on {{Distributed Computing}} and {{Networking}}}}. \bibinfo{publisher}{ACM}, \bibinfo{address}{Varanasi India}, \bibinfo{pages}{1--6}.
\newblock
\showISBNx{978-1-4503-6397-6}
\href{https://doi.org/10.1145/3170521.3170534}{doi:\nolinkurl{10.1145/3170521.3170534}}


\bibitem[Yu et~al\mbox{.}(2006)]%
        {yu_framework_2006}
\bibfield{author}{\bibinfo{person}{Shiqi Yu}, \bibinfo{person}{Daoliang Tan}, {and} \bibinfo{person}{Tieniu Tan}.} \bibinfo{year}{2006}\natexlab{}.
\newblock \showarticletitle{A {{Framework}} for {{Evaluating}} the {{Effect}} of {{View Angle}}, {{Clothing}} and {{Carrying Condition}} on {{Gait Recognition}}}. In \bibinfo{booktitle}{\emph{18th {{International Conference}} on {{Pattern Recognition}} ({{ICPR}}'06)}}, Vol.~\bibinfo{volume}{4}. \bibinfo{pages}{441--444}.
\newblock
\showISSN{1051-4651}
\href{https://doi.org/10.1109/ICPR.2006.67}{doi:\nolinkurl{10.1109/ICPR.2006.67}}


\bibitem[Zeng et~al\mbox{.}(2016)]%
        {zeng_wiwho_2016}
\bibfield{author}{\bibinfo{person}{Yunze Zeng}, \bibinfo{person}{Parth~H. Pathak}, {and} \bibinfo{person}{Prasant Mohapatra}.} \bibinfo{year}{2016}\natexlab{}.
\newblock \showarticletitle{{{WiWho}}: {{WiFi-Based Person Identification}} in {{Smart Spaces}}}. In \bibinfo{booktitle}{\emph{2016 15th {{ACM}}/{{IEEE International Conference}} on {{Information Processing}} in {{Sensor Networks}} ({{IPSN}})}}. \bibinfo{publisher}{IEEE}, \bibinfo{address}{Vienna}, \bibinfo{pages}{1--12}.
\newblock
\showISBNx{978-1-5090-0802-5}
\href{https://doi.org/10.1109/IPSN.2016.7460727}{doi:\nolinkurl{10.1109/IPSN.2016.7460727}}


\bibitem[Zhang et~al\mbox{.}(2022)]%
        {zhang_learning_2022}
\bibfield{author}{\bibinfo{person}{Ziyuan Zhang}, \bibinfo{person}{Luan Tran}, \bibinfo{person}{Feng Liu}, {and} \bibinfo{person}{Xiaoming Liu}.} \bibinfo{year}{2022}\natexlab{}.
\newblock \showarticletitle{On {{Learning Disentangled Representations}} for {{Gait Recognition}}}.
\newblock \bibinfo{journal}{\emph{IEEE Transactions on Pattern Analysis and Machine Intelligence}} \bibinfo{volume}{44}, \bibinfo{number}{1} (\bibinfo{date}{Jan.} \bibinfo{year}{2022}), \bibinfo{pages}{345--360}.
\newblock
\showISSN{0162-8828, 2160-9292, 1939-3539}
\href{https://doi.org/10.1109/TPAMI.2020.2998790}{doi:\nolinkurl{10.1109/TPAMI.2020.2998790}}


\bibitem[Zhao et~al\mbox{.}(2019)]%
        {zhao_mid_2019}
\bibfield{author}{\bibinfo{person}{Peijun Zhao}, \bibinfo{person}{Chris~Xiaoxuan Lu}, \bibinfo{person}{Jianan Wang}, \bibinfo{person}{Changhao Chen}, \bibinfo{person}{Wei Wang}, \bibinfo{person}{Niki Trigoni}, {and} \bibinfo{person}{Andrew Markham}.} \bibinfo{year}{2019}\natexlab{}.
\newblock \showarticletitle{{{mID}}: {{Tracking}} and {{Identifying People}} with {{Millimeter Wave Radar}}}. In \bibinfo{booktitle}{\emph{2019 15th {{International Conference}} on {{Distributed Computing}} in {{Sensor Systems}} ({{DCOSS}})}}. \bibinfo{publisher}{IEEE}, \bibinfo{address}{Santorini Island, Greece}, \bibinfo{pages}{33--40}.
\newblock
\showISBNx{978-1-7281-0570-3}
\href{https://doi.org/10.1109/DCOSS.2019.00028}{doi:\nolinkurl{10.1109/DCOSS.2019.00028}}


\bibitem[Zheng et~al\mbox{.}(2022)]%
        {zheng_gait_2022}
\bibfield{author}{\bibinfo{person}{Jinkai Zheng}, \bibinfo{person}{Xinchen Liu}, \bibinfo{person}{Wu Liu}, \bibinfo{person}{Lingxiao He}, \bibinfo{person}{Chenggang Yan}, {and} \bibinfo{person}{Tao Mei}.} \bibinfo{year}{2022}\natexlab{}.
\newblock \showarticletitle{Gait {{Recognition}} in the {{Wild}} with {{Dense 3D Representations}} and {{A Benchmark}}}. In \bibinfo{booktitle}{\emph{2022 {{IEEE}}/{{CVF Conference}} on {{Computer Vision}} and {{Pattern Recognition}} ({{CVPR}})}}. \bibinfo{publisher}{IEEE}, \bibinfo{address}{New Orleans, LA, USA}, \bibinfo{pages}{20196--20205}.
\newblock
\href{https://doi.org/10.1109/cvpr52688.2022.01959}{doi:\nolinkurl{10.1109/cvpr52688.2022.01959}}


\bibitem[Zivkovic(2004)]%
        {10.5555/1018428.1020644}
\bibfield{author}{\bibinfo{person}{Zoran Zivkovic}.} \bibinfo{year}{2004}\natexlab{}.
\newblock \showarticletitle{Improved Adaptive Gaussian Mixture Model for Background Subtraction}. In \bibinfo{booktitle}{\emph{Proceedings of the Pattern Recognition, 17th International Conference on (ICPR'04) Volume 2 - Volume 02}} \emph{(\bibinfo{series}{ICPR '04})}. \bibinfo{publisher}{IEEE Computer Society}, \bibinfo{address}{USA}, \bibinfo{pages}{28–31}.
\newblock
\showISBNx{0769521282}


\end{thebibliography}

\appendix

\section{Demographics}\label{sec:demographics}
Details on the categorical demographic information that we collected, specifically nationality, gender, hair and skin color, can be found in \cref{tab:demo-categorical}.

\begin{table}[b]
  \caption{Categorical demographics of our participants.}
  \label{tab:demo-categorical}
  \centering
  \small
  \begin{tabular}{@{}lr@{}}
    \toprule
    Nationality & \%\\
\midrule
German & 73.4\% \\
Turkish & 2.5\% \\
Hungarian & 2.0\% \\
Chinese & 2.0\% \\
Peruvian & 1.5\% \\
Greek & 1.5\% \\
Ukrainian & 1.0\% \\
Belarusian & 1.0\% \\
Romanian & 1.0\% \\
Bulgarian & 1.0\% \\
Austrian & 1.0\% \\
Iranian & 1.0\% \\
Ecuadorian & 1.0\% \\
Vietnamese & 1.0\% \\
Brazilian & 1.0\% \\
Lebanese & 0.5\% \\
Kazakhstani & 0.5\% \\
Spanish & 0.5\% \\
Sri Lankan & 0.5\% \\
Polish & 0.5\% \\
Indian & 0.5\% \\
Argentine & 0.5\% \\
Lithuanian & 0.5\% \\
Czech & 0.5\% \\
Togolese & 0.5\% \\
Colombian & 0.5\% \\
Belgian & 0.5\% \\
Mexican & 0.5\% \\
Sudanese & 0.5\% \\
Korean & 0.5\% \\
\midrule
other or n/a & 0.5\% \\
    \bottomrule
  \end{tabular}
  \qquad\qquad
  \begin{tabular}{@{}lr@{}}
  \toprule
Gender & \%\\
\midrule
male & 58.8\% \\
female & 39.7\% \\
\midrule
other or n/a & 1.5\% \\
\bottomrule
\\
\\
\\
\\
  \toprule
Hair Color & \%\\
\midrule
brown & 32.2\% \\
darkblond & 21.1\% \\
black & 13.6\% \\
darkbrown & 13.1\% \\
blond & 9.0\% \\
lightbrown & 8.5\% \\
redbrown & 1.0\% \\
\midrule
other or n/a & 1.5\% \\
\bottomrule
\\
\\
\\
\\
\toprule
Skin Color & \%\\
\midrule
light & 47.2\% \\
light intermediate & 34.2\% \\
dark intermediate & 9.5\% \\
dark & 4.5\% \\
very light & 3.0\% \\
very dark & 0.5\% \\
\midrule
other or n/a & 1.0\% \\
\bottomrule
  \end{tabular}
\end{table}

\section{Additional Results}
This section serves to report the full results from our identity inference benchmark.
For our single and multi-session experiments they can be found in \cref{tab:ss} and \cref{tab:ms}, respectively.
Additionally, the full results of our attribute inference validation can be found in \cref{tab:attributes}.

\begin{table*}
  \caption{Measured accuracies for all sensors, all recognition systems and all perspectives in our single-session experiment.}
  \label{tab:ss}
  \centering
  \small
  \begin{tabular}{@{}llcccc@{}}
    \toprule
    Sensor & Rec. System & Center-Low & Center-High & Left & Right \\
    \midrule
    video & GaitBase \cite{fan_opengait_2023} & $100.0\% \pm 0.0$ & $99.9\% \pm 0.1$ & $99.8\% \pm 0.3$ & $100.0\% \pm 0.0$	\\
	& DeepGaitv2 \cite{fan_opengait_2025} & $99.4\% \pm 0.8$ & $99.9\% \pm 0.0$ & $96.8\% \pm 2.0$ & $99.9\% \pm 0.1$	\\
	& GaitPart \cite{fan_gaitpart_2020} & $99.9\% \pm 0.1$ & $99.9\% \pm 0.0$ & $96.7\% \pm 0.4$ & $99.4\% \pm 0.2$	\\
	& GaitSet \cite{chao_gaitset_2019} & $81.6\% \pm 1.2$ & $76.2\% \pm 2.5$ & $41.2\% \pm 0.7$ & $38.3\% \pm 4.1$	\\
	& GaitGL \cite{lin_gait_2021} & $100.0\% \pm 0.0$ & $99.9\% \pm 0.0$ & $99.9\% \pm 0.0$ & $100.0\% \pm 0.0$	\\
	& GEINet \cite{shiraga_geinet_2016} & $99.6\% \pm 0.1$ & $99.3\% \pm 0.2$ & $98.9\% \pm 0.2$ & $99.5\% \pm 0.1$	\\
    \midrule
    depth & GaitBase \cite{fan_opengait_2023} & $100.0\% \pm 0.0$ & $100.0\% \pm 0.0$ & $100.0\% \pm 0.0$ & $100.0\% \pm 0.0$	\\
	& DeepGaitv2 \cite{fan_opengait_2025} & $100.0\% \pm 0.0$ & $99.9\% \pm 0.1$ & $99.9\% \pm 0.0$ & $99.9\% \pm 0.0$	\\
	& GaitPart \cite{fan_gaitpart_2020} & $100.0\% \pm 0.0$ & $99.8\% \pm 0.0$ & $99.7\% \pm 0.0$ & $99.8\% \pm 0.1$	\\
	& GaitSet \cite{chao_gaitset_2019} & $71.8\% \pm 3.9$ & $74.4\% \pm 3.3$ & $86.5\% \pm 5.1$ & $86.2\% \pm 2.0$	\\
	& GaitGL \cite{lin_gait_2021} & $100.0\% \pm 0.0$ & $99.6\% \pm 0.0$ & $99.9\% \pm 0.0$ & $99.9\% \pm 0.0$	\\
	& GEINet \cite{shiraga_geinet_2016} & $98.8\% \pm 0.3$ & $96.1\% \pm 0.4$ & $98.5\% \pm 0.4$ & $97.8\% \pm 0.3$	\\
    \midrule
    NIR & GaitBase \cite{fan_opengait_2023} & $100.0\% \pm 0.0$ & $100.0\% \pm 0.0$ & $100.0\% \pm 0.0$ & $99.8\% \pm 0.1$	\\
	& DeepGaitv2 \cite{fan_opengait_2025} & $100.0\% \pm 0.0$ & $100.0\% \pm 0.0$ & $99.9\% \pm 0.1$ & $99.6\% \pm 0.0$	\\
	& GaitPart \cite{fan_gaitpart_2020} & $100.0\% \pm 0.0$ & $100.0\% \pm 0.0$ & $99.7\% \pm 0.1$ & $99.3\% \pm 0.1$	\\
	& GaitSet \cite{chao_gaitset_2019} & $96.0\% \pm 0.4$ & $95.0\% \pm 0.9$ & $81.2\% \pm 5.1$ & $69.9\% \pm 3.3$	\\
	& GaitGL \cite{lin_gait_2021} & $100.0\% \pm 0.0$ & $100.0\% \pm 0.0$ & $99.7\% \pm 0.0$ & $98.6\% \pm 0.1$	\\
	& GEINet \cite{shiraga_geinet_2016} & $99.0\% \pm 0.2$ & $98.4\% \pm 0.2$ & $91.5\% \pm 1.4$ & $83.2\% \pm 1.2$	\\
    \midrule
    LWIR & GaitBase \cite{fan_opengait_2023} & $100.0\% \pm 0.0$ & $93.3\% \pm 0.4$ & $99.7\% \pm 0.1$ & $99.6\% \pm 0.0$	\\
	& DeepGaitv2 \cite{fan_opengait_2025} & $99.3\% \pm 0.2$ & $7.3\% \pm 4.1$ & $99.2\% \pm 0.4$ & $99.3\% \pm 0.1$	\\
	& GaitPart \cite{fan_gaitpart_2020} & $99.7\% \pm 0.0$ & $66.3\% \pm 1.0$ & $99.0\% \pm 0.1$ & $98.8\% \pm 0.1$	\\
	& GaitSet \cite{chao_gaitset_2019} & $100.0\% \pm 0.0$ & $77.4\% \pm 1.7$ & $99.1\% \pm 0.2$ & $99.1\% \pm 0.0$	\\
	& GaitGL \cite{lin_gait_2021} & $100.0\% \pm 0.0$ & $76.4\% \pm 0.6$ & $99.1\% \pm 0.1$ & $99.0\% \pm 0.1$	\\
	& GEINet \cite{shiraga_geinet_2016} & $98.1\% \pm 0.3$ & $98.4\% \pm 0.2$ & $92.4\% \pm 0.4$ & $89.7\% \pm 0.7$	\\
    \midrule
    radar & mmGaitNet \cite{meng_gait_2020} & $23.3\% \pm 16.9$ & --- & $57.6\% \pm 22.5$ & $24.0\% \pm 24.2$	\\
	   & SRPNet \cite{cheng_person_2022} & $83.4\% \pm 3.8$ & --- & $18.9\% \pm 1.8$ & $31.4\% \pm 1.2$	\\
          & mID \cite{zhao_mid_2019} & $2.1\% \pm 0.3$ & --- & $0.7\% \pm 0.5$ & $0.8\% \pm 0.3$	\\
    \midrule
    lidar & LidarGait \cite{shen_lidargait_2023} & $100.0\% \pm 0.0$ & $100.0\% \pm 0.0$ & $99.7\% \pm 0.0$ & $99.8\% \pm 0.0$	\\
	& GEINet \cite{shiraga_geinet_2016} & $94.2\% \pm 1.5$ & $98.3\% \pm 0.3$ & $47.1\% \pm 2.1$ & $44.7\% \pm 1.9$	\\
	& LidarGait++ \cite{shen_lidargait_2025} & $91.5\% \pm 3.2$ & $95.3\% \pm 1.7$ & $76.7\% \pm 5.4$ & $85.3\% \pm 1.6$	\\
    \midrule
    CSI & BFId \cite{todt_BFId_2025} & $83.8\% \pm 2.7$ & $86.0\% \pm 1.5$ & $85.8\% \pm 1.3$ & $87.9\% \pm 0.9$	\\
	& LW-WiID \cite{cao_lightweight_2021} & $99.6\% \pm 0.1$ & $97.9\% \pm 0.9$ & $98.7\% \pm 0.1$ & $98.4\% \pm 0.1$	\\
	& CAUTION \cite{wang_caution_2022} & $58.6\% \pm 2.0$ & $63.9\% \pm 2.3$ & $55.4\% \pm 2.3$ & $79.1\% \pm 0.3$	\\
	& FreeSense \cite{xin_freesense_2016} & $35.0\%$ & $31.3\%$ & $18.7\%$ & $30.3\%$ \\
    \midrule
    BFI & BFId \cite{todt_BFId_2025} & $99.3\% \pm 0.5$ & $98.5\% \pm 0.4$ & $99.0\% \pm 0.5$ & $97.2\% \pm 0.4$	\\
    \bottomrule
  \end{tabular}
\end{table*}

\begin{table*}
  \caption{Measured accuracies for all sensors, all recognition systems and all perspectives in our multi-session experiment.}
  \label{tab:ms}
  \centering
  \small
  \begin{tabular}{@{}llcccc@{}}
    \toprule
    Sensor & Rec. System & Center-Low & Center-High & Left & Right \\
    \midrule
    video & GaitBase \cite{fan_opengait_2023} & $58.6\% \pm 1.3$ & $59.3\% \pm 1.0$ & $73.2\% \pm 0.4$ & $64.1\% \pm 1.8$	\\
	& DeepGaitv2 \cite{fan_opengait_2025} & $48.7\% \pm 1.8$ & $48.1\% \pm 1.3$ & $41.1\% \pm 2.1$ & $42.1\% \pm 3.5$	\\
	& GaitPart \cite{fan_gaitpart_2020} & $46.0\% \pm 0.3$ & $45.4\% \pm 0.4$ & $43.0\% \pm 1.5$ & $48.7\% \pm 0.6$	\\
	& GaitSet \cite{chao_gaitset_2019} & $11.5\% \pm 0.6$ & $8.2\% \pm 0.1$ & $4.6\% \pm 0.2$ & $5.0\% \pm 0.3$	\\
	& GaitGL \cite{lin_gait_2021} & $48.9\% \pm 0.5$ & $56.3\% \pm 0.2$ & $50.1\% \pm 0.4$ & $52.9\% \pm 0.7$	\\
	& GEINet \cite{shiraga_geinet_2016} & $35.6\% \pm 1.5$ & $27.7\% \pm 1.0$ & $45.6\% \pm 1.1$ & $44.4\% \pm 1.1$	\\
    \midrule
    depth & GaitBase \cite{fan_opengait_2023} & $76.7\% \pm 0.5$ & $72.5\% \pm 0.7$ & $65.0\% \pm 0.4$ & $65.3\% \pm 1.4$	\\
	& DeepGaitv2 \cite{fan_opengait_2025} & $67.1\% \pm 0.6$ & $60.0\% \pm 2.6$ & $48.9\% \pm 1.0$ & $51.5\% \pm 3.6$	\\
	& GaitPart \cite{fan_gaitpart_2020} & $65.0\% \pm 0.7$ & $54.2\% \pm 0.7$ & $47.9\% \pm 0.4$ & $57.1\% \pm 0.9$	\\
	& GaitSet \cite{chao_gaitset_2019} & $30.9\% \pm 1.0$ & $27.4\% \pm 1.4$ & $30.5\% \pm 2.5$ & $29.7\% \pm 3.1$	\\
	& GaitGL \cite{lin_gait_2021} & $67.1\% \pm 0.7$ & $60.4\% \pm 1.0$ & $47.6\% \pm 0.3$ & $57.4\% \pm 1.0$	\\
	& GEINet \cite{shiraga_geinet_2016} & $48.4\% \pm 2.1$ & $45.8\% \pm 1.8$ & $43.8\% \pm 1.5$ & $41.1\% \pm 0.8$	\\
    \midrule
    NIR & GaitBase \cite{fan_opengait_2023} & $50.7\% \pm 0.9$ & $47.6\% \pm 0.3$ & $47.9\% \pm 0.5$ & $40.6\% \pm 0.5$	\\
	& DeepGaitv2 \cite{fan_opengait_2025} & $53.8\% \pm 1.2$ & $50.4\% \pm 2.2$ & $48.1\% \pm 0.4$ & $41.4\% \pm 1.1$	\\
	& GaitPart \cite{fan_gaitpart_2020} & $47.3\% \pm 0.7$ & $43.3\% \pm 0.8$ & $43.8\% \pm 0.4$ & $32.9\% \pm 0.7$	\\
	& GaitSet \cite{chao_gaitset_2019} & $24.8\% \pm 0.3$ & $25.5\% \pm 1.4$ & $19.5\% \pm 1.2$ & $14.9\% \pm 1.5$	\\
	& GaitGL \cite{lin_gait_2021} & $45.9\% \pm 0.4$ & $42.7\% \pm 0.2$ & $44.6\% \pm 0.7$ & $39.1\% \pm 0.8$	\\
	& GEINet \cite{shiraga_geinet_2016} & $33.4\% \pm 1.7$ & $32.4\% \pm 0.8$ & $29.3\% \pm 2.0$ & $22.5\% \pm 1.0$	\\
    \midrule
    LWIR & GaitBase \cite{fan_opengait_2023} & $71.7\% \pm 0.6$ & $45.5\% \pm 0.8$ & $65.2\% \pm 0.9$ & $62.6\% \pm 0.4$	\\
	& DeepGaitv2 \cite{fan_opengait_2025} & $55.3\% \pm 5.7$ & $7.9\% \pm 4.4$ & $59.4\% \pm 0.8$ & $59.1\% \pm 0.7$	\\
	& GaitPart \cite{fan_gaitpart_2020} & $53.4\% \pm 0.5$ & $22.2\% \pm 0.6$ & $49.4\% \pm 0.8$ & $49.7\% \pm 0.9$	\\
	& GaitSet \cite{chao_gaitset_2019} & $62.7\% \pm 0.2$ & $26.2\% \pm 0.6$ & $55.2\% \pm 1.0$ & $49.8\% \pm 0.6$	\\
	& GaitGL \cite{lin_gait_2021} & $58.5\% \pm 0.6$ & $22.8\% \pm 0.5$ & $52.3\% \pm 0.9$ & $51.1\% \pm 0.5$	\\
	& GEINet \cite{shiraga_geinet_2016} & $50.5\% \pm 1.4$ & $50.2\% \pm 1.4$ & $35.0\% \pm 1.1$ & $32.0\% \pm 1.4$	\\
    \midrule
    radar & mmGaitNet \cite{meng_gait_2020} & $4.0\% \pm 1.5$ & --- & $4.9\% \pm 0.7$ & $3.4\% \pm 1.5$	\\
	& SRPNet \cite{cheng_person_2022} & $1.4\% \pm 0.4$ & --- & $9.1\% \pm 0.4$ & $7.6\% \pm 0.7$	\\
        & mID \cite{zhao_mid_2019} & $2.3\% \pm 0.7$ & --- & $2.3\% \pm 1.9$ & $1.2\% \pm 1.4$	\\
    \midrule
    lidar & LidarGait \cite{shen_lidargait_2023} & $74.5\% \pm 0.5$ & $77.4\% \pm 0.1$ & $74.3\% \pm 0.6$ & $78.4\% \pm 0.6$	\\
	& GEINet \cite{shiraga_geinet_2016} & $52.9\% \pm 2.6$ & $61.5\% \pm 3.7$ & $28.3\% \pm 0.6$ & $28.6\% \pm 0.7$	\\
	& LidarGait++ \cite{shen_lidargait_2025} & $41.9\% \pm 2.0$ & $50.4\% \pm 3.3$ & $35.1\% \pm 1.6$ & $38.7\% \pm 1.5$	\\
    \midrule
    CSI & BFId \cite{todt_BFId_2025} & $1.9\% \pm 0.2$ & $1.6\% \pm 0.8$ & $1.7\% \pm 0.5$ & $0.2\% \pm 0.3$	\\
	& LW-WiID \cite{cao_lightweight_2021} & $0.4\% \pm 0.4$ & $1.0\% \pm 0.5$ & $2.1\% \pm 0.6$ & $0.1\% \pm 0.1$	\\
	& CAUTION \cite{wang_caution_2022} & $0.4\% \pm 0.4$ & $1.0\% \pm 0.5$ & $2.1\% \pm 0.6$ & $0.1\% \pm 0.1$	\\
	& FreeSense \cite{xin_freesense_2016} & $1.8\%$ & $0.6\%$ & $1.4\%$ & $1.4\%$ \\
    \midrule
    BFI & BFId \cite{todt_BFId_2025} & $0.7\% \pm 0.4$ & $1.2\% \pm 0.7$ & $1.1\% \pm 0.8$ & $0.7\% \pm 1.1$	\\
    \bottomrule
  \end{tabular}
\end{table*}

\begin{table*}
  \caption{Measured balanced accuracies for all sensors, all tested attributes, and all perspectives in our attribute recognition experiment.}
  \label{tab:attributes}
  \centering
  \small
  \begin{tabular}{@{}llcccc@{}}
    \toprule
    Sensor & Attribute & Center-Low & Center-High & Left & Right \\
    \midrule
video & gender & $91.9\% \pm 0.2$ & $94.4\% \pm 0.2$ & $97.5\% \pm 0.2$ & $97.3\% \pm 0.3$	\\
	& age & $24.6\% \pm 1.7$ & $21.9\% \pm 2.7$ & $18.6\% \pm 1.0$ & $19.3\% \pm 0.4$	\\
	& weight & $34.0\% \pm 2.1$ & $26.0\% \pm 3.3$ & $23.7\% \pm 3.9$ & $26.4\% \pm 1.7$	\\
	& height & $36.9\% \pm 2.3$ & $29.0\% \pm 3.6$ & $24.9\% \pm 3.7$ & $32.8\% \pm 5.4$	\\
	\midrule
depth & gender & $99.2\% \pm 0.1$ & $94.8\% \pm 0.2$ & $98.1\% \pm 0.3$ & $99.7\% \pm 0.2$	\\
	& age & $21.4\% \pm 0.7$ & $15.9\% \pm 1.1$ & $21.8\% \pm 1.3$ & $19.4\% \pm 1.3$	\\
	& weight & $45.0\% \pm 1.7$ & $37.4\% \pm 6.0$ & $35.8\% \pm 0.9$ & $38.5\% \pm 1.4$	\\
	& height & $44.8\% \pm 0.9$ & $47.7\% \pm 5.8$ & $46.7\% \pm 1.1$ & $52.2\% \pm 1.7$	\\
	\midrule
NIR & gender & $96.6\% \pm 0.3$ & $98.4\% \pm 0.3$ & $97.1\% \pm 0.2$ & $97.2\% \pm 0.3$	\\
	& age & $21.3\% \pm 0.5$ & $23.1\% \pm 0.3$ & $19.3\% \pm 0.5$ & $18.9\% \pm 0.4$	\\
	& weight & $34.0\% \pm 1.0$ & $31.6\% \pm 0.8$ & $35.4\% \pm 0.9$ & $38.9\% \pm 0.7$	\\
	& height & $52.5\% \pm 0.7$ & $48.3\% \pm 0.7$ & $43.8\% \pm 1.0$ & $44.9\% \pm 0.8$	\\
	\midrule
LWIR & gender & $93.4\% \pm 0.1$ & $95.2\% \pm 0.3$ & $99.6\% \pm 0.0$ & $98.0\% \pm 0.1$	\\
	& age & $24.3\% \pm 0.6$ & $22.8\% \pm 1.2$ & $23.1\% \pm 1.0$ & $21.2\% \pm 0.5$	\\
	& weight & $40.7\% \pm 0.7$ & $33.0\% \pm 1.1$ & $40.4\% \pm 0.7$ & $41.6\% \pm 0.6$	\\
	& height & $57.7\% \pm 0.2$ & $58.4\% \pm 0.5$ & $45.8\% \pm 0.7$ & $51.8\% \pm 0.5$	\\
	\midrule
lidar & gender & $95.5\% \pm 0.5$ & $98.6\% \pm 0.3$ & $99.5\% \pm 0.1$ & $99.7\% \pm 0.1$	\\
	& age & $24.6\% \pm 0.4$ & $24.9\% \pm 0.6$ & $21.5\% \pm 0.9$ & $22.6\% \pm 0.3$	\\
	& weight & $47.4\% \pm 0.9$ & $48.5\% \pm 0.9$ & $35.6\% \pm 1.8$ & $43.4\% \pm 0.4$	\\
	& height & $60.0\% \pm 0.6$ & $60.4\% \pm 1.0$ & $47.2\% \pm 2.0$ & $54.6\% \pm 0.0$	\\
	\midrule
radar & gender & $48.4\% \pm 2.9$ & --- & $67.9\% \pm 1.0$ & $70.8\% \pm 1.0$	\\
	& age & $32.0\% \pm 5.1$ & --- & $20.9\% \pm 1.5$ & $20.0\% \pm 1.0$	\\
	& weight & $21.9\% \pm 2.4$ & --- & $31.6\% \pm 1.2$ & $28.4\% \pm 0.7$	\\
	& height & $26.4\% \pm 1.8$ & --- & $37.5\% \pm 9.2$ & $35.6\% \pm 1.3$	\\
	\midrule
CSI & gender & $72.9\% \pm 7.5$ & $78.8\% \pm 5.7$ & $63.2\% \pm 5.9$ & $55.5\% \pm 2.2$	\\
	& age & $22.3\% \pm 4.1$ & $17.5\% \pm 3.6$ & $16.9\% \pm 1.1$ & $21.3\% \pm 1.6$	\\
	& weight & $26.2\% \pm 4.1$ & $21.3\% \pm 3.9$ & $21.9\% \pm 3.2$ & $21.2\% \pm 1.6$	\\
	& height & $52.4\% \pm 5.0$ & $53.6\% \pm 5.1$ & $24.7\% \pm 2.8$ & $20.4\% \pm 2.0$	\\
    \midrule
BFI & gender & $60.2\% \pm 3.3$ & $51.1\% \pm 4.9$ & $59.0\% \pm 8.8$ & $44.3\% \pm 2.0$	\\
	& age & $18.4\% \pm 4.3$ & $21.4\% \pm 5.0$ & $21.9\% \pm 5.2$ & $19.7\% \pm 2.5$	\\
	& weight & $16.8\% \pm 4.8$ & $16.4\% \pm 2.8$ & $19.0\% \pm 2.1$ & $24.9\% \pm 2.8$	\\
	& height & $19.7\% \pm 3.4$ & $17.9\% \pm 6.6$ & $19.3\% \pm 3.4$ & $17.5\% \pm 3.0$	\\
    \bottomrule
  \end{tabular}
\end{table*}

\section{Dataset Release}\label{sec:datasetrelease}
Researchers interested in obtaining the dataset will need to sign a release agreement and send it to the dataset administrators.
Individuals that want to obtain the dataset and are not currently or have recently been employed by a research institution will also need to briefly outline their research intention.
This was done to address the trade-off between usability of our dataset (incl. reproducibility of our results) and the sensitivity of the biometric data of our study participants, in coordination with our data protection officer.
You can find the dataset release agreement below:

\paragraph{Release Agreement}
\begin{enumerate}
    \item The dataset may not be, entirely or partially, distributed, published, copied, or disseminated in any form whatsoever, whether for profit or not. This includes further distributing, copying, or disseminating to a different facility or organizational unit within the requesting university, organization, or company. All users of the dataset must sign this document, send it to the dataset administrators, and be granted access by them.
    \item The dataset may only be used for academic research. Use, either entirely or partially, for commercial purposes is strictly prohibited.
    \item The dataset may not be modified.
    \item When including more than 10 still frames or a clip from the dataset in publications, researchers must obtain approval in writing from the dataset administrators. Applicants must blackout the faces of individuals in the frames and/or clip. In no case should frames be included in publications in such a way that could cause the original subject embarrassment or mental anguish.
    \item Any publication that reports on research the uses the dataset must acknowledge its use by including a citation to "\textit{Julian Todt, Felix Morsbach, Philip Dissert, and Thorsten Strufe. 2026. MultiGait:  Multi-Sensor Multi-Perspective Multi-Session Biometric Inference Benchmark and its Dataset.}"
    \item A copy of any publication that reports on research the uses the dataset should be sent to the dataset administrators.
    \item Researchers agree to indemnify, defend, and hold harmless Karlsruhe Institute of Technology and its officers, employees, and agents, individually and collectively, from any and all losses, expenses, damages, demands, and/or claims based upon any injury or damage (real or alleged) related to, and shall pay all damages, claims, judgments or expenses from, the researchers' use of the dataset.
\end{enumerate}

\paragraph{Consent}
The researcher agrees to the above usage rules for the MultiGait dataset.
If the requesting researcher is a student or postdoctoral researcher without a permanent contract at a research institute, the signature of a permanent member (e.g., professor) is additionally required.

\end{document}